\documentclass[nofootinbib,superscriptaddress,eqsecnum,tightenlines,11pt,notitlepage,longbibliography]{revtex4-1}
\usepackage[linktocpage]{hyperref}
\usepackage{graphicx}
\usepackage{amsmath,amssymb,amsfonts,amsthm,stmaryrd,mathtools,mathtools}
\usepackage{yfonts}
\usepackage{multirow,bigdelim}
\usepackage{dsfont}
\usepackage{color}
\usepackage{graphicx,color,epstopdf,epsfig,psfrag}
\usepackage[dvipsnames]{xcolor}
\usepackage{ mathrsfs }

\def\be{\begin{equation}}
\def\ee{\end{equation}}
\def\ba{\begin{eqnarray}}
\def\ea{\end{eqnarray}}

\def\SU{\text{SU}}

\newcommand{\bd}{\mathbf d}
\newcommand\nn{\nonumber}
\newcommand\sss{\scriptstyle}
\newcommand{\q}{\quad}
\newcommand{\RN}[1]{%
 \textup{\uppercase\expandafter{\romannumeral#1}}%
}

\hypersetup{
    colorlinks=true,       % false: boxed links; true: colored links
    linkcolor=MidnightBlue,          % color of internal links (change box color with linkbordercolor)
    citecolor=BrickRed,        % color of links to bibliography
    filecolor=BrickRed,      % color of file links
    urlcolor=BrickRed           % color of external links
}

\begin{document}

\title{Towards a phase diagram for spin foams}

\author{Clement Delcamp}\email{cdelcampATperimeterinstituteDOTca}
\affiliation{Perimeter Institute for Theoretical Physics,\\ 31 Caroline Street North, Waterloo, Ontario, Canada N2L 2Y5}
\affiliation{Department of Physics $\&$ Astronomy and Guelph-Waterloo Physics Institute \\  University of Waterloo, Waterloo, Ontario N2L 3G1, Canada}
\author{Bianca Dittrich}\email{bdittrichATperimeterinstituteDOTca} 
\affiliation{Perimeter Institute for Theoretical Physics,\\ 31 Caroline Street North, Waterloo, Ontario, Canada N2L 2Y5}

\begin{abstract}
	{ One of the most pressing issues for loop quantum gravity and spin foams is the construction of the continuum limit.
		In this paper, we propose a systematic coarse--graining scheme for three--dimensional lattice gauge models including spin foams. This scheme is based on the concept of decorated tensor networks, which have been introduced recently. Here we develop an algorithm applicable to gauge theories with non--Abelian groups, which for the first time allows for the application of tensor network coarse--graining techniques to proper spin foams.  The procedure deals efficiently with the large redundancy of degrees of freedom resulting from gauge symmetry. 
		The algorithm is applied to 3D spin foams defined on a cubical lattice which, in contrast to a proper triangulation, allows for non--trivial simplicity constraints. This mimics the construction of spin foams for 4D gravity.  For lattice gauge models based on a finite group we  use the algorithm to obtain phase diagrams, encoding the continuum limit of a wide range of these models. We find  phase transitions for various families of models carrying non--trivial simplicity constraints. }
\end{abstract}

\maketitle

\renewcommand{\baselinestretch}{0.90}\normalsize
\small
\tableofcontents
\renewcommand{\baselinestretch}{1.0}\normalsize
\normalsize

\section{Introduction}

Spin foam models propose a covariant, background independent and non-perturbative quantization of general relativity \cite{foams3,reisenbergerSF,Barrett-Crane,eprl,fk,bf-theory}, based on insights from loop quantum gravity \cite{foams1,thomasbook}. Being non--perturbative, the models rely on the definition of a path integral regularized via an auxiliary discretization.  Since this discretization breaks the  diffeomorphism invariance \cite{dittrich08,dittrich-broken} underlying general relativity, one cannot take this discrete formulation as fundamental. One rather has to construct a continuum limit, in which one hopes to restore diffeomorphism symmetry \cite{improved1,review14}. Diffeomorphism invariance is a very powerful symmetry, whose restoration should also ensure the independence of the chosen discretization \cite{improved2,dittrich12a}. Requiring this symmetry also resolves ambiguities in the construction of the discrete models \cite{measure1,BahrSteinhausPRL}.

Most importantly, only if diffeomorphism symmetry is correctly implemented into the path integral, can one hope that the path integral does act as a projector on the so--called physical states  \cite{HartleHalliwell}, that is states that satisfy the Wheeler--DeWitt (constraint) equation which encodes the dynamics of (quantum) general relativity.  The breaking of diffeomorphism symmetry by the discretization leads also to a violation of the constraints determining the Wheeler--DeWitt equation \cite{dittrich-broken, hoehn1}. The construction of the continuum limit in diffeomorphism invariant systems is therefore tantamount to solving the dynamics \cite{dittrich12a,timeevol,review14} as is illustrated in \cite{improved2}.

Unfortunately not much is known about the behaviour of spin foam models in the `many body' or `thermodynamic' regime, which is the regime of interest for the construction of the continuum limit. This is mainly due to the highly complex amplitudes encoding the non--linear dynamics of general relativity. Additionally, a conceptual understanding of renormalization and coarse--graining -- which is used to construct the continuum limit -- needed to be developed for background independent systems. But here much progress has been achieved \cite{dittcyl,timeevol, BahrSFR,review14}, involving as a tool the  inductive limit construction for Hilbert spaces. The latter is used in loop quantum gravity to define a continuum Hilbert space as an inductive limit of a family of discrete Hilbert spaces \cite{thomasbook,ashtekar-isham}. A crucial role is then played by so--called embedding maps which relates coarse configurations to physically equivalent configurations on a finer discretization. This is done by putting the additional degrees of freedom on the finer discretization into a vacuum state. In the constructions of the (so--called kinematical) continuum Hilbert spaces for LQG \cite{ashtekar-isham,ashtekar-lewan2,bf-vacuum,bfform3} this vacuum state is pre--chosen, without input from the dynamics of the system, and is therefore not a physical state.\footnote{The exception is 3D quantum gravity in which the vacuum states used in \cite{bf-vacuum, bfform3} and \cite{DGtqft} turn out to be physical for a dynamics without and with a cosmological constant respectively.} A fundamental goal of the program to  construct the continuum limit of spin foams is the extraction of a {\it physical} embedding map, that would describe a physical vacuum state. 

This is an ambitious aim and can only be hoped to be achieved in an approximation scheme.  Luckily the coarse--graining schemes developed in \cite{dittcyl, timeevol, review14}, which aim at an understanding of the continuum limit, provide such an approximation scheme. For instance, to lowest order, one aims at the understanding of the sector of the theory describing homogeneous states. Furthermore, these coarse--graining schemes can be realized by concrete algorithms, known as Tensor Network coarse--graining algorithms \cite{levin, gu-wen, vidal-evenbly, evenbly2}. Such techniques produce a recursive coarse--graining of the amplitudes of a given model, resulting in effective amplitudes for large building blocks or rather for building blocks carrying a very refined boundary. But one expresses these amplitudes only as functions of coarse boundary data, which is determined by the order of the truncation. The relevant observables which emerge as coarse boundary data are determined by the dynamics of the system. In fact this information is encoded in (now dynamically determined) embedding maps, which can be extracted from the coarse--graining algorithm.

However, tensor network algorithms have been so far mostly been developed for 2D systems, whereas we are ultimately interested in 4D spin foams.\footnote{There are a few exceptions, {\it e.g.} \cite{xie2012}, but to the best knowledge of the authors none that has been developed for gauge systems in higher dimensions yet.}
Additionally, given the high complexity of spin foams, which can be seen as more complicated versions of lattice gauge theories \cite{holonomy1}, there is not much known about the behaviour of these models in the continuum limit. In particular, it is not known which ingredients in the constructions are relevant or not. Here a key role is played by simplicity constraints \cite{dittrich-ryan08,ryanSimp,constraints2,foams3,Barrett-Crane,eprl,fk,BarOr,Maite}, in the imposition of which the models differ. An important open question is how these simplicity constraints behave under coarse--graining. Furthermore we wish to identify relevant couplings, parametrizing the simplicity constraints, and the various ways of implementing them.

Given this state of affairs, a two--aimed program has been developed: one aim is to develop (tensor network) coarse--graining algorithms applicable to spin foams, another one is to understand the behaviour of the simplicity constraints under coarse--graining. To this end a family of 2D analogue models, called spin net models, has been constructed, which do carry a notion of simplicity constraints \cite{eckert1,eckert2,merce,qgroup,BCspinnets} . Tensor network techniques have been developed that can deal with the symmetry structure of these models and the phase diagram for various models has been constructed in \cite{eckert1,merce,qgroup,SteinhausMatter,BCspinnets}. For instance \cite{qgroup} revealed a rich phase diagram for (quantum group based) so--called spin net models. \cite{BCspinnets} considers spin nets which have algebraically the same simplicity constraints as the full gravitational spin foam models (again based on a quantum group), dealing successfully -- via a redesign of the algorithm -- with the challenging task of increasing computational requirements.

The next step is to consider proper spin foam models which require tensor network algorithms applicable to lattice gauge models. In \cite{decorated} was proposed the first tensor network coarse--graining algorithm, dealing with the problem of redundancy in variables, arising from the gauge symmetry.\footnote{Systems with gauge symmetry have been addressed in a framework where tensor networks are used to construct states for a variational  procedure \cite{vidalgauge,osborn2,tagglia,schwingermodel}.} This required the introduction of so--called Decorated Tensor Networks. The algorithm proposed in \cite{decorated} has been tested successfully for a 3D Abelian (finite group) gauge model. Such Abelian gauge models do however not allow for (sufficiently interesting) simplicity constraints, these rather require non--Abelian groups.

In this work we propose a first (Decorated) Tensor Network coarse--graining algorithm, which can be applied to (3D) lattice gauge models based on non--Abelian groups. Here we will deal with lattice gauge models based on finite groups \cite{finitesf}, as the computational resources needed, scale with the size of the group. Lie groups can be dealt with in principle, either by implementing a further truncation, or by using also semi--analytical tools, to which the Decorated Tensor Network schemes are amenable \cite{decorated}.  An alternative is to involve quantum groups at root of unity, which are finite and describe, or are conjectured to describe,  Euclidean gravity with a positive cosmological constant \cite{qgroupmodels1,qgroupmodels2,qgroupmodels3,qgroupmodels4,qgroupmodels5,qgroupmodels6,qgroupmodels7}. This strategy has also been used in \cite{merce,qgroup,SteinhausMatter,BCspinnets} for spin nets.

We will test this Decorated Tensor Network algorithm  on lattice gauge models based on a finite non--Abelian group, namely the permutation group ${\cal S}_3$. Since this group allows for (non--trivial) simplicity constraints, we present here the first coarse--graining results for such models. Furthermore, as mentioned above, spin net models have been constructed such that these capture key ingredients of spin foams. The hope was that the behaviour under coarse--graining for spin nets and spin foams is similar. We will compare the phase diagrams which we obtain for the spin foam models  to the phase diagrams obtained for spin nets in \cite{merce}. 

Tensor network algorithms come now in a wide variety, {\it e.g.} \cite{levin, vidal-evenbly, gu-wen, beijing, looptnr}. Given that this is a first tensor network algorithm for 3D non--Abelian gauge theories, we will use a version which requires the least computational resources, and is sufficiently reliable to identify the phase structure and thus possible phase transitions.  Algorithms, implementing a so--called (short range) entanglement filtering procedure \cite{vidal-evenbly, evenbly2, looptnr} are better suited to characterize the conformal theories arising at (second order) phase transitions. But these more involved algorithms require larger computational resources and have been developed, so far, only for 2D systems. 

The main question, we are interested in, is to identify possibly new phases, which might only arise with spin foam models, but not within the standard form of lattice gauge theories. These phases often correspond to topological field theories, {\it e.g.} BF theory.  New phases could correspond to new topological field theories, from which one can design new Hilbert space representations based on vacua defined by the topological field theories \cite{timeevol}. With a phase diagram we can also identify potential phase transitions, which are needed to describe a continuum limit with propagating degrees of freedom.  
Another question is to confirm or not the similarity of the phase diagrams for spin nets and spin foams.

As we will discuss in detail, the main challenge in the design of the algorithm is due to the non--Abelian structure of the gauge group. In a certain sense gauge invariance is not preserved under coarse--graining for non--Abelian gauge groups.  The reason is that under coarse--graining effective degrees of freedom appear which describe a violation of the Gau\ss~constraints (or torsion), these constraints implementing gauge invariance for wave functions \cite{eteraCG3,bfform2}. This happens because the definition of `effective' Gau\ss~constraints for larger regions require, for non--Abelian  groups, a parallel transport involving a connection. In the presence of curvature this leads to a deformation\footnote{One expects that in some cases, in particular for the dynamics of 3D gravity with cosmological constant, this can be indeed described by a deformation of the gauge group to a quantum group \cite{lqg-lambda3}.} of the Gau\ss~constraints \cite{eteraCG2}, which means that the original form of gauge invariance is not satisfied anymore.  This effect means that the coarse--graining flow could take the system out of the space of (gauge invariant) lattice gauge models. Thus we would have in principle to allow for a much larger space of models which can accommodate for torsion degrees of freedom. We will here however assume\footnote{For lattice Yang--Mills systems this is part of the confinement conjecture.} that torsion degrees of freedom are not--relevant for the dynamics and project the amplitude to a gauge invariant form. We will comment in the discussion on frameworks in which the torsion degrees of freedom can be taken into account, and for which theories one would need such a framework.

~\\
This paper is organized as follows. In section \ref{sec_latticegauge} we briefly summarize the structure of lattice gauge models. We will in particular detail the structure of spin foam models as a constrained BF theory and how they can be reformulated as (decorated) tensor networks. We then review the Decorated Tensor Network coarse--graining algorithm for 2D systems in section \ref{sec_DTNR}. The next section \ref{sec_3dalgo} details the coarse--graining algorithm for 3D lattice gauge models with a non--Abelian structure group. 
 In section \ref{sec_results} we describe the space of models to which we apply the coarse--graining algorithm and the details of their parametrization. We also present the results of the coarse--graining algorithms. We close with a discussion and outlook in section \ref{discussion}. The appendices collect technical material needed for the development of the coarse--graining algorithm and its application to models with structure group ${\cal S}_3$.

\section{Lattice gauge models}\label{sec_latticegauge}

\subsection{Definitions}

Here we will shortly explain the class of models we will be considering and their connection to gravity. A more detailed introduction, highlighting the connections to statistical models, can be found in \cite{finitesf}. We will in particular consider first order formulations of gravity whose definitions involve connection variables. In order to arrive at a well--defined path integral, one discretizes the underlying action for these systems. This yields spin foam models, which in their basic kinematical inputs, have a lot in common with lattice gauge theories. We will refer to both spin foam models and lattice gauge theories as lattice gauge models.

In the following we will review shortly the construction of such lattice gauge models. We begin with the action for the corresponding continuum theory.

Let ${\cal G}$ be a compact Lie group and $\mathfrak{g}$ the corresponding Lie algebra. Given a four--dimensional manifold ${\cal M}$, we consider the Plebanski--action \cite{plebanski}: 
\be\label{plebaction}
S_P[B,\omega]\,=\, \int_\mathcal{M} \text{tr}\big(B \wedge F(\omega)\big)  + \phi_{IJ} \, B^I \wedge B^J  \; .
\ee 
Here  $B$ denotes a $\mathfrak{g}$-valued ($d-2$)-form and $F$ the curvature of the $\mathfrak{g}$-connection $\omega$.  The Lagrange multipliers $\phi$ (carrying two Lie--algebra indices) impose the so--called simplicity constraints. 
The simplicity constraints ensure that for the four--dimensional case $d=4$, the two--form $B$ is actually `simple', that is of the form\footnote{This is the solution for the so--called gravitational sector for the simplicity constraints, another sector is given by $B=\pm e\wedge e$.} $B=\pm \star e\wedge e$, where $e$ are the tetrad variables and $\star$ indicates a dualization acting on the Lie algebra indices.
For $d=4$ and ${\cal G}$ given by $SO(4)$ this action (\ref{plebaction}) describes general relativity in a first order formulation. The first term in (\ref{plebaction})
\ba\label{BFaction}
S_{BF}[B,\omega]\,=\, \int_\mathcal{M} \text{tr}\big(B \wedge F(\omega)\big) 
\ea
describes a topological theory, known as BF theory \cite{horowitz}. The discretization and related path integral  of this topological theory can be constructed without the many ambiguities coming with the discretization of interacting theories. This is therefore the starting point for spin foam models. In a second step one has however to implement the simplicity constraints, which so--far is a process with many ambiguities \cite{dittrich-ryan08,ryanSimp,constraints2,foams3,BarOr,Maite}. The simplicity constraints are however key for implementing the correct dynamics, as their role is to turn a topological theory into one with propagating degrees of freedom. It is therefore important to understand how the different implementations and forms of the simplicity constraints differ in their influence on the dynamics arising in the continuum limit.  This is a long term goal of a line of research \cite{merce,qgroup,BCspinnets}, in which this paper constitutes an important step forward.

The four--dimensional spin foam models are of high complexity. Extracting their continuum limit requires the development of appropriate tools and also a better understanding of the impact of simplicity constraints in general. In this work we will therefore consider three--dimensional models, with the aim to investigate these numerically. These have to be understood as `analogue' models, that is we will impose simplicity constraints -- following the procedure for the four--dimensional models -- and investigate the impact of these simplicity constraints for the continuum limit. A similar strategy, but using two--dimensional (non--gauge) models, led already to interesting insights. In fact we will confirm here the close relationship between these two--dimensional so--called spin net models and (so--far) three--dimensional spin foam models. 

Note that the actual theory of three--dimensional general relativity is topological, that is without propagating degrees of freedom. This is described by the BF action (\ref{BFaction}), with the choice ${\cal G}=\SU(2)$ for Euclidean signature space--times\footnote{Here we do not mean 'Euclidean gravity' which usually refers to employing a Wick--rotation and path integral weights $e^{-S}$. Here we will employ complex weights $e^{iS}$. This makes Monte--Carlo methods non--applicable and requires therefore the need to develop alternative methods.}. Being topological, the theory has a `trivial' continuum limit. In fact a triangulation independent discretization can be constructed, and BF theory will appear as one fixed point in the coarse--graining flow. 

Other theories related to BF theory are Yang Mills theory in first order formulation
\ba
	S= \int  \text{tr} ( B \wedge F  + g^2\, B \wedge \star B)
\ea
with $\star$ being the (metric--dependent) Hodge star operator and $g$ the Yang--Mills coupling.
Furthermore 3D general relativity with a cosmological constant $\Lambda$ is described by
\ba
	S_\Lambda \,=\,  \int \text{tr}(B \wedge F + \frac{\Lambda}{6} B \wedge B) \; .
\ea
Lattice gauge theories, which provide lattice versions of Yang Mills theory, will constitute a subset of the phase space of models we will be considering.

As for the construction for spin foam models we consider now a discretization and quantization of BF theory.  The BF partition function is given by
\be
\label{path1}
\mathcal{Z}_{BF} \,=\, \int \mathcal{D}B \, \mathcal{D}\omega \; \text{exp}\Big( i \int_{\mathcal{M}}\text{tr}\big(B \wedge F(\omega)\big)\Big) \; .
\ee
This expression is only formal and a discretization is necessary to make it well defined.

We denote by $\Delta$ the discretization and by $\Delta^*$ the dual 2--complex. The discretization $\Delta$ is constructed by gluing $d$--dimensional building blocks along their $(d-1)$--dimensional boundaries (often referred to as faces $f$). The $d$--dimensional building blocks also have $(d-2)$--dimensional `corners' or `hinges' in their boundaries shared by several $d$--dimensional building blocks. In the case of a triangulation, all these building blocks are given by simplices of the appropriate dimension.
Here we will also consider other discretizations such as a cubical lattice.

The dual complex $\Delta^*$ consists of dual vertices $v^\ast$ for every $d$--dimensional building block of $\Delta$, connected by dual edges $e^*$. We therefore have a dual edge $e^*$ for every face $f$ of $\Delta$. These dual edges bound dual faces $f^*$, where we have one dual face $f^*$ for every $(d-2)$--dimensional hinge of the discretization $\Delta$.

The dual complex carries the variables for the discretized path integral (\ref{path1}). The connection degrees of freedom will be encoded into holonomies, that is group variables $g_{e^*}\in {\cal G}$, associated to the dual edges $e^*$. The curvature $F(\omega)$ of the connection can then be approximated by the holonomies along the smallest available loops -- which are given by the boundaries of the dual faces $f^*$. After choosing a source and target vertex $v^*(f^*)$ for the loop, we define such holonomies as
\ba
g_{f^*}= g_{e^*_p} \cdots g_{e^*_1}
\ea
where $e^*_1,\dots,e^*_p$ are the ordered and oriented edges, starting at the choosen source vertex $v^*(f^*)$, bounding the dual face $f^*$.  

We furthermore discretize the $\mathfrak{g}$--valued  $(d-2)$--form $B$ by associating to each dual face $f^*$ a $\mathfrak{g}$--valued variable $B_{f^*}$. This represents the $B$--field integrated over the dual face. As a $\mathfrak{g}$--valued entity, it also needs a frame. We choose the one attached to the dual vertex $v^*(f^*)$, the same as for the face holonomy $g_{f^*}$. Thanks to these definitions, we can define the discretized path integral for the BF action as
\be\label{BFpathint2}
	\mathcal{Z}_{BF}(\Delta) = \int_{{\cal G}^{|e^*|} }   \prod_{e^* \in \Delta^*} {\bf d}g_{e^*}  \int_{{\mathfrak g}^{|f^*|}}\prod_{f^* \in \Delta^*} \bd B_{f^*} \;\;  \exp \big(i\, \text{tr} (B_{f^*} g_{f^*}) \big)
\ee
where ${\bf d}g_{e^*}$ denotes the Haar measure in ${\cal G}$ and $\bd B_{f^*}$ a measure invariant under adjoint action on the Lie algebra $\mathfrak{g}$.

Note that in \eqref{BFpathint2}, the fields $B_{f^*}$ appear only linearly in the exponential. Therefore it can be be integrated out, at least formally. This leads to delta functions which enforce the face holonomies to vanish
\be\label{BF3}
	 \mathcal{Z}_{BF}(\Delta) = \int_{{\cal G}^{|e^*|} } \Big( \prod_{e^* \in \Delta^*} {\bf d}g_{e^*} \Big) \prod_{f^* \in \Delta^*}  \delta (g_{f^*})
	= \int_{{\cal G}^{|e^*|} } \Big( \prod_{e^* \in \Delta^*} {\bf d}g_{e^*} \Big) \prod_{f^* \in \Delta^*}  \delta \Big(\prod_{e^* \subset f^*}^{\leftarrow} g_{e^*}\Big)\; .
\ee
Thus the partition function for BF theory implements an integral over the space of flat connections on the (discretized) manifold ${\cal M}$. 

The starting point for spin foam models is obtained by Fourier transforming the BF  partition function (\ref{BF3}). That is, we rewrite this partition function as a sum over group representations $\rho$, replacing the integral over group variables $g$. This is achieved by using the following expression for the group delta function
\be
	\delta(g) = \sum_{\rho}d_{\rho} \, \chi_{\rho}(g)
\ee
where the sum is over a complete set of (representatives of) irreducible unitary representations of the group ${\cal G}$ and $\chi_{\rho}, d_\rho$ are the corresponding characters and dimensions, respectively.

 The partition function becomes 
\be\label{BF4}
	\mathcal{Z}_{BF}(\Delta) = \int_{{\cal G}^{|e^*|} } \Big( \prod_{e^* \in \Delta^*} {\bf d}g_{e^*} \Big) \prod_{f^* \in \Delta^*}  \sum_{\rho_{f^*}} d_{\rho_{f^*}} \chi_{\rho_{f^*}}\Big( \prod_{e^* \subset f^*}^{\leftarrow} g_{e^*}\Big)  \; .
\ee
After performing the group integrals we obtain
\be\label{BF5}
	\mathcal{Z}_{BF}(\Delta) = \sum_{\rho_{f^*}}\prod_{f^* \in \Delta^*} d_{\rho_{f^*}} \prod_{e^* \in \Delta^*} \big(\mathbb{P}_{e^*}\big)^{\{ n_{f^*} \}_{f^* \supset e^*}}_{\{ m_{f^*} \}_{f^* \supset e^*}}\big(\{\rho_{f^*}\}_{f^* \supset e^*}\big)
\ee 
where $\mathbb{P}_{e^*}$ defines the Haar projector. This is a map
\begin{align}\label{HaarPro}
	\mathbb{P}_{e^*}  \;:\; \text{Inv}\Big( \bigotimes _{f^* \supset e^*} V_{\rho_{f^*}}\Big) &\rightarrow \text{Inv}\Big( \bigotimes _{f^* \supset e^*} V_{\rho_{f^*}}\Big) 
\end{align}
defined by\footnote{ Here we  assume for simplicity that the orientation of the dual edge and the dual faces adjacent to it agree. If this is not the case one just changes the representation of the dual face to the contragredient representation.}
\ba
\big(\mathbb{P}_{e^*}\big)^{\{ n_{f^*} \}_{f^* \supset e^*}}_{\{ m_{f^*} \}_{f^* \supset e^*}}\big(\{\rho_{f^*}\}_{f^* \supset e^*}\big) =    \int_{\cal G} \bd g \bigotimes_{i=1}^p \left[D^{\rho_{f^*_i}}(g) \right]^{n_{f^*_i}}_{m_{f^*_i}}
\ea	
with $p$ the number of dual faces $f^*$ meeting at the dual edge $e^*$.  We denote by $D^\rho(g)$ the representation matrices. Note that the group integrations in (\ref{BF4}) have been absorbed into the Haar projectors (\ref{HaarPro}).  The Haar projector is invariant under both left and right action (that is action on the $n$ or $m$ indices) for linear operators on the representation space $\bigotimes _{f^* \supset e^*} V_{\rho_{f^*}}$. It can be thus written as
\ba
\big(\mathbb{P}_{e^*}\big)^{\{ n_{f^*} \}_{f^* \supset e^*}}_{\{ m_{f^*} \}_{f^* \supset e^*}} &=& \sum_\iota  
{^{\{n_{f^*}\}_{f^* \supset e^*}}}\!| \iota \rangle \langle \iota |_{\{m_{f^*}\}_{f^* \supset e^*}}
\ea
where $\{|\iota\rangle\}_\iota$ is a orthonormal basis of intertwiners (invariant tensors) on the representation space  $\bigotimes _{f^* \supset e^*} V_{\rho_{f^*}}$.

The form (\ref{BF5}) for the BF partition function is given as a sum over representations (which would be `spins' for $SO(4)$ or $\SU(2)$) and in this sense describes a spin foam model. It is however just one particular example of this class of models, which -- due to the topological nature of BF theory -- also happens to be triangulation invariant.\footnote{For Lie groups this requires a proper regularization of the divergencies, which can be seen to result from redundant delta functions in (\ref{BF3}).}  A larger class of models is achieved by:
\begin{itemize}
\item Changing the (dual)  face weights, which in the case of (\ref{BF5}) are given by $d_{\rho_{f^*}}$. In fact, a discretization of Yang Mills theory is given by replacing $d_{\rho_{f^*}}$ with appropriate functions $\omega(\rho_{f^*})$.  Partition functions which can be written in this form with general face weights associated to the dual faces and the Haar projectors to the dual edges, will be called of lattice gauge theory form.  One also encounters different face weights in spin foam models. In fact it appears to be a relevant parameter for the continuum limit \cite{Christensen,BahrSteinhausPRL,BCspinnets} and also heavily influences the divergence structure for spin foams based on Lie groups \cite{perini,aldo, bonzomdittrich,Linqin}.
\item The imposition of the simplicity constraints leads in particular to a replacement of the Haar projector with a map $\mathbb{P}_{e^*}'$ projecting onto a smaller invariant subspace in  $\bigotimes _{f^* \supset e^*} V_{\rho_{f^*}}$.  We will call models with such a non--trivial restriction of these invariant subspaces, models with non--trivial simplicity constraints, or `proper spin foam models'. In general the choice of such invariant maps imposing non--trivial projections is quite large, compared to just changing the face weights. There are however set--ups in which non--trivial simplicity constraints are not possible (or rather artificial), {\it e.g.} when dealing with a multiplicity--free group together with a three--dimensional triangulation. In this case we would have to consider invariant tensors on a triple tensor product of representations. (The reason is that in the dual complex each dual edge is shared by three dual faces, reflecting the fact that each triangle has three edges.) These are unique, that is the Haar projectors map onto one-- or zero--dimensional spaces. A further restriction is only possible by forbidding some a priori allowed combinations of representations. Forbidding a particular representation to appear altogether can be also imposed via the face weights, and does therefore not count as proper spin foam model. We will avoid this feature, by choosing a cubical lattice. This leads to a quadruple tensor product of representations (as squares are bounded by four edges), which also agrees with the case resulting from four--dimensional triangulations (reflecting the fact that tetrahedra are bounded by four triangles). We can thus test the effect of simplicity constraints also in three--dimensional models by working with a cubical discretization. 

In the case one considers spin foams based on $\SU(2)$, {\it i.e.} models with a geometric interpretation, one can also choose simplicity constraints which carry a geometric meaning. For instance one can impose that the squares of the cubical lattice are flat, that is that the four edges making up the square span only a  plane. 
\end{itemize}

This defines the space of models we will be considering. We will later specify in more detail how we restrict and parametrize the choice of simplicity constraints. In order to make the models accessible for numerical treatment we will consider a finite group.  The integral with respect to the invariant measure becomes $\int_{\cal G} {\bf d}g \rightarrow \frac{1}{|{\cal G}|}\sum_g=: \sum'_g$ where $|{\cal G}|$ is the order of the group.

\subsection{Reformulations of lattice gauge models}\label{reformGauge}

\begin{figure}[h]
 	\centering
 	\includegraphics[scale = 0.7]{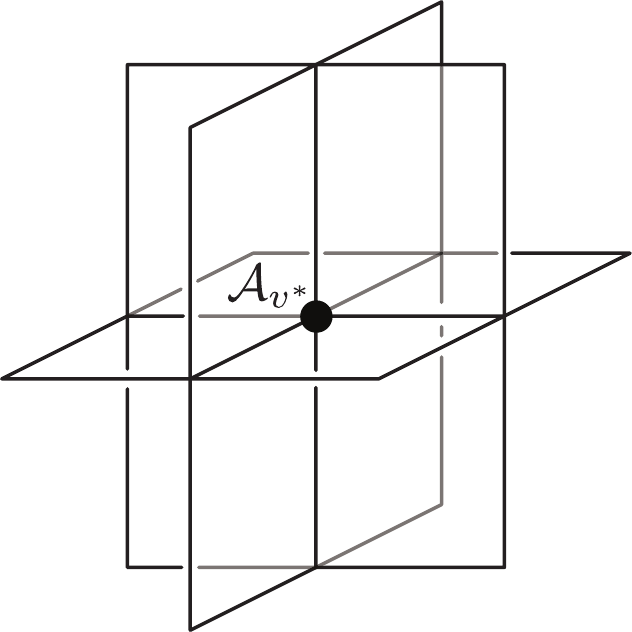}	\q \q \q 
 	\includegraphics[scale = 0.7]{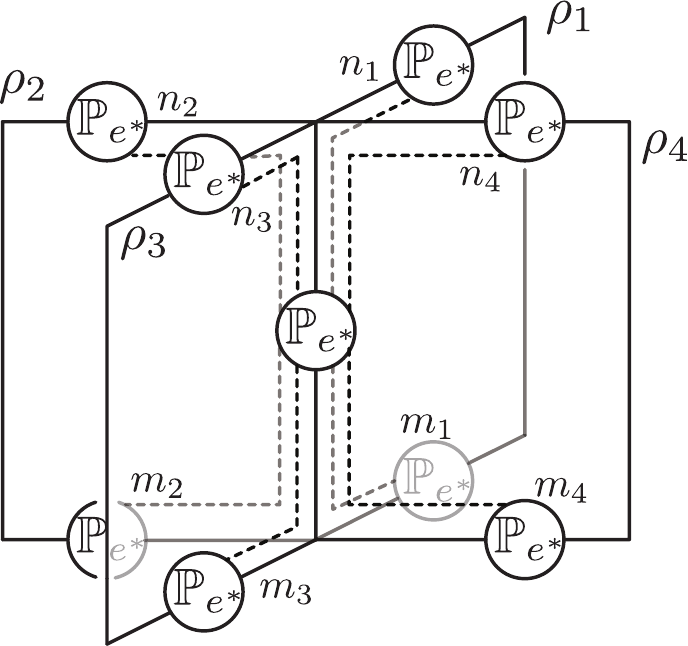}
 	\caption{The left panel depicts a fundamental cell in the 2--complex centered on a dual vertex. Each one of these blocks in the lattice is associated with an amplitude $\mathcal{A}_{v^*}$ carried by the dual vertex. The projetors $\mathbb{P}_{e^*}'$ are associated to each edge of the 2--complex. Projectors sharing a face are contracted to each other according to the pattern represented on the right panel.}
 	\label{fig:haar0}
 \end{figure}

We will now consider specifically a three--dimensional cubical lattice as discretization. Thus the dual complex is also a cubical lattice. We arrived at the following form of the partition function for lattice gauge models 
\ba\label{parti16}
	\mathcal{Z} = \sum_{\rho_{f^*}}\prod_{f^* \in \Delta^*} \omega({\rho_{f^*}}) \prod_{e^* \in \Delta^*} \big(\mathbb{P}'_{e^*}\big)^{\{ n_{f^*} \}_{f^* \supset e^*}}_{\{ m_{f^*} \}_{f^* \supset e^*}}\big(\{\rho_{f^*}\}_{f^* \supset e^*}\big) \; ,
\ea
 where the face weights $\omega$ are associated to the dual faces (or plaquettes) and the invariant tensors $\mathbb{P}'$ to the dual edges. These tensors are contracted among each other according to the pattern depicted in figure \ref{fig:haar0}.
 
In the following we will rewrite the partition function into a more local form, namely such that we can associate an amplitude to the cubes of the direct lattice (or alternatively the dual vertices).  To this end, the tensors $\mathbb{P}'$ are required to split as follows
\ba
	\label{decomposition2}
	 \big(\mathbb{P}_{e^*}'\big)^{\{ n_{f^*} \}_{f^* \supset e^*}}_{\{ m_{f^*} \}_{f^* \supset e^*}} = \sum_{\iota }\beta_\iota \,\, {^{\{n_{f^*}\}_{f^* \supset e^*}}}| \iota \rangle \langle \iota |_{\{m_{f^*}\}_{f^* \supset e^*}} \; ,
\ea
where $\{|\iota\rangle\}$ is a basis for the space $\text{Inv}\big(\bigotimes_{f \supset e}V_{\rho_f}\big)$, that is a basis of invariant tensors (intertwiners) for $\bigotimes_{f \supset e}V_{\rho_f}$.

The fact that such a form for $\mathbb{P}_{e^*}'$ exists follows from the fact that $\mathbb{P}_{e^*}'$ is invariant under both the left and right action of the group. Thus, with respect to any basis of intertwiners $\{|\iota\rangle\}$, it is of the form
\ba
\mathbb{P}_{e^*}' \,=\, \sum_{\iota,\iota'} \alpha_{\iota \iota'}  \,\, |\iota \rangle \langle \iota' |  \; .
\ea
Assuming that the matrix $\alpha_{\iota\iota'}$ is diagonalizable (which indeed is the case if $\mathbb{P}_{e^*}'$ is a projector), leads to the form (\ref{decomposition2}) of $\mathbb{P}_{e^*}'$. Note however that the basis $\{|\iota\rangle\}$ is not necessarily free to choose. In the case that $\mathbb{P}_{e^*}'$ is indeed a projector, {\it i.e.} $\mathbb{P}_{e^*}' \circ \mathbb{P}_{e^*}'=\mathbb{P}_{e^*}'$ we can reach a form (\ref{decomposition2}) with the coefficients $\beta_\iota$ equal to one or zero (if $\{|\iota\rangle\}$ is an orthonormal basis).  

For a regular lattice one can  absorb the face weights $\omega$ into the maps $\mathbb{P}_{e^*}'$, which however affects the projector conditions. In our models we will entirely shift the parametrization of the models towards the choice of the maps $\mathbb{P}_{e^*}'$.

\begin{figure}[h]
	\centering
	\includegraphics[scale = 0.7]{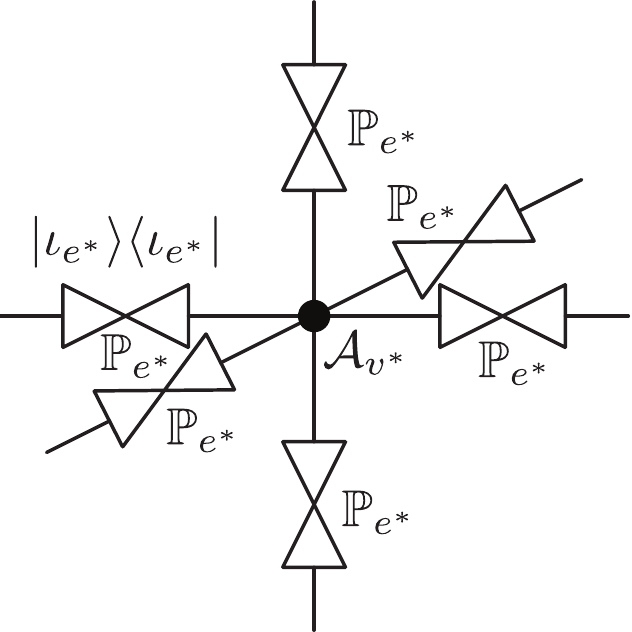}	\q \q \q
	\includegraphics[scale = 0.7]{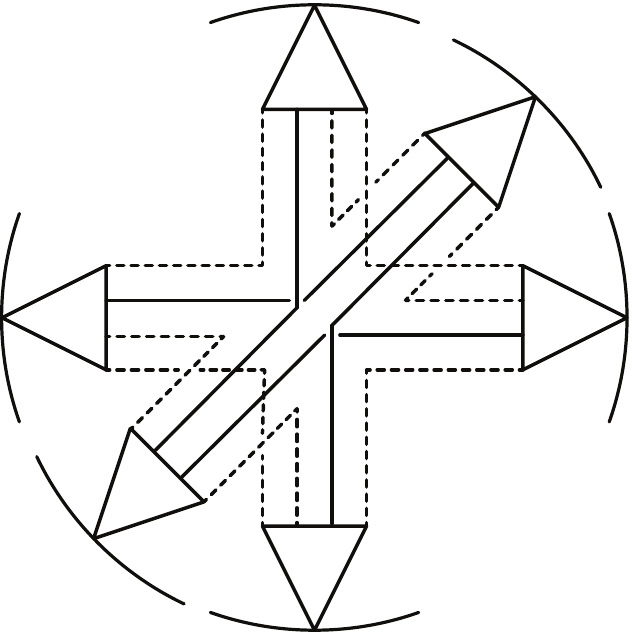}
	\caption{The left panel represents the decomposition of the maps $\mathbb{P}_{e^*}'$ in the intertwiner basis. The dual vertex amplitude is defined by contracting the intertwiners associated to this dual vertex according to the pattern represented on the right panel.}
	\label{fig:haar}
\end{figure}

Performing the splitting (\ref{decomposition2}) for each of the tensors $\mathbb{P}'_{e^*}$, we can associate an intertwiner variable $\iota$ to each dual half--edge (see figure \ref{fig:haar}). The intertwiner labels at the two halves of a dual edge have to agree. Note also that the intertwiner is between a set of representations $\{\rho\}$, associated to the dual faces hinging at the dual edge. 
The magnetic indices coming with the intertwiners $|\iota\rangle$ associated to the dual half--edges ending at one dual vertex $v^*$ contract all among themselves (see figure \ref{fig:haar}). That is for each dual vertex $v^*$ we can define an amplitude
\ba
{\cal A}_{v^*}( \{\rho_{f^*}\}_{f^* \supset v^*}, \{\iota_{e^*}\}_{e^* \supset v^*})  \,=\,      \prod_{ e^* \supset v^*} \sqrt{\beta_\iota}   \prod_{e^* \supset v^*}  |\iota_{e^*} \rangle \; ,
\ea
where the contraction of the intertwiners is implicit. The dependence on the representation labels $\rho_{f^*}$ is via the dependence of the intertwiners $\iota$ on these representations.  (Here we assumed that the face weights $\omega$ have been absorbed into the maps $\mathbb{P}_{e^*}'$.)

Furthermore, if we change the viewpoint from the dual to the direct lattice, the amplitude ${\cal A}_{v^*}={\cal A}_{\rm cube}$ can now be associated to a cube. The contraction pattern for the magnetic indices of the intertwiners is the same as for the evaluation of a four--valent spin network  on the boundary of the cube. The underlying graph is dual to the surface of the cube. In the direct lattice the representation labels $\rho_{f^*}=\rho_e$  are now associated to the edges of the direct lattice and the intertwiners to the faces $\iota_{e^*}=\iota_f$. 

Putting everything together, we obtain the form of the partition function we will be working with, namely as a gluing of amplitudes associated to cubes:  
\ba
\mathcal{Z}(\Delta) = \sum_{\rho_e,\iota_f}   \prod_{\rm cubes} {\cal A}_{\rm cube}( \{\rho_e\}_{e \subset {\rm cube}}, \{ \iota_f\}_{f\subset {\rm cube}}) \; .
\ea
The gluing proceeds by summing over the representation and intertwiner labels associated to the shared edges and faces.

\begin{figure}[h]
	\centering
	\includegraphics[scale = 0.7]{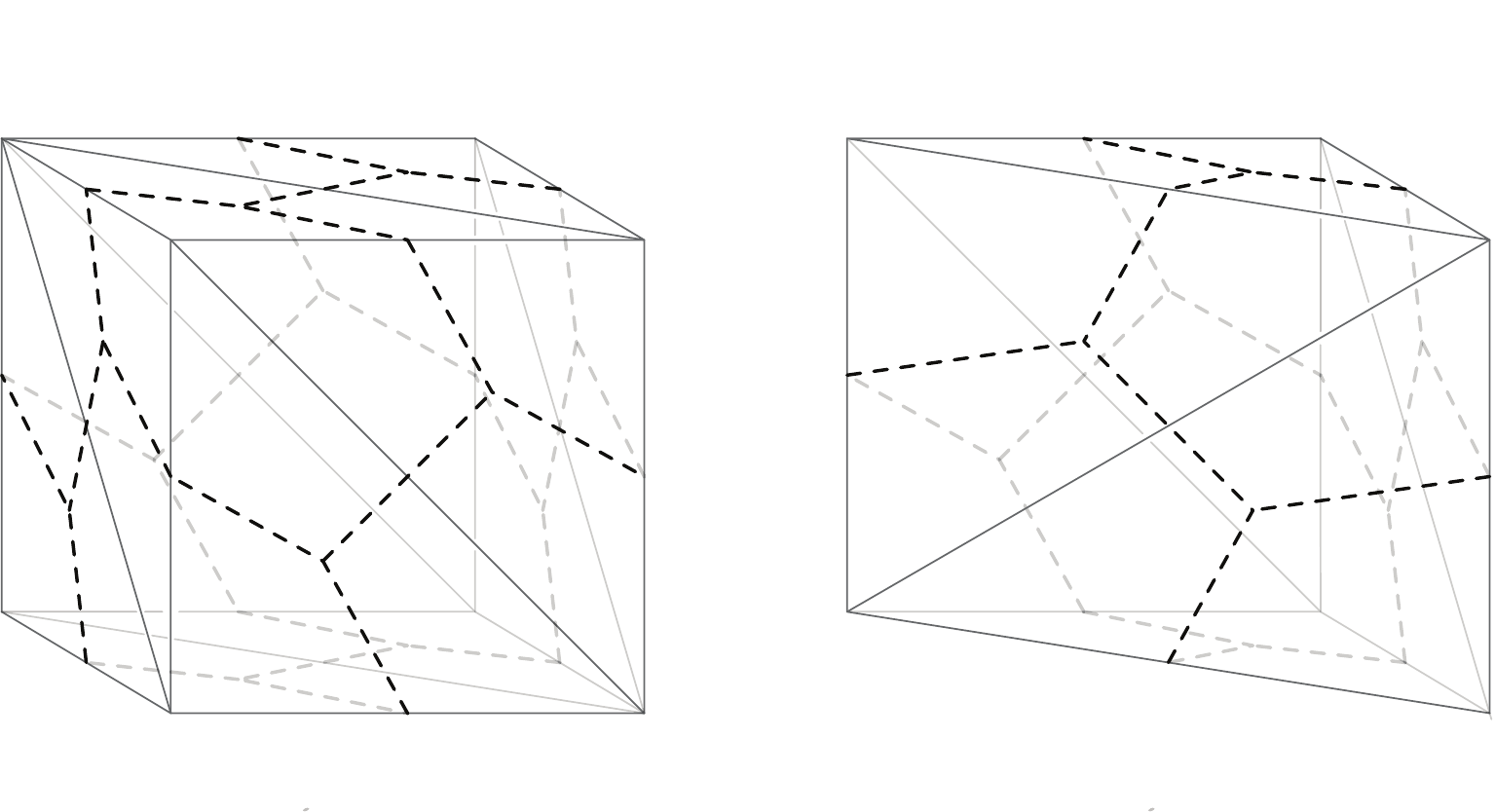}	
	\caption{The (expanded) graphs on the surface of a cube and a prism. This graph carries the variables. These are either holonomies, that is group elements $g_l$ associated to the links of the graph. Or, by a group Fourier transform, we can also use representation labels as variables. We assume here that the three--valent intertwiners are unique and we have therefore no more (after expansion of the four--valent nodes into three--valent nodes) intertwiner labels decorating the graph.}
	\label{fig:bdry}
\end{figure}

We just mentioned that the representation labels and intertwiners can also be thought of as being associated to a spin network, that is a contraction of intertwiner tensors along a pattern given by the network, on the boundary of the cubes. The gluing of cubes also translates into a gluing of boundaries with embedded spin networks -- by summing over representation labels and intertwiners. As explained in appendix \ref{app_fourier}, using again the (inverse) group Fourier transform we can implement a variable transformation and replace the sum over representations and intertwiners by a sum (in the case of finite groups) over group elements. These group elements are holonomies associated to the boundary graph underlying the spin network (see figure \ref{fig:bdry}). The amplitude for a cube is then expressed as a gauge invariant functional of these holonomies $A_{\rm cube} ( \{ g_l\})$ where $l$ denotes the links of the boundary graph. We therefore rewrite the partition function as
\be
	\mathcal{Z}(\Delta) = \sum'_{g_l}\prod_{\rm cubes} {\cal A}_{\rm cube} (\{g_l\}) \; .
\ee

The cube amplitudes are invariant under the action of the gauge group at the nodes $n$ of the boundary graph. That is for a set of gauge group elements $\{G_n\}$ associated to the nodes $n$ of the boundary graph, we have
\ba
{\cal A}_{\rm cube} (\{ G_{s(l)}  g_l  G^{-1}_{t(l)}  \}) \,=\, {\cal A}_{\rm cube} (\{ g_l\})  \; ,
\ea
where $s(l)$ denotes the source node of the link $l$ and $t(l)$ the target node. 
This gauge invariance implies that the set of variables $\{g_l\}$ provides an over--parametrization of the configurations. In order to reach an effective coarse--graining algorithm it is important to avoid this over--parametrization. To this end we will employ a gauge fixing procedure which will be detailed in section \ref{Aoverview}. (The representation labels $\rho$ and the intertwiners $\iota$ constitute a gauge invariant labelling. However this set of data is not preserved under coarse--graining, as discussed in section \ref{Aoverview}.)

The gauge invariance of the amplitudes allows us also to perform certain changes of the boundary graph. We can for instance expand the four--valent vertices into pairs of three--valent ones, and arrive at a boundary graph for the cubes as depicted in figure \ref{fig:bdry}. This change does not introduce  nor removes any  gauge invariant data.  These gauge invariant data can be constructed as follows: one chooses a set of independent cycles of the graph, all with the same source and target node. The holonomies associated to this set of closed cycles represent almost gauge invariant data. The only gauge action that is left is a global adjoint action of the gauge group on this set of cycle holonomies.  As we will later explain in more detail, a set of independent cycles can be found by choosing a rooted connected and spanning tree\footnote{A tree is a subset of links (together with the adjacent nodes) in a graph, so that these links do not form a cycle. A spanning tree is a tree that includes all the nodes of the graph. A root is a preferential node in the tree. If the tree is spanning and connected, every node in the graph can be reached by a unique path of tree--links from the root node.}  in the graph. The remaining links, which are not part of the tree, are called leaves. These determine a set of independent cycles. 

An expansion of a four--valent graph into three--valent graphs does not change the number of leaves $|\ell|$, as it is determined by the difference of the number $|l|$ of links and the number $|n|$ of nodes:
\be
|\ell|=|l|-|n|+1  \; .
\ee
We can thus expand higher--valent nodes into three--valent nodes without changing the amount of gauge invariant information in a given amplitude. In order to find the amplitude for the extended graph, we only need to ensure its gauge invariance at all new nodes. For instance, given a four--valent node with incoming links and associated holonomies $g_1, \ldots, g_4$, we introduce a new link $l_5$ connecting the target nodes of $l_1,l_2$ with the target nodes of $l_3,l_4$. The amplitude ${\cal A}'$ with respect to the expanded graph is then given by
 \ba\label{expGr}
 {\cal A}'( g_1,g_2,g_3,g_4,g_5) \,=\,  {\cal A}( g_5 g_1, g_5 g_2,  g_3,  g_4)  \; .
 \ea
Gauge invariance of the extended amplitude at the new nodes is due to the gauge invariance of the original amplitude and by construction.

To coarse-grain our lattice gauge models we will work with the cube amplitudes ${\cal A}_{\rm cube}$ mostly in the holonomy representation. The basic philosophy \cite{dittcyl} is to glue several cubes together, by integrating over the shared holonomy variables, associated to the matching links on the shared faces. (This might require the subdivision of links into half--links. The corresponding extension of the amplitude can be constructed in the same way as above.)  

The resulting building block will carry on its boundary a more complicated boundary graph. It will also carry more gauge invariant information than the graph on the original building blocks, thus the amplitude of the resulting building block will depend on more (gauge invariant) variables.  To avoid an (exponential) growth of this number of variables we have to truncate back the number of variables to its original number. This can be done by constructing a so--called embedding map
\ba
{\cal E}:  {\cal C}_{\rm coarse} \rightarrow {\cal C}_{\rm fine}
\ea
from the space of (coarse) configurations on the original building blocks to the space of (finer) configurations on the larger building block. This allows to pull back the amplitude of the larger building block to the coarser configurations
\ba
{\cal A}'_{\rm cube} ( \{ g_l\} )   \,=\, {\cal A}_{\rm larger} (  {\cal E}(  \{ g_l\}) )
\ea
and thus define the new amplitude ${\cal A}'_{\rm cube}$ for the same amount of data as for the original building block.

The construction of this truncation, provided by the embedding map, is the key step of such a coarse--graining procedure. In so--called Tensor Renormalization Group (TRG) algorithms, such a truncation is determined from the dynamics of the system. This is done with the aim to minimize the truncation error in the partition function. In the following we will describe a variant of such tensor network algorithms, the Decorated Tensor Network algorithm \cite{decorated}. It offers more flexibility, in particular regarding the treatment of gauge models. 

Note that the gluing and truncation in these algorithms are organized differently from the description above. The truncation is rather implemented first, via a procedure that splits building blocks to smaller pieces. These pieces are glued to a bigger building block in the second step. In the following we will explain the Decorated Tensor Network algorithm. First, we will review it for a 2D (non--gauge) system, then we will develop the algorithm for 3D lattice gauge models.

\section{ (Decorated) Tensor Network Renormalization}\label{sec_DTNR}

Levin and Nave \cite{levin} suggested the first coarse--graining algorithm, named Tensor Renormalization Group (TRG) algorithm, for 2D statistical models involving tensor networks. Gu and Wen \cite{gu-wen}  proposed another variant, applicable to statistical models defined on a square lattice.

There are two main points for TRG methods: firstly one reformulates the partition function of the (local) statistical model as a tensor network contraction. Consider for instance a vertex model, that is the partition function is given as a sum over variables $x_e$ associated to the edges $e$ of the lattice
\ba
Z\,=\,  \sum_{\{x_e\}}  \prod_v w_v( \{x_e\}_{v \subset e})
\ea
whereas the weights $w_v$ are associated to the vertices of the lattice and have as arguments the variables $x_e$ associated to the adjacent edges.  Each variable $x_e$ appears in two vertex weights (for a 2D lattice without boundaries) and therefore the partition function can naturally be interpreted as a contraction of a tensor network
\ba
Z\,=\, \sum_{\{x_e\}} \prod_v   T^v_{x_{e_1(v)} x_{e_2(v)} x_{e_3(v)} x_{e_4(v)}  }  \; .
\ea
Here we assumed a square lattice and $T^v_{x_{e_1(v)} x_{e_2(v)} x_{e_3(v)} x_{e_4(v)}  }  = w_v( \{x_e\}_{v \subset e}$ (with some ordering prescription for the edges adjacent to $v$). 

A coarse--graining move proceeds by blocking several vertices into one coarser vertex. Thus we have to sum over all variables shared by vertices belonging to the same  coarse vertex. This is described by a certain  contraction of tensors into a new tensor.  The coarser vertex will then have more adjacent edges than the original vertices. Likewise the new tensors are of higher rank than the original ones. One can summarize several indices of a given tensor into one index, so that the new tensor has the same rank as the original one. This does however increase the index range.

This is where the second main point of the TRG algorithm comes in, the truncation. The idea is to keep the number of adjacent edges constant, or equivalently the range of the corresponding indices fixed. This fixed index range is usually referred to as the bond dimension $\chi$.  The guideline of how to do this is as follows: the edges represent summation over variables shared between the coarser vertices. One wishes to reduce the summation range by neglecting non--relevant variables, that is modes which do not contribute substantially to the sum. To identify such modes one uses a singular value decomposition (SVD), and neglects the modes associated to the smallest singular values (see appendix \ref{app_svd}). 

All (local) statistical 2D models can be reformulated into a tensor network. However it turns out that for (higher dimensional) lattice gauge theories it is rather difficult to find an efficient encoding into a tensor network. (Note that it is always possible to find a tensor network description, see \cite{eckert1,decorated}, but these have a large initial bond--dimension arising from the need to double variables.) This is the motivation for the introduction of so--called Decorated Tensor Networks, that allow more flexibility in the design of the algorithm. They do also offer additional advantages, for instance a more straightforward access to expectation values for observables \cite{decorated}.

 In the following we will explain this algorithm for  a 2D `edge model'  or a 2D scalar field. These models can be rewritten into a tensor model, but we use their original form in order to illustrate the Decorated Tensor Network algorithm. The same algorithm will be used for the 3D gauge models.

We again assume a square lattice, but this time the variables $x_v$ are associated to the vertices of the lattice.  We then associate to each plaquette an amplitude $\mathcal{A}_{\text{square}}$ which in the case of the `edge models' can be written 
\ba
{\cal A}_{\rm square}(x_{v_1}, x_{v_2},x_{v_3}, x_{v_4})  \,=\,  \sqrt{w_{e_{12}}(x_{v_1},x_{v_2}) w_{e_{23}}(x_{v_2},x_{v_3}) w_{e_{34}}(x_{v_3},x_{v_4})w_{e_{41}}(x_{v_4},x_{v_1})} \; ,
\ea
where the weights $w_e(x_{v_i},x_{v_j})$ are associated to the edges.
The partition function is finally defined as
\ba
Z\,=\, \sum_{\{x_v\}} \prod_{\rm squares} {\cal A}_{\rm square}(x_{v_1}, x_{v_2},x_{v_3}, x_{v_4})  \; .
\ea

\begin{figure}[h]
	\centering
	\includegraphics[scale = 0.7]{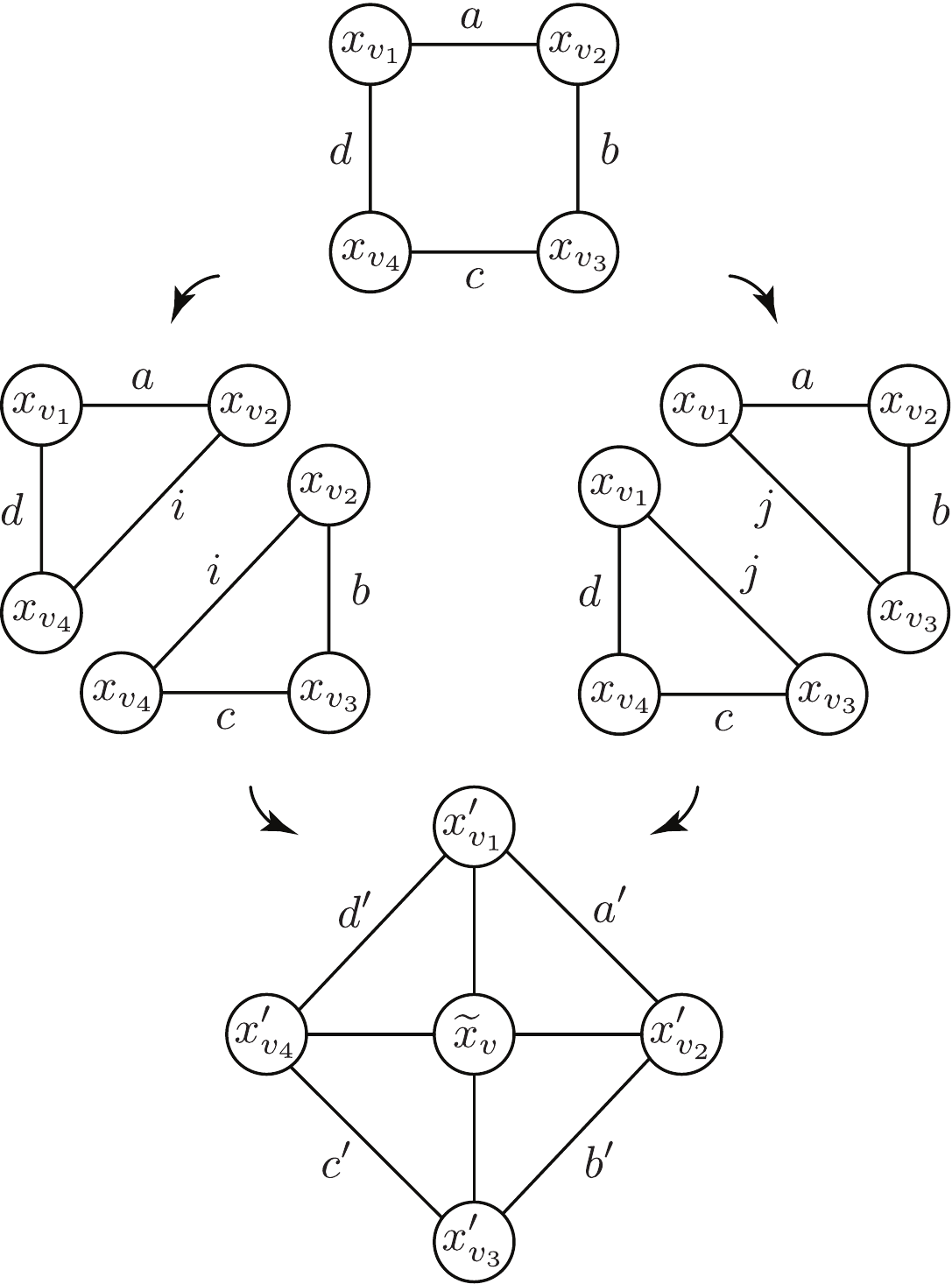}
	\caption{Example of renormalization of a 2D decorated tensor network where the coarse-grained amplitude is associated to squares. Thanks to the translational symmetry of the lattice, this is possible to focus on a single square. The initial square is split along two different diagonals. Each splitting produces two triangles. The resulting four triangles are glued back together to form a larger square rotated by $45^\circ$. }
	\label{2dalgo}
\end{figure}

We can glue neighbouring squares to larger effective squares, by integrating over shared variables in the bulk of the effective squares. Again these effective squares will in general have more boundary variables than the original squares along the edges.  We can take this into account by allowing more variables $a,b,c,d$ associated to the four edges of the effective squares (see figure \ref{2dalgo}). In fact these variables can be understood as indices belonging to a tensor which sits at the centre of the square and whose edges are perpendicular to the edges of the square. (Alternatively these indices can be interpreted as values of the original scalar field, arising as described above.)  Thus we have a tensor network `decorated' with additional variables $x_v$.

We therefore assume an effective square amplitude of the form
\ba
{\cal A}_{\rm square}(x_{v_1}, x_{v_2},x_{v_3}, x_{v_4}; a,b,c,d) \; .
\ea
The coarse--graining algorithm takes such a square amplitude as starting point and in each iteration constructs a coarse-grained effective amplitude ${\cal A}'_{\rm square}$.

We will now describe this coarse--graining algorithm, which is a `decorated' version of the algorithm in \cite{gu-wen}. The algorithm consists of two steps, splitting squares into two triangles, and gluing four triangles back to a square. The splitting step implements the truncation via a singular value decomposition. The gluing step implements the blocking or coarse--graining step. 

To describe the splitting and associated SVD, we refer to figure \ref{2dalgo}. Neighbouring squares are split along the two different diagonals. This allows to glue back these triangles back into bigger squares which are however rotated by $45^\circ$, see figure \ref{2dalgo}. To split the amplitude associated to a square into two amplitudes associated to the two triangles, we first form super--indices $A=(x_{v_1},a,d)$ and $B=(x_{v_3},b,c)$ as well as $C=(x_{v_2},x_{v_4})$. This allows us to define a family of  matrices
\be
	\label{splittingSq}
	M^C_{AB}= {\cal A}_{\rm square}(x_{v_1}, x_{v_2},x_{v_3}, x_{v_4}; a,b,c,d) 
\ee
labelled by the index $C$. Using a singular value decomposition (see appendix \ref{app_svd}), we can approximate each matrix in this family by a product over two matrices
\ba
M^C_{AB}\,\approx\, \sum^\chi_{i=1} (S_1)^C_{A,i} (S_2)^C_{B,i} \; .
\ea
The matrices $S$ now define the amplitude for the triangles (see figure \ref{2dalgo}), for instance
\ba
(S_1)^C_{A,i} \,=\, {\cal A}_{\rm triang. 1} (x_{v_1},x_{v_2},x_{v_4};a,i,d) \; .
\ea
The splitting of the squares along the two different diagonals gives four types of triangles, which are then glued back to larger squares according to figure \ref{2dalgo}:
\begin{align}
	&{\cal A}'_{\rm square}(\tilde{x}_v,x_{v_1}',x_{v_2}',x_{v_3}',x_{v_4}';a',b',c',d')\\[0.3em]
	& \q  =  \sum_{a,b, c,d}\;
	{\cal A}_{\text{triang.1}}(\tilde{x}_v,x_{v_2}',x_{v_3}';a,b',d)\,
	{\cal A}_{\text{triang.2}}(\tilde{x}_v,x_{v_1}',x_{v_4}';b,d',c)\\[-0.3em]
	& \q \q \q \;\; \times {\cal A}_{\text{triang.3}}(\tilde{x}_v,x_{v_4}',x_{v_3}';c,c',d)\,
	{\cal A}_{\text{triang.4}}(\tilde{x}_v,x_{v_1}',x_{v_2}';a,a',b)
	  \; .	
\end{align} 
This finishes one coarse--graining step and (after a possible rescaling of the amplitude) one can now iterate the procedure. 

Let us add two remarks: The Decorated Tensor Network algorithm comes with one essential difference to the TNG algorithm \cite{gu-wen}, which is that the SVD splitting procedure is performed for an entire family of matrices, parametrized by an additional index $C$. This index $C$ summarizes variables which are carried by both triangles arising from this splitting. We will also have such variables in the 3D algorithm for gauge models. Furthermore, note that the lowest possible approximation is given by choosing $\chi=1$. This trivializes the indices $a,b,c,\ldots$ of the actual tensor networks so that we are only dealing with the original variables $x_v$.  In this case the coarse--graining flow is described by a family of square amplitudes $A^{(k)}_{\rm square}(\{x_v\})$, where $k$ indicates the iteration number.

\section{The 3D algorithm for gauge models}\label{sec_3dalgo}

\subsection{Overview}\label{Aoverview}

\begin{figure}[h]
	\centering
	\includegraphics[scale = 0.68]{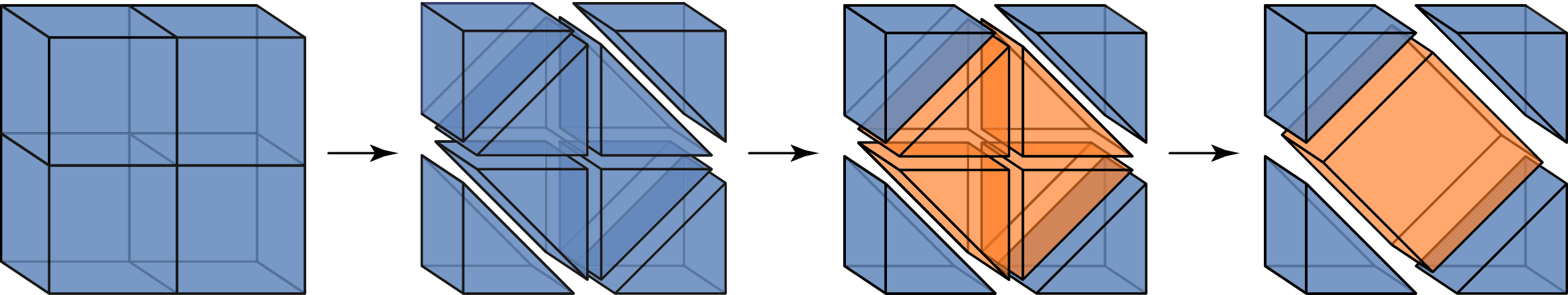}
	\caption{The algorithm we use is a 3D generalisation of the 2D  Decorated Tensor Network algorithm. The coarse--graining is a succession of two steps, the splitting of cubes into prisms and then the gluing of prisms to form bigger cubes.  This is shown in the last step, where four prisms obtained by splitting four blue cubes are glued together to form a new orange cube. The blue prisms are also glued to cubes, that appear adjacent to the orange cube. One then rotates the entire lattice to repeat this procedure in an orthogonal plane. (Note that the coarser building blocks are geometrically not cubes anymore, but we will nevertheless refer to them as cubes.) }
	\label{TRGcube}
\end{figure}

As explained in section \ref{reformGauge} we can re-formulate the partition function for $d$--dimensional lattice gauge models and spin foams as a gluing over $d$--dimensional building blocks. That is the amplitude is associated to these building blocks, which are characterized by boundary data.   For the 3D algorithm we will work with cubes (and prisms) as building blocks. The basic steps of  the coarse--graining algorithm, namely splitting and gluing building blocks, then proceed in a way similar to the 2D case. Indeed we apply the same coarse--graining geometry as in 2D in alternating planes of the 3D lattice (see figure \ref{TRGcube}).   

For gauge models the boundary data is encoded in variables associated to a graph embedded into the boundary of the building block.  We can, for example, choose holonomies  (that is group elements) associated to the links of the graph. Gauge symmetry then forces the amplitude to be invariant under gauge transformations acting on the nodes of the graph. Alternatively, we can use a (gauge invariant) spin network basis, to express the amplitude. This will be however not convenient for several reasons as explained further.

An important consequence of the gauge symmetry is that physical, that is gauge invariant, degrees of freedom are de--localized. Consider for instance the boundary data given in terms of holonomies $g_l$ associated to the links of the boundary graph. The set of these holonomies $\{g_l\}_l$ is redundant due to the gauge symmetry at the vertices.   If no special attention is paid to this redundant information, we would obtain a very inefficient algorithm since computational resources would be committed to this redundant data. It turns out that the physical, {\it i.e.} gauge invariant, boundary data is encoded in the traces of closed holonomies obtained from the link holonomies. It is however highly non--trivial to find an {\it independent and complete} set of fully gauge invariant variables.\footnote{An alternative is the fusion basis \cite{fusbas} which provides a non--local encoding of the degrees of freedom. This basis diagonalizes holonomy operators while the spin network basis rather diagonalizes the observables dual to traces of Wilson loops. Furthermore, in contrast to the spin network basis, the fusion basis is stable under coarse--graining.}

We can however obtain an almost gauge invariant set of observables by choosing a root node and considering the  loop--holonomies associated to a set of independent cycles, starting and ending at the root node.  (These variables are still not completely invariant, as they transform under the adjoint action resulting from gauge transformations at the root node.) The choice of such a set of independent cycles is equivalent to the choice of a connected spanning tree in the graph.  Links of the graph which are not part of the tree are called leaves. The set of leaves is in one--to--one correspondence with a set of independent loops. Given a leaf there is a unique loop that visits the root vertex once and traverses only this leaf and tree--edges. The set of loops determined from the leaves is independent, as each loop in this set traverses a different leaf and the corresponding holonomies define the set of loop--holonomies.

\begin{figure}[h]
	\centering
	\includegraphics[scale = 0.7]{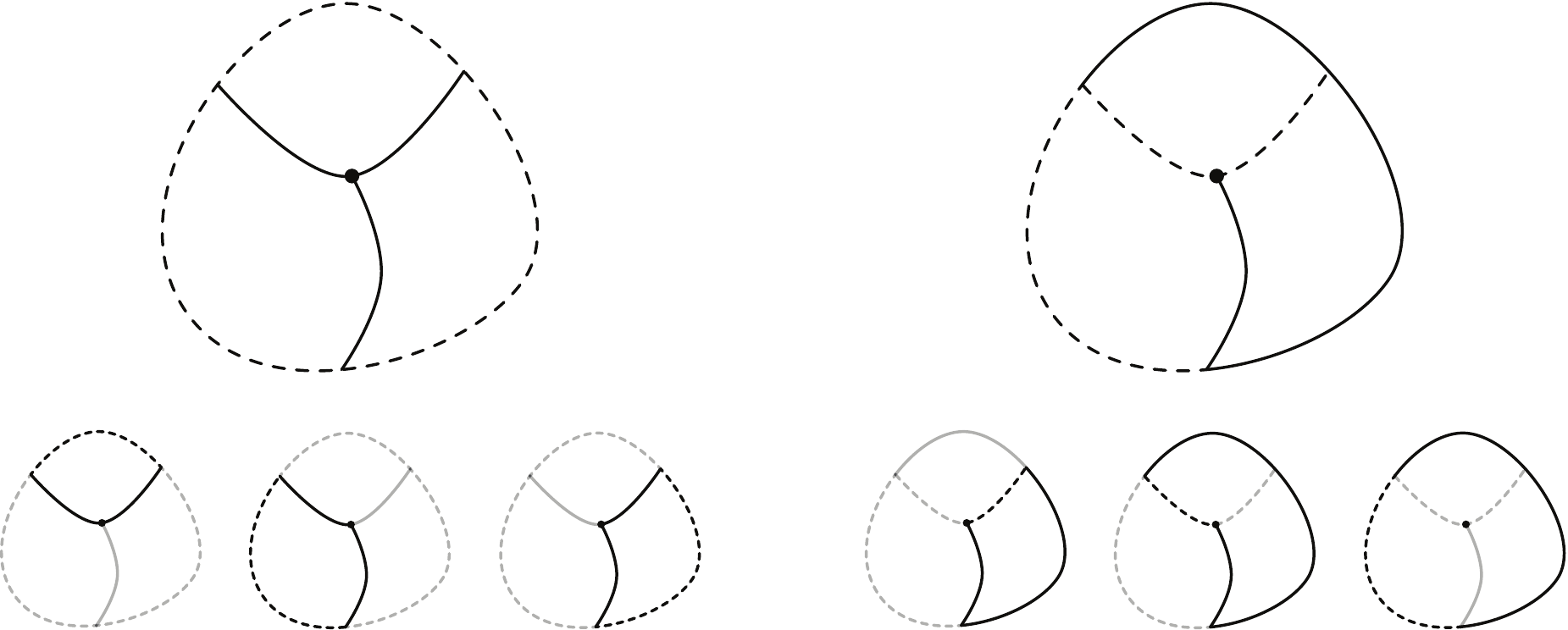}
	\caption{Two examples of a choice of rooted spanning tree for a given underlying graph. The branches are depicted by solid lines while the leaves are represented by the dashed lines. The leaves are in one--to--one correspondence with the independent cycles of the underlying graph which are represented in the lower panels. }
	\label{gaugefixing}
\end{figure}

The choice of tree can be understood as choosing a set of (almost gauge invariant) observables as well as localizing them (see figure \ref{gaugefixing}). Furthermore we can gauge fix the amplitude so as to obtain a functional of leaf-holonomies only. Thus, to obtain the gauge fixed amplitude we have to set the holonomies associated to links of the tree to be trivial:
\ba
{\cal A}^{\rm gf}( \{ g_\ell\}_\ell) \,=\, {\cal A}(  \{ g_\ell\}_\ell , \{g_l = \mathbb{I} \}_{l \subset {\cal T}})
\ea
where $\ell$ labels the leaves with respect to the tree ${\cal T}$. The gauge fixed amplitude has one remaining invariance, namely under adjoint action: 
${\cal A}^{\rm gf}( \{ g_\ell\}_\ell) = {\cal A}^{\rm gf}( \{ G g_\ell G^{-1}\}_\ell)$.

This gauge fixing will play an important role in our algorithm, as we need to localize the degrees of freedoms in a certain way for the gluing and splitting procedures.  For instance, after gluing building blocks we will have a natural choice of tree resulting from the trees associated to the original building blocks. This choice will in general not be appropriate for the next splitting step, and therefore a tree transformation will have to be performed.

\begin{figure}[h]
	\centering
	\includegraphics[scale = 0.7]{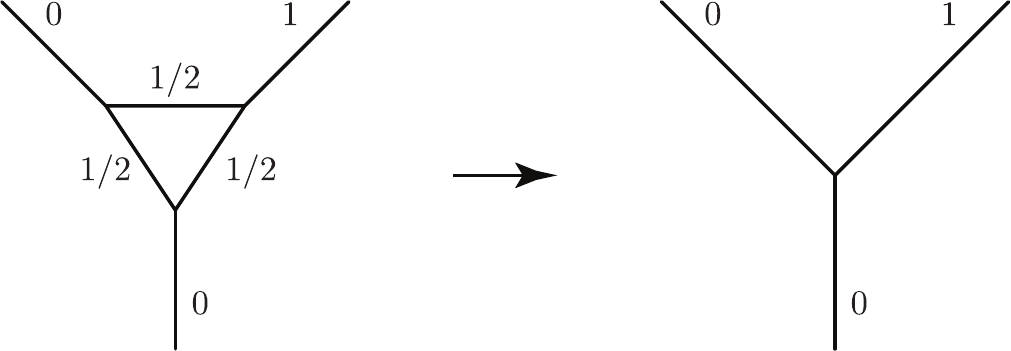}
	\caption{Graphical representation of the 3-1 Pachner move performed on a spin network basis state. The effective node obtained from the Pachner move represents an incompatible coupling between irreducible representations for which there is no intertwiner. This illustrates the fact that the spin network basis is not stable under coarse--graining.}
	\label{CGspins}
\end{figure}

An alternative to working with the holonomy basis and a gauge fixing of it, would be to work with a gauge invariant spin network basis. There are however two disadvantages in doing this. Firstly, the gauge invariant spin network basis is not preserved under coarse--graining \cite{eteraCG3,bfform2}. An example can be seen in figure \ref{CGspins}, showing an `effective node', representing a coupling between the representations associated to the adjacent links, for which no (`bare') intertwiner exist. One would therefore find a way to project such configurations out, or enlarge the configurations space to gauge covariant spin networks. The second issue is an algorithmic one: assume we work only with gauge invariant spin networks one would also make use of the associated reduction of required memory space. This makes a considerable difference: not taking the coupling conditions into account means that the amplitude requires a memory scaling with $|\rho|^{24}$ where $|\rho|$ denotes the number of representations. Taking the coupling conditions into account requires a memory scaling with a number smaller than $|{\cal G}|^9$.  This can be done by introducing super--indices, see {\it e.g.} \cite{qgroup}. These would be non--local, as the super--indices would take the coupling conditions for the entire boundary spin network into account. This would complicate very much the entire algorithm, and we therefore rather work with a gauge fixing, which as mentioned above allows as to localize degrees of freedom in a certain way.  Using this gauge fixing we will employ both the holonomy representation, and a (gauge fixed) spin network representation.

\subsection{Technical preliminaries \label{subsec_technic}}
There are three main steps to our algorithm, namely splitting, gluing and tree transformations. Splitting and gluing are best performed in the Fourier transformed picture (that is with gauge fixed spin network function) since in this picture these operations appear as matrix operations. In contrast, the tree transformation is best performed in the holonomy representation, as it only requires a relabelling of variables in this case. (In the spin network representation, it would require a large matrix multiplication.) Therefore we also need to include a group Fourier transform and its inverse in--between the different steps. 

Here we will describe the details of each one of these necessary steps. We start with the description of the boundary graphs and spanning trees.

~\\
{\bf {\small Boundary graph:}}\\~
The partition function for the gauge models discussed here can be rewritten as a gluing of cubical building blocks. The building blocks are equipped with an oriented  boundary graph that carries group variables. The amplitude associated to a building block is a gauge invariant functional of these group variables. The basic features of this boundary graph are as follows:
\begin{itemize}
\item The boundary graph arising from the rewriting of the partition function has {\it a priori} the following structure. Each face of the cube carries a four--valent node. The nodes of two neighbouring faces on a given cube are connected by a link. These links are thus crossing the edges of the cube.  One can introduce two--valent nodes which partition these links into half--links belonging to a definite face. We will furthermore expand each four--valent node in the middle of a face into two three--valent nodes connected by a new link, as described in equation (\ref{expGr}). This expansion of four--valent nodes to three--valent ones leads to a dual triangulation of the boundary of the cube. We will later split the cube into a prism along the edges of this triangulation.

After one step of the coarse--graining procedure the boundary graph will be refined in the following way: on two opposite faces we will have an additional cycle of graph links, see figure \ref{blue_split}. Further iterations will keep this boundary graph stable.\footnote{One could also adopt a higher truncation scheme, involving a even more refined graph.}

The boundary graph for the prisms are similar: the triangular faces carry a three--valent node whereas the quadrilateral faces carry two three--valent nodes arising from expanding four--valent ones. But one quadrilateral face will carry an additional cycle of the graph, inherited from the cubic building block. This face will have four three--valent nodes, see figure \ref{blue_split}.

\item We want to have the same boundary graph for every cube in the lattice.  Since gluing neighbouring cubes requires pairs of links which are identified to have the same orientation, the orientations of the (half--) links on opposite faces of the cube have to match. 

\end{itemize}

A boundary graph for the cube is depicted in figure \ref{blue_split}, using a planar representation of the boundary of the cube. For the definition of the Fourier transform (see appendix \ref{app_fourier}) we have to introduce a further convention for which we need to colour the faces of the building blocks with two colours, namely white and grey. In the gluing process a white face needs to be matched to a grey face, thus opposite faces on a cube need to be coloured differently. 

~\\
{\bf {\small Spanning tree:}}\\~
Each part of the algorithm requires a specific choice of rooted and connected spanning tree. In particular, this implies that between the gluing and splitting steps, a change of spanning tree will be necessary as these steps require a different choice. Such a tree transformation is described in the next subsection. For instance, the gluing of two building blocks requires that the tree for the glued building blocks, arising from the trees of the initial building blocks, is connected and spanning again. On the other hand, for splitting a cubic building block into two prisms we demand that the same number of physical degrees of freedom are distributed between both prisms.  Counting the number of independent cycles for the cube and the prism (see figure \ref{blue_split}), we see that a spanning tree for the cube leads to 9 leaves whereas a spanning tree for the prism leads to 6 leaves.  To split 9 degrees of freedom into two sets of 6 degrees of freedoms, we need to copy 3 of the initial 9 to both building blocks.\footnote{Remember that the same happens for the decorated tensor network algorithm for the 2D scalar field: the scalar field associated to two corners of the diagonal along which the squares are split, are copied to both triangles, and furthermore serve as parameters in the SVD procedure.}

The previous requirement implies that the distribution of the leaves over the cube must be such that three leaves will be associated to one prism and three other leaves to the other prism. Furthermore, we need three `shared leaves', that is three links of the graph across the boundary along which the cube is cut into prisms. These `shared leaves' are the ones which are copied to both prisms. But cubes are cut in two different ways into prisms. To satisfy the requirement of having three leaves of `sharing type' we need two different trees for the two different splittings. We will refer to the cubes cut in the two different ways as red and blue cubes. 

A similar counting argument applies for the gluing procedure which determines the required number of leaves of `sharing type'. 

Most of the links of the boundary graph are crossing an edge of the corresponding building block, {\it i.e.} they are included in two faces. These links can be cut into two half links. If the initial link happens to be a leaf we can extend the tree by choosing one of the half links as tree--link. Note that this does not change the assignment of a full link as `shared leaf'. This will be made more obvious in the discussion on the splitting procedure. 

~\\
{\bf {\small Group Fourier transform:}}\\~
We will perform the gluing and splitting in the Fourier transformed picture since these operations boil down to matrix multiplications in this case. As explained in appendix \ref{app_fourier}, this comes from the fact that the gluing of two amplitudes along two matching graph--links obtained by integrating over a group variable translates into a summation over the representation labels in the Fourier transformed picture. However, for such a translation to hold, it is necessary to introduce two definitions of the Fourier transform. These two definitions differ by a complex conjugation. We colour faces in white or grey to encode which convention applies and impose that only face of different colours can be glued together. For a functional which depends on a single group variable. the Fourier transform is defined as
\begin{align}	
	\label{Ftrans0}
	\begin{cases}
	&\widetilde{\cal A} \big(\rho,m,n \big) = \sum'_{ G_{\text{grey}}}  {\cal A} \big(G_{\text{grey}} \big)  \sqrt{d_{\rho}} \; \overline{D^{\rho}_{mn}(G_{\text{grey}})} \\[1em]
	&\widetilde{\cal A} \big(\rho,m,n \big)=  \sum'_{ G_{\text{white}}}  {\cal A} \big(G_{\text{white}} \big) \sqrt{d_{\rho}} \; D^{\rho}_{mn}(G_{\text{white}})
	\end{cases} 
\end{align}
 where the primed sum includes a normalization $\sum'_G=\frac{1}{|{\cal G}|} \sum_G$, and $D^\rho_{mn}(G)$ are the representation matrix elements for an irreducible unitary representation $\rho$. 
The inverse Fourier transform is given as
\begin{align}	
	\label{Ftrans}
	\begin{cases}
	&{\cal A} \big(G_{\text{grey}} \big) = \sum_{\rho,m,n} \widetilde{{\cal A}}(\rho,m,n) \sqrt{d_{\rho}} \;D^{\rho}_{mn}(G_{\text{grey}}) \\[1em]
	&{\cal A} \big(G_{\text{white}} \big) = \sum_{\rho,m,n} \widetilde{{\cal A}}(\rho,m,n) \sqrt{d_{\rho}} \; \overline{D^{\rho}_{mn}(G_{\text{white}})}
	\end{cases} \; .
\end{align}
 Note that  $n$ and $m$ are  associated to the source and target node of the leaf respectively.

\subsection{Basic steps of the algorithm}

Here we discuss some basic procedures needed for the coarse--graining algorithm.

~\\
{\bf {\small Change of spanning tree:}}  \\
As discussed in the previous section, it is sometimes necessary to change the spanning tree between two steps of the algorithm. This change of spanning tree is more easily performed in the group representation. Let $\{G_{\ell}\}$ denote the loop--holonomies with respect to the old spanning tree and $\{G'_{\ell'}\}$ the loop--holonomies with respect to the new spanning tree. Both trees will have the same root vertex. Remember that the lopp--holonomies are the gauge fixed representatives for the holonomies associated to a basis set of (rooted) cycles, determined by the spanning tree. Since both sets of cycles form a basis, we can express the loop--holonomies of one set as a combination\footnote{This combination is given as follows: Let $c_\ell$ be the rooted cycle associated to a leaf $\ell$ with respect to the old spanning tree.  It will traverse some number of leaves $\ell'$ of the new spanning tree, in the order $\{(\ell'_1)^{s_1},(\ell'_2)^{s_2},\ldots\}$ where $s_k=\pm 1$ denotes the orientation in which the leaf $\ell'_k$ is traversed. One then has $G_\ell= \dots (G'_{\ell'_2})^{s_2}  (G'_{\ell'_2})^{s_1}$.} of the loop--holonomies from the other set: 
$G_\ell\,=\,G_\ell(\{G'_{\ell'}\})$. Denoting by ${\cal A}$ the amplitude expressed in the old loop--holonomies basis and ${\cal A'}$ the amplitude expressed in the new loop--holonomies basis, we have
\ba
{\cal A}'( \{ G'_{\ell'}\}) &=& {\cal A} \big(  \big\{ G_\ell( \{ G'_{\ell'}\})\big\} \big) \; .  \label{Ttrans}
\ea

~\\
{\bf {\small Gluing building blocks:}}  \\
Gluing two building blocks means identifying the two faces along which the blocks are glued. For the boundary graph, this requires a matching of the pieces of oriented graphs associated to these faces such that leaves are matched to leaves. There are then two situations depending if the support of the leaf is a link which is fully embedded in the face or a link which is only half embedded in the face.

\begin{figure}[h]
	\centering
	\includegraphics[scale = 0.7]{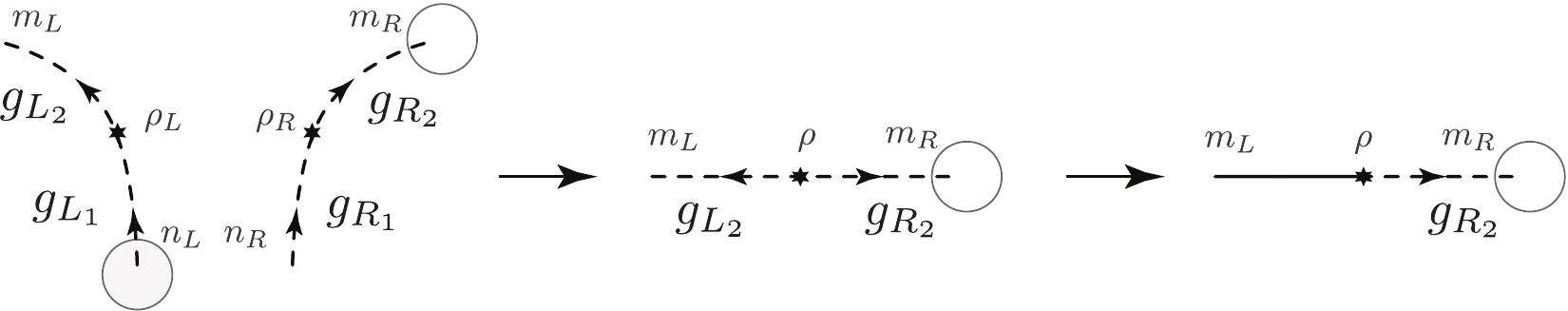}
	\caption{The gluing can be performed both in the `group' representation and the `Fourier transformed' representation. These two operations are related by the Plancherel theorem \eqref{Plancherel}. In the `group' representation, the gluing is obtained by tracing over holonomies associated to two half--links while it boils down to a contraction of magnetic indices in the `Fourier transformed' representation}
	\label{toy}
\end{figure}

In the first case, we integrate the product of the amplitudes of the two blocks ${\cal A}_L$ and ${\cal A}_R$ over the identified loop--holonomy $G_\ell$:
\ba
{\cal A}_{\rm glued}\,=\, \sum'_{G_\ell}     {\cal A}_L(G_\ell)   {\cal A}_R(G_\ell)  \,=\,    \sum_{\rho,m,n}  \widetilde{\cal A}_L(\rho,m,n)  \widetilde{\cal A}_R(\rho,m,n) \; .
\ea
When only half the support of the leaf is embedded in the glued face,we can subdivide the corresponding link into pairs of half--links. We then only integrate over the holonomy associated with the half link which is embedded in the glued face. Consider for instance figure \ref{toy}, in which this division into half--links has already taken place. The `left' and `right' holonomies can thus be written
\ba
G_L\,=\,  g_{L_2} g_{L_1}   \q ,\q  G_R\,=\,  g_{R_2} g_{R_1}   \; .
\ea
The orientation of the half--links to be glued does coincide and we can thus just sum over the shared group element $g_{L_1}=g_{R_1}$:
\ba\label{GlueB2}
	{\cal A}_{\rm glued }(g_{L_2}, g_{R_2}) \,=\, \sum_g {\cal A}_L( g_{L_2} g) {\cal A}_R( g_{R_2} g)  \; .
\ea
The glued amplitude is a function of two holonomies associated to two half--links with opposite orientation. This realizes a gluing of the leaves as the support of these holonomies is now contiguous. Furthermore, note that the glued amplitude inherits a gauge invariance at the common node such that
\ba
	{\cal A}_{\rm glued }(g_{L_2} h , g_{R_2} h ) \,=\, {\cal A}_{\rm glued}( g_{L_2 }, g_{R_2})  \; .
\ea
We can therefore perform a gauge fixing, {\it e.g.} $g_{L_2}= \mathbb{I}$ so that we deal with the glued amplitude ${\cal A}'_{\rm glued}(g_{R_2}):={\cal A}_{\rm glued}(\mathbb{I}, g_{R_2})$ which depends on a single group variable. Note that we could have applied this gauge fixing also before gluing the links. The choice of half--link to gauge fix then determines the orientation of the resulting link. Of course, the final orientation of the edge can always be changed and the amplitude accordingly transformed.

The gluing procedure in terms of holonomy variables (\ref{GlueB2}) is cumbersome to implement numerically since it represents a type of convolution. Such a convolution involves a group multiplication as compared to a more direct summation over variables. Indeed it turns out that the Fourier transformed picture offers a more efficient way of computing such a gluing of leaves.

As mentioned before, the Fourier transformation must follow the conventions encoded in the grey and white colouring of the faces. For the example presented in figure \ref{toy}, the colouring imposes  
\ba
{\cal A}_L(  g_{L_1} ) &=& \sum_{\rho_L,m_L,n_L}  \widetilde{\cal A}_L(\rho_L,m_L,n_L) \sqrt{d_{\rho_L}}\,\,  D^{\rho_L}_{m_L n_L} (g_{L_1}) \; , \nn\\
 {\cal A}_L(  g_{R_2}  g_{R_1}  ) &=& \sum_{\rho_R,m_R,n_R} \sum_k   \widetilde{\cal A}_R(\rho_R,m_R,n_R)     \sqrt{d_{\rho_R}} \,\, \overline{D^{\rho_R}_{m_R k} (g_{R_2}) }  \overline{D^{\rho_R}_{k  n_R} (g_{R_1}) } \; ,
\ea
where the gauge fixing has already been used to set $g_{L_2}=\mathbb{I}$. Summing over $g_{L_1}=g_{R_1}$  gives 
\ba\label{GlueB4}
{\cal A}'_{\rm glued}(g_{R_2})&=& \sum'_{g} {\cal A}_L(  g )  {\cal A}_L(  g_{R_2}  g  )  \nn\\
&=& \sum_{ \stackrel{\rho_L,m_R,n_R}{\rho_R,m_R,n_R}}\sum_k  \delta_{\rho_L,\rho_R} \delta_{k,m_L} \delta_{n_R,n_L}  \widetilde{\cal A}_L(\rho_L,m_L,n_L)  \widetilde{\cal A}_R(\rho_R,m_R,n_R) \overline{D^{\rho_R}_{m_R k} (g_{R_2}) } \nn\\
&=& \sum_{\rho,m,n}\left(  \sum_{l}  \frac{1}{\sqrt{d_\rho}} \widetilde{\cal A}_L(\rho,n,l)  \widetilde{\cal A}_R(\rho,m,l) \right)   \sqrt{d_\rho}\,\, \overline{D^{\rho}_{m n} (g_{R_2}) }  \; .
\ea
Thus we can read off the Fourier transformed glued amplitude
\ba\label{GlueB5}
\widetilde{{\cal A}}'_{\rm glued}(\rho,m,n)&=&  \sum_{l}  \frac{1}{\sqrt{d_\rho}} \widetilde{\cal A}_L(\rho,n,l)  \widetilde{\cal A}_R(\rho,m,l)  \; .
\ea
Notice that the Fourier transform convention we use for the glued amplitude is the one for white faces. This is consistent with the fact that we have chosen the gauge fixing $g_{L_1}=\mathbb{I}$. Indeed the remaining leaf is associated to the half--link embedded in the white face. Therefore, in the Fourier transformed picture we obtain the glued amplitude (modulo a rescaling by a dimensional factor) by identifying the representations such that $\rho \equiv \rho_L = \rho_R $ and summing over the magnetic index associated to the node sitting in the glued face. The other magnetic indices are copied over for the new amplitude, that is $n \equiv m_L$ is associated to the source node of the new leaf and $m \equiv m_R$ is associated to the target node (see figure \ref{toy}). 

This procedure can be generalized to all other cases (of differing orientations or differing colouring of faces).  The precise contraction rule in the Fourier transformed picture can be  derived with a calculation as in (\ref{GlueB4}). Alternatively one can use a graphical derivation as in figure \ref{toy}. The main point is that the glued Fourier transformed amplitude arises from summing the initial amplitudes over the magnetic index associated to the glued face. The gluing of building blocks is obtained by repeating the same operation for every leaf whose support is fully or partly embedded in the glued face. The latter ones are the so--called `shared leaves'. 

~\\
{\bf {\small Splitting:}}  \\
At the beginning of each coarse--graining step, neighbouring cubes are split along two different planes. Depending on the choice of cutting plane, the cubes will be referred to as `blue' and `red' cubes. For a given cube, the plane goes through the diagonals of two opposite faces. But the boundary graph is chosen such that its dual triangulation is consistent with the splitting. This means that the cutting plane proceeds only along edges of the dual triangulation without intersecting any.

\begin{figure}[h]
	\centering
	\includegraphics[scale = 0.7]{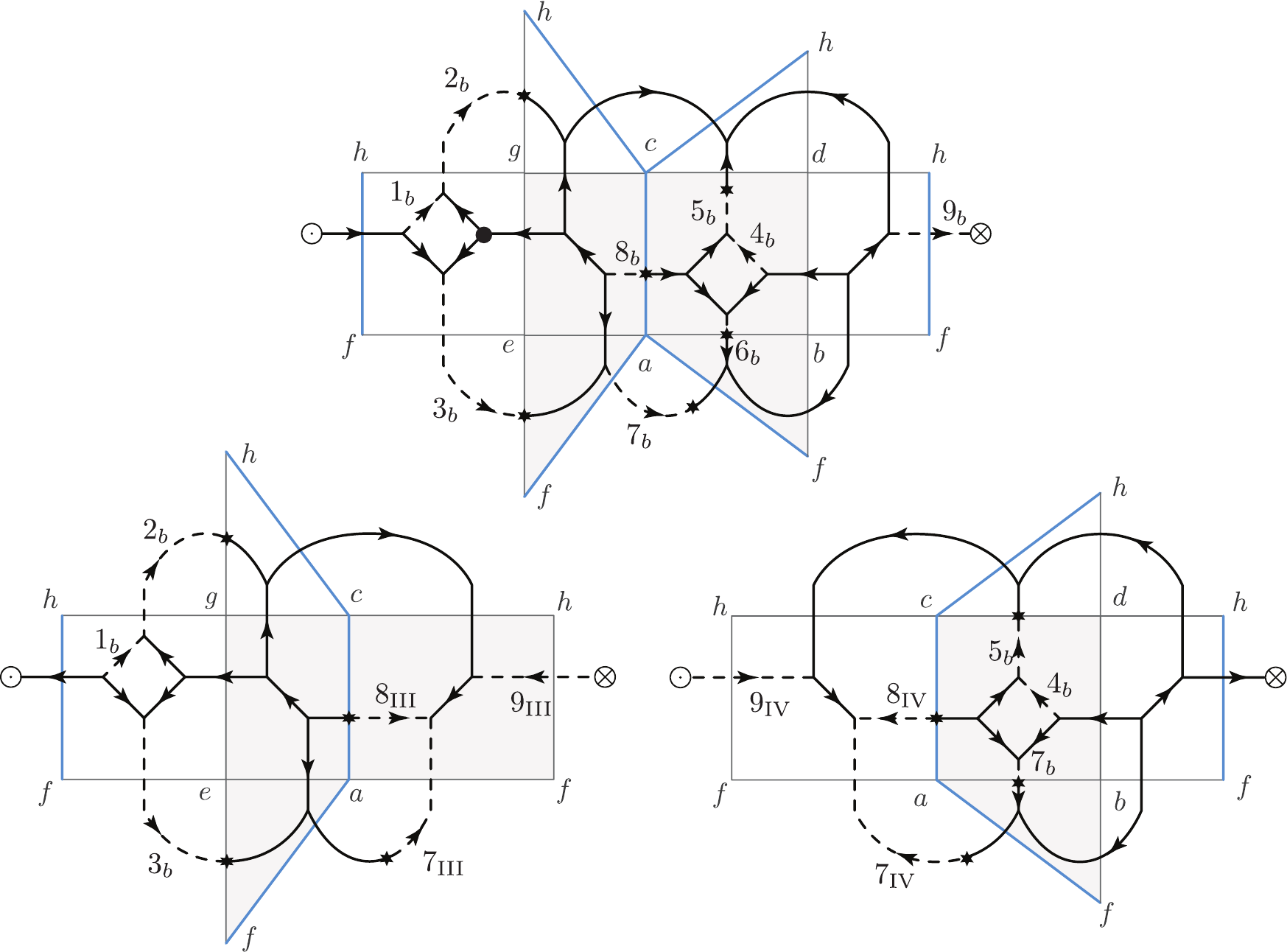}
	\caption{
	The top panel shows the boundary of the (blue) cube, together with the boundary graph, in a planar representation. The tree is given by the solid links, the leaves are the dashed links. The root is indicated by a black dot. The corners of the cube are labelled with small latin letters, and match the labelling in figure \ref{over1}. 
	 The blue cube is split via an SVD along the plane $(h \, c \, a \, f)$ (indicated by a blue line in the top panel) into two prisms ${\rm III}$ and ${\rm IV}$.  As required this plane intersects three shared leaves namely $G_{7_b}$, $G_{8_b}$ and $G_{9_b}$. To find the result of the splitting one should think of this step as the inverse of the gluing. The boundary graphs associated to the prisms are such that they give the boundary graph of the cube if glued back together.}
	\label{blue_split}
\end{figure}

As discussed in \ref{subsec_technic}, the choice of spanning tree localizes the degrees of freedom. This localization is such that the same number of physical degrees of freedom are distributed between both prisms. We exlained earlier that it requires having three sheared leaves which will be copied for both prisms. These shared leaves are actually the ones intersected by the cutting plane. For instance for the blue cube, an appropriate spanning tree is depicted in figure \ref{blue_split}. We have chosen a planar representation for the boundary of the cube and the blue line indicates the plane along which this boundary is cut.

The holonomy variables associated to the shared leaves are denoted by $G_{7_b}$, $G_{8_b}$ and $G_{9_b}$. As for the gluing, the splitting is better performed in the Fourier transformed picture. However it is only necessary to perform such a transformation for the shared leaves. Following the convention encoded in the grey/white colour of the faces, the transformation reads
\begin{align}\label{blueFT1}
	& \tilde A_{\rm blue} ( G_{1_b},\ldots, G_{6_b},  (\rho_{7_b},m_{7_b},n_{7_b}), (\rho_{8_b},m_{8_b},n_{8_b}),(\rho_{9_b},m_{9_b},n_{9_b})) \nn\\
	& \q = \sum'_{G_{7_b}, G_{8_b}, G_{9_b}} A_{\rm blue} ( G_{1_b},\ldots, G_{6_b}, G_{7_b}, G_{8_b}, G_{9_b}) \nn\\[-0.5em]
	&\q\q \q \q \q \q \times \sqrt{d_{7_b} d_{8_b}d_{9_b}} \,\overline{D^{\rho_{7_b}}_{m_{7_b}n_{7_b}}\big(G_{7_b}\big)} \, \overline{D^{\rho_{8_b}}_{m_{8_b}n_{8_b}}\big(G_{8_b}\big)} D^{\rho_{9_b}}_{m_{9_b}n_{9_b}}\big(G_{9_b}\big)  \; .
\end{align}
We can now perform the splitting which can be seen as an inverse procedure to the gluing. In particular, if we glue the amplitude obtained from the splitting we wish to re-obtain (approximately) the original amplitude. 

In the discussion about the gluing procedure, we saw the representation labels $\{\rho\}$ associated with the leaves which are glued need to agree on the two building blocks. For the splitting procedure, it means the representation labels $\rho_{7_b}, \rho_{8_b}$  and $\rho_{9_b}$ will act as parameters in the SVD--based splitting procedure. These are the analogue of the label $C=(x_2,x_4)$ in (\ref{splittingSq}), where we discussed the 2D decorated tensor network coarse--graining procedure. In this case $C=(x_2,x_4)$ encoded the value of the scalar field on the corners of the to-be-split square, that were copied to both triangles. 

From the gluing procedure we can also see how to split the magnetic indices $(m_{l_b},n_{l_b})$ for $l_b=7,8,9$. These are distributed to the left and right prisms depending on the direction of the links carrying these leaves. If the target node of the (full) link $l$ belongs to the `left' half of the cube we associate the index $m_{l_b}$  to the left prism, otherwise to the right. Similarly for the source node and the index $n_{l_b}$. 

Finally we have to distribute the holonomy variables $G_{1_b}, \dots,G_{6_b}$ to the two resulting prisms. This is determined by the position of the associated leaf: according to figure \ref{blue_split} we attribute $G_{1_b}, \dots,G_{3_b}$ to the left prism and $G_{4_b}, \dots, G_{6_b}$ to the right prism. 

With these preliminaries settled we can define a family of matrices 
\ba
M^C_{AB} \equiv \sqrt{d_{\rho_{7_b}}d_{\rho_{8_b}}d_{\rho_{9_b}}}  
\tilde A_{\rm blue} \big( G_{1_b},\ldots, G_{6_b},  (\rho_{7_b},m_{7_b},n_{7_b}), (\rho_{8_b},m_{8_b},n_{8_b}),(\rho_{9_b},m_{9_b},n_{9_b}) \big)
\ea
parametrized by an index $C=(\rho_{7_b}, \rho_{8_b},\rho_{9_b})$, and with $A=(G_{1_b}, G_{2_b}, G_{3_b}, n_{7_b},n_{8_b}, m_{9_b})$ and $B=(G_{4_b}, G_{5_b}, G_{6_b}, m_{7_b},m_{8_b}, n_{9_b})$. Using a singular value decomposition (see appendix \ref{app_svd}), we write this family of matrices as a family of products of matrices
\ba\label{splitB1}
M^C_{AB} \,\approx \, \sum_{i=1}^\chi (S_1)^C_{Ai} (S_2)^C_{Bi} \; .
\ea
The $(S_1)^C_{Ai}$ and $(S_2)^C_{Bi}$ matrices encode the amplitudes for the left and right prisms respectively. We will comment on how to fix the value of $\chi$ in a moment. 

We again wish to understand these amplitudes as functionals of holonomies defined on the boundary graphs of the prisms.  We therefore need to choose a boundary graph for each of the two prisms, together with a white/grey colouring convention for the new face on each building block. Of course the choice for the two building blocks has to match, that is we must be able to glue back the two building blocks along the new face. This means that one of the new face must be white and the other one grey. Nevertheless, some freedom remains about which face to colour in white for instance. The choice we make is such that no change of colours is required in any of the subsequent steps.

For the boundary graph on the new face we choose the coarsest one possible which connects the four links entering or leaving the new face and which is three--valent. Therefore we only need to introduce an additional link (see figure \ref{blue_split}). This completes the boundary graph for the prisms and determines the number of leaves to be six. These six leaves are already determined by the three shared leaves crossing into the new face and the other three leaves distributed over the remaining part of the prism. Thus the additional link added to the new face of a given prism is a tree--link. This fixes the spanning tree for the prisms. 

We choose the minimal\footnote{
Note that it is possible to accommodate a higher order approximation, that is choose a larger $\chi$. This can be either done by introducing a proper tensor network, whose vertices sit inside the building blocks and whose edges are dual to the faces of these building blocks. This means that these faces carry additional indices, labelling different copies of the boundary graph data. See \cite{decorated} for a detailed discussion. An alternative procedure is to reconstruct a finer boundary graph on the new face, {\it e.g.} we could have introduced an additional cycle on the new face. This would eventually lead to a finer boundary graph on the full cube.}  
bond dimension $\chi$, so that we can understand the amplitude encoded in the matrices $(S_1)^C_{Ai}$ and $(S_2)^C_{Bi}$ as a gauge fixed (and then partially Fourier transformed) function of holonomies associated to the boundary graph. Consider for instance the left prism. The indices $A$ and $C$ represent already part of the variables or boundary data, which we need for the given boundary graph. What is missing are three magnetic indices\footnote{
The somewhat peculiar naming of indices is due to having to name uniquely all variables appearing in one full coarse--graining cycle. The Roman numerals differentiate between the four prisms arising from the two cutting procedures for the red and blue cube respectively.} 
  $k_{7_{\rm III}}, k_{8_{\rm III}},k_{9_{\rm III}}$ to be associated to the nodes of the shared leaves in the new face. Thus the index $i$ in (\ref{splitB1}) has to account for these three magnetic indices. The range of the latter is determined by the representation labels $\rho_{7_b},\dots,\rho_{9_b}$  we therefore need 
\ba
\chi \,=\, d_{\rho_{7_b}} d_{\rho_{8_b}} d_{\rho_{9_b}} \; .
\ea
We also need to distribute the $\chi$ index values $i$ onto the $\chi$ index combinations $(k_{7_{\rm III}}, k_{8_{\rm III}},k_{9_{\rm III}})$ for the left prism and $(k_{7_{\rm IV}}, k_{8_{\rm IV}},k_{9_{\rm IV}})$ for the right prism.  We do this by choosing  matrices $R^{\rm III}_{i,\{k_{\rm III}\}}$ and $R^{\rm IV}_{i,\{k_{\rm IV}\}}$ such that
\ba\label{isocond1}
\sum_{\{k\}}  R^{\rm III}_{i,\{k_{\rm III}\}} R^{\rm IV}_{i',\{k_{\rm IV}\}} &=& \delta_{i,i'} \; .
\ea
This guarantees that 
\ba
\tilde {\cal A}_{\rm III}(A,C,\{k_{\rm III}\} ) = \sum_{i=1}^\chi  (S_1)^C_{A,i} R^{\rm III}_{i,\{k_{\rm III}\}} \q ,\q \tilde {\cal A}_{\rm IV}(B,C,\{k_{\rm IV}\} ) = \sum_{i=1}^\chi  (S_2)^C_{B,i} R^{\rm III}_{i,\{k_{\rm IV}\}} 
\ea
define an equally good splitting as the $S_1$ and $S_2$ matrices. Using the previously defined encoding of the indices $A,B,C$  we have thus defined the (partially Fourier transformed) amplitudes for the left and right prism respectively.

In addition to   (\ref{isocond1})   we also require to choose $R^{\rm III}$ and $R^{\rm IV}$ such that the resulting coarse--graining procedure preserves the fixed point given by BF theory. For this fixed point we will have a maximal number of non--vanishing singular values, which makes the action of  $R^{\rm III}$ and $R^{\rm IV}$ non--trivial for any combination of representations $\rho_{7_b}$, $\rho_{8_b}$ and $\rho_{9_b}$. In contrast for the strong coupling fixed point, we would only obtain a non--vanishing singular value for the trivial representations, and the corresponding requirement for the $R$ matrices would be empty.

This finishes the discussion for the splitting of the blue cubes. The same procedure applies for the splitting of any building block.

~\\
{\bf {\small Gauge averaging and rescaling of the amplitude, rotation of the coarse--graining plane:}}  \\
The algorithm proceeds by splitting cubes into prisms, and gluing these prisms into cubes again. At the end of each coarse--graining cycle we have to perform three operations, before starting the cycle again.

The first operation is a gauge averaging procedure.  This is to make sure that violations of gauge invariance of the amplitude introduced by our procedure are projected out again. We use a gauge fixing via a spanning tree, so the only remaining gauge invariance is the invariance under a common adjoint action, corresponding to a gauge transformation at the root node of the spanning tree.  Thus after having completed the gluing of the prisms to a new cube amplitude ${\cal A}'_{\rm cube}$ and applying all the necessary inverse Fourier transforms to express this amplitude in terms of holonomy variables, we apply an averaging
\ba
{\cal A}''_{\rm cube} =\sum'_{h} {\cal A}'_{\rm cube} ( h G_{\ell_1} h^{-1}, \ldots, h G_{\ell_9} h^{-1}) \; .
\ea
Secondly, we will impose a normalization condition. That is after each coarse--graining step we will rescale the cube amplitude, such that a certain normalization condition is satisfied. This will avoid divergent partition functions, as arise for instance in the BF case. As a normalization condition, we will choose to keep the total average of the cube amplitude 
\ba
\sum_{G_1, \ldots G_9} {\cal A}_{\rm cube}(G_1, \ldots, G_9) \,=\, \text{const.}
\ea
to be constant, or in the Fourier transformed picture, to fix the value of $\tilde A_{\rm cube}$ for all representation labels $\rho$ being trivial.  Note that the rescaling factors should be saved if one wishes to evaluate the partition function. This partition function would be given by a product of the rescaling factors from each coarse--graining step multiplied by a factor arising from the gluing of the final cube (amplitude) to a closed 3--torus.

Finally, as explained in section \ref{Aoverview} , as the coarse--graining happens to change the scales only in a two--dimensional plane, after each step of coarse--graining we have to rotate the plan in which the coarse--graining takes place.  This can be implemented by a change of the spanning tree and a renaming of variables.

\subsection{Detailed algorithm}

\begin{figure}[h]
	\centering
	\includegraphics[scale = 0.7]{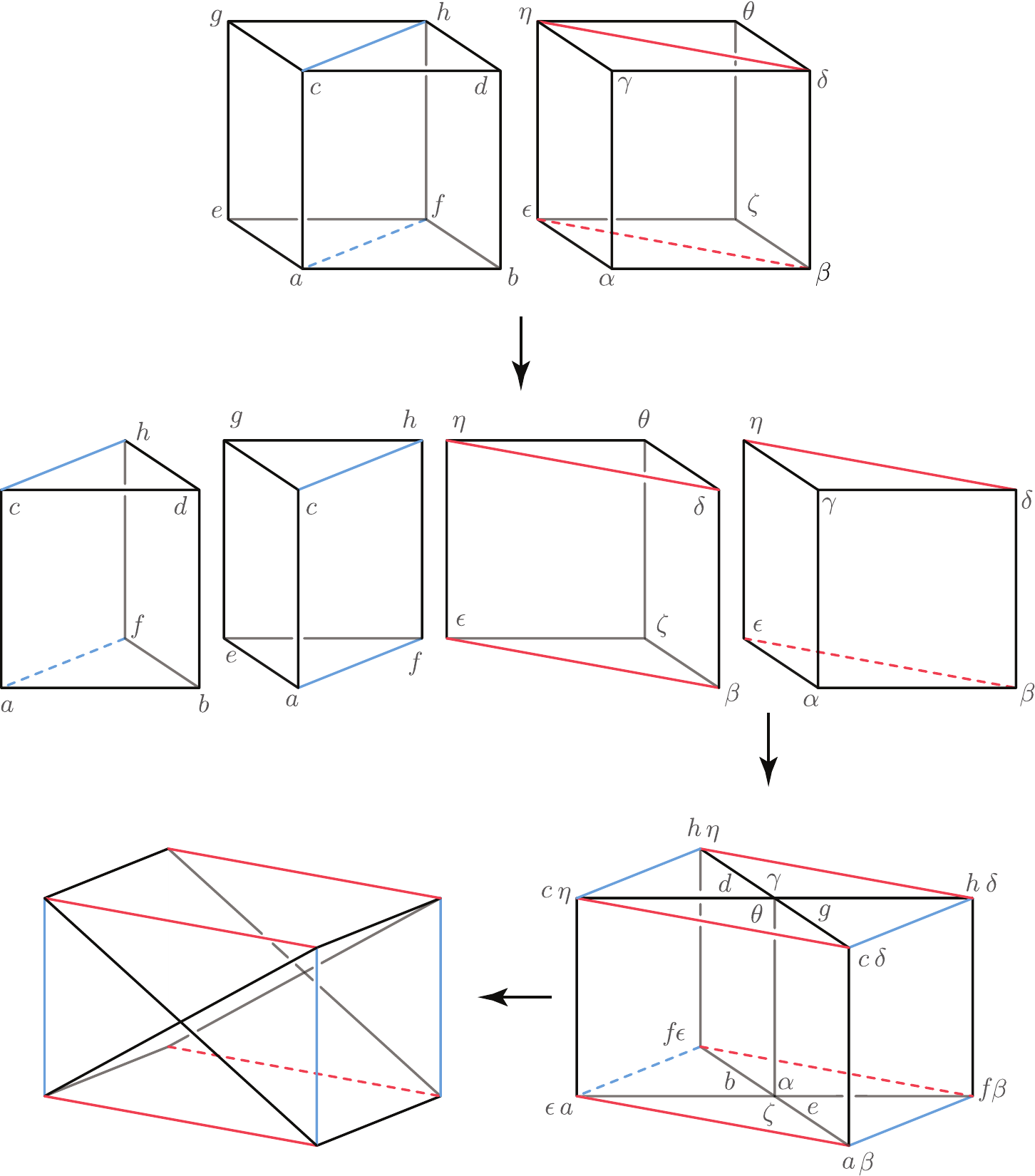}
	\caption{Thanks to the translational symmetry of the cubical lattice, it is possible to focus on a single cube. First a copy of the initial cube is made so as to get two identical cubes. These cubes are referred as the blue cube and the red cube. Both cubes are split in halves with respect to two transversal planes respectively depicted by the blue and red lines. This operation gives four prisms. The four prisms are then put back together in order to get a bigger cube. The last step of the iteration is a rotation so that the next splitting will happen in a orthogonal plane.}
	\label{over1}
\end{figure}

We have discussed all the different operations appearing in the coarse--graining algorithm. We will now describe the succession of steps for one full coarse--graining iteration. This is most easily described in the planar boundary graph representations, as already used in figure \ref{blue_split}. %Note that after a full iteration of the algorithm, the coarse-grained shape is not a cube anymore. % but a parallelepiped. Similarly, an additional iteration turns it into a cuboid. 
%However we will keep referring to the resulting shapes as `cubes'.  
Figure \ref{over1} presents the main steps of the algorithms: first a splitting of a `blue' and `red' cube along different planes into prisms, and then a gluing of four prisms into a larger cube. (The resulting shape is actually not a cube anymore, but we will nevertheless refer to it as such.)

\begin{figure}[!h]
	\centering
	\includegraphics[scale = 0.7]{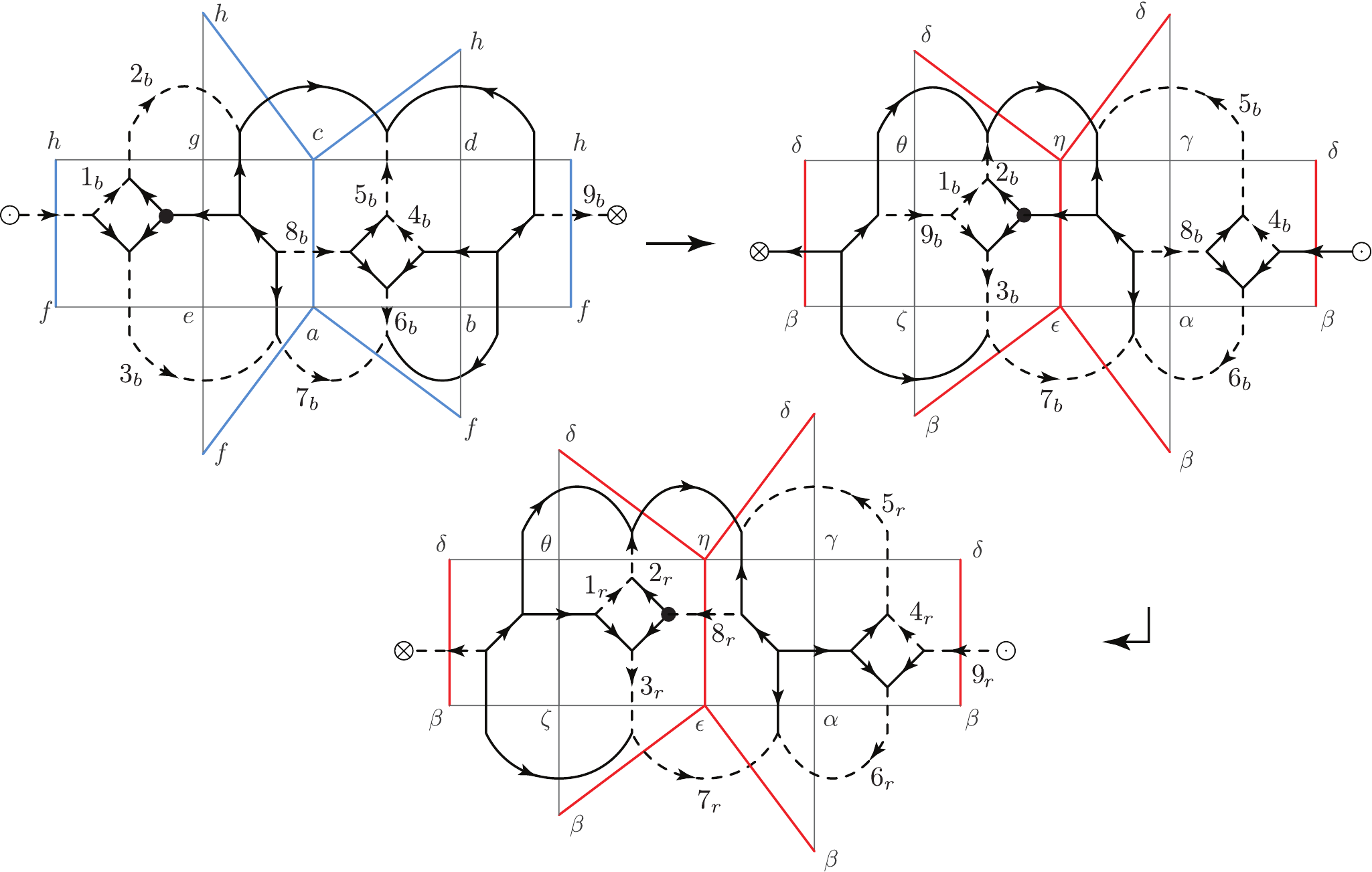}
	\caption{Instead of working directly with the 3d shape and its boundary graph, we look at the 2d net obtained by unfolding the cube. The choice of spanning tree for the blue cube is made so as to have three shared leaves. The shared leaves labelled by $7_b$, $8_b$ and $9_b$ are the ones which are cut during the splitting. To obtain the red cube from the blue cube, a change of boundary triangulation is performed corresponding to the different plane along which the cube is cut. The bottom panel represents the result of the tree transformation  necessary to ensure that the red cube has three shared leaves labelled by $7_r$, $8_r$ and $9_r$.}
	\label{red0}
\end{figure}

\newpage
~\\
{\bf {\small Amplitudes for blue and red cubes:}} \\
We assume that we are given the amplitude for the `blue' cube in terms of holonomy variables. The colour indicates along which plane we cut the cube into prisms.  We choose a spanning tree that is adapted to this cutting procedure, namely so that we have three `shared leaves', as depicted in figure \ref{red0}.

The `red' cube is cut along a different plane than the initial `blue' cube. This requires two changes. Firstly, we need to change the triangulation of the surface of the cube. More precisely, the diagonals for two of the faces need to be changed in order to match the new cutting plane. Correspondingly the boundary graph is changed. One can however check that this does not affect the cycles as defined by the spanning tree for the blue cube. Therefore, the corresponding loop holonomies remain unchanged and the amplitude is not affected. 

The second change, which needs to be performed when going from the blue cube to the red cube, is a modification of the spanning tree. This is necessary to satisfy the requirement that the cutting plane goes through three `shared leaves' as for the blue cube. The spanning trees for the blue and red cubes are depicted in figure \ref{red0}.  The corresponding transformation of the amplitude follows from the general formula \eqref{Ttrans} and is given by
\begin{align}
{\cal A}_{\rm red} \big( G_{1_r}, \ldots G_{9_r} \big) &=  
{\cal A}_{\rm blue} \big(G_{1_r},(G_{8_r}G_{2_r}),(G_{8_r}G_{3_r}),(G_{8_r}G_{9_r}G_{4_r}G_{9_r}^{-1}G_{8_r}^{-1}),
	(G_{8_r}G_{5_r}G_{9_r}G_{8_r}^{-1}),  \nn\\ &\q\q\q \q (G_{8_r}G_{6_r}G_{9_r}G_{8_r}^{-1}),(G_{8_r}G_{7_r}G_{8_r}^{-1}),(G_{8_r}G_{9_r}^{-1}G_{8_r}^{-1}),G_{8_r}^{-1}\big) \; .
\end{align}

\begin{figure}[h]
	\centering
	\includegraphics[scale = 0.7]{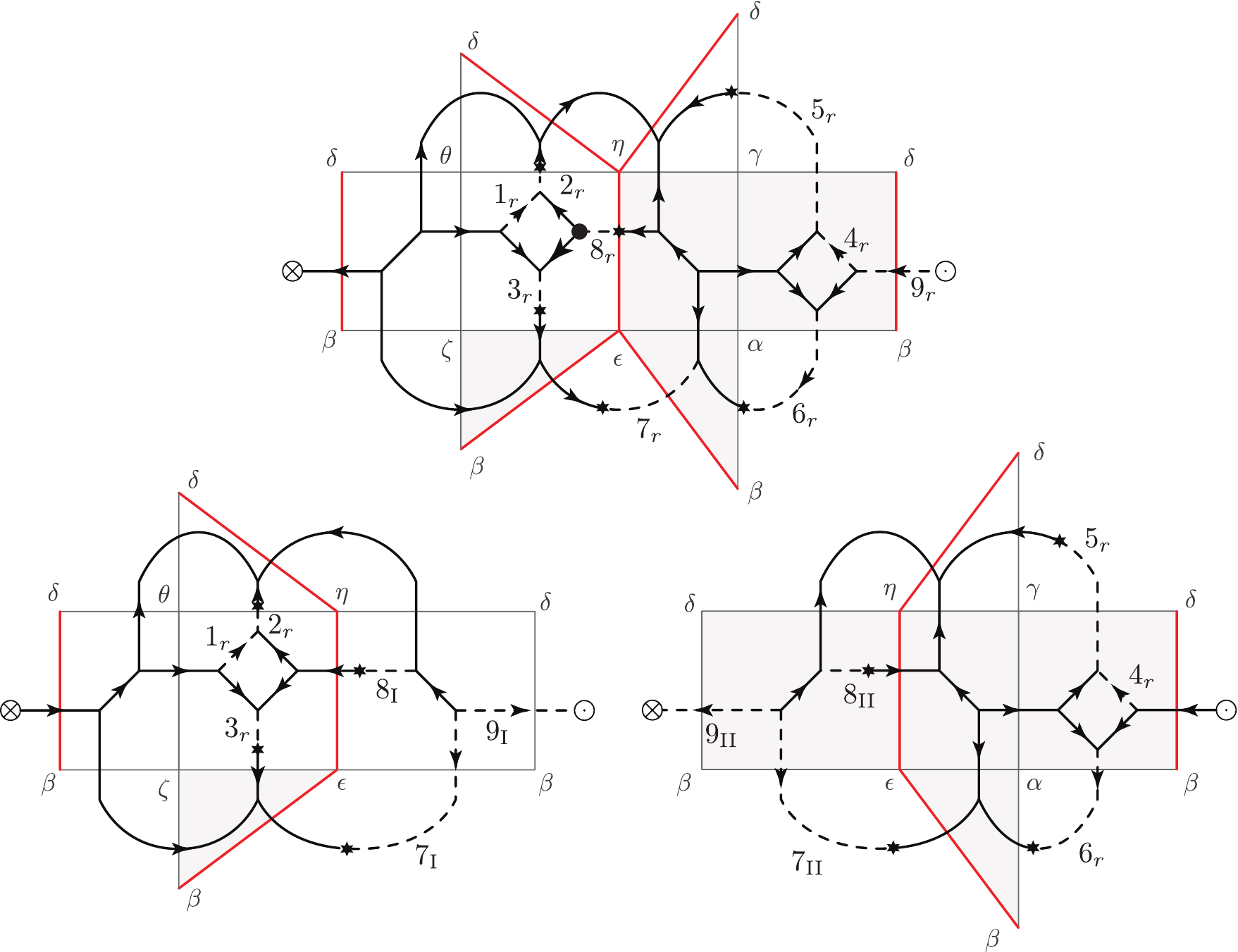}
	\caption{The red cube is split via an SVD along the plane $(\delta \, \eta \, \epsilon \, \beta)$ into the prisms $\RN{1}$ and $\RN{2}$. As required this plane intersects three shared leaves, namely $G_{7_b}$, $G_{8_b}$ and $G_{9_b}$.}
	\label{red_split}
\end{figure}

~\\
{\bf {\small Splitting of the cubes:}}\\
We can now proceed to split the blue and red cubes into two prisms each. To this end we first have to perform a group Fourier transform for the set of `shared leaves', as prescribed for the blue cube in \eqref{blueFT1}. We then perform the splitting as described in detail in the previous section. The procedure for the blue cube is depicted figure \ref{blue_split}. The splitting operation for the red cube proceeds in the same way and the choice of conventions is displayed in figure \ref{red_split}.

This procedure leads to four prisms and the corresponding amplitudes are denoted by ${\cal A}_{\RN{1}},\cdots, {\cal A}_{\RN{4}}$ where $(\RN{1},\RN{2})$ label the prisms resulting from the splitting of red cube and $(\RN{3},\RN{4})$ the ones resulting from the splitting of the blue cube. The same labeling is used for the leaves which have been split.

\newpage

~\\
{\bf {\small Gluing of the prisms into larger prisms:}}\\
We continue the coarse--graining procedure with gluing the pair of prisms ${\RN{2}}$ and $\RN{3}$ to a larger prism labeled by $\RN{2}{\sss \times}\RN{3}$.  Similarly we glue the prisms $\RN{1}$ and $\RN{4}$ to a larger prism labeled by $\RN{1}{\sss \times}\RN{4}$. 

Let us describe the gluing of the prisms ${\rm II}$ and ${\rm III}$ in more details. First we need to perform a Fourier transform for the leaves which are about to be glued. For instance, these are the leaves $G_{1_b}$, $G_{2_b}$ and $G_{9_\RN{3}}$ for the prism ${\rm III}$. We then apply the procedure described in the previous section for each  of the three pairs of shared leaves. In particular, for a given pair of leaves, this means summing over one of the magnetic indices and distributing the remaining ones over the new leaf. This distribution of indices is performed according to the orientation of the corresponding links as well as the colour of the faces which are identified. Furthermore, this gluing step involves the matching of a full leaf (corresponding to a cycle inside the glued face) for which we identify and sum over the labels $(\rho,m,n)$ associated to the full leaf.

\begin{figure}[!h]
	\centering
	\includegraphics[scale = 0.7]{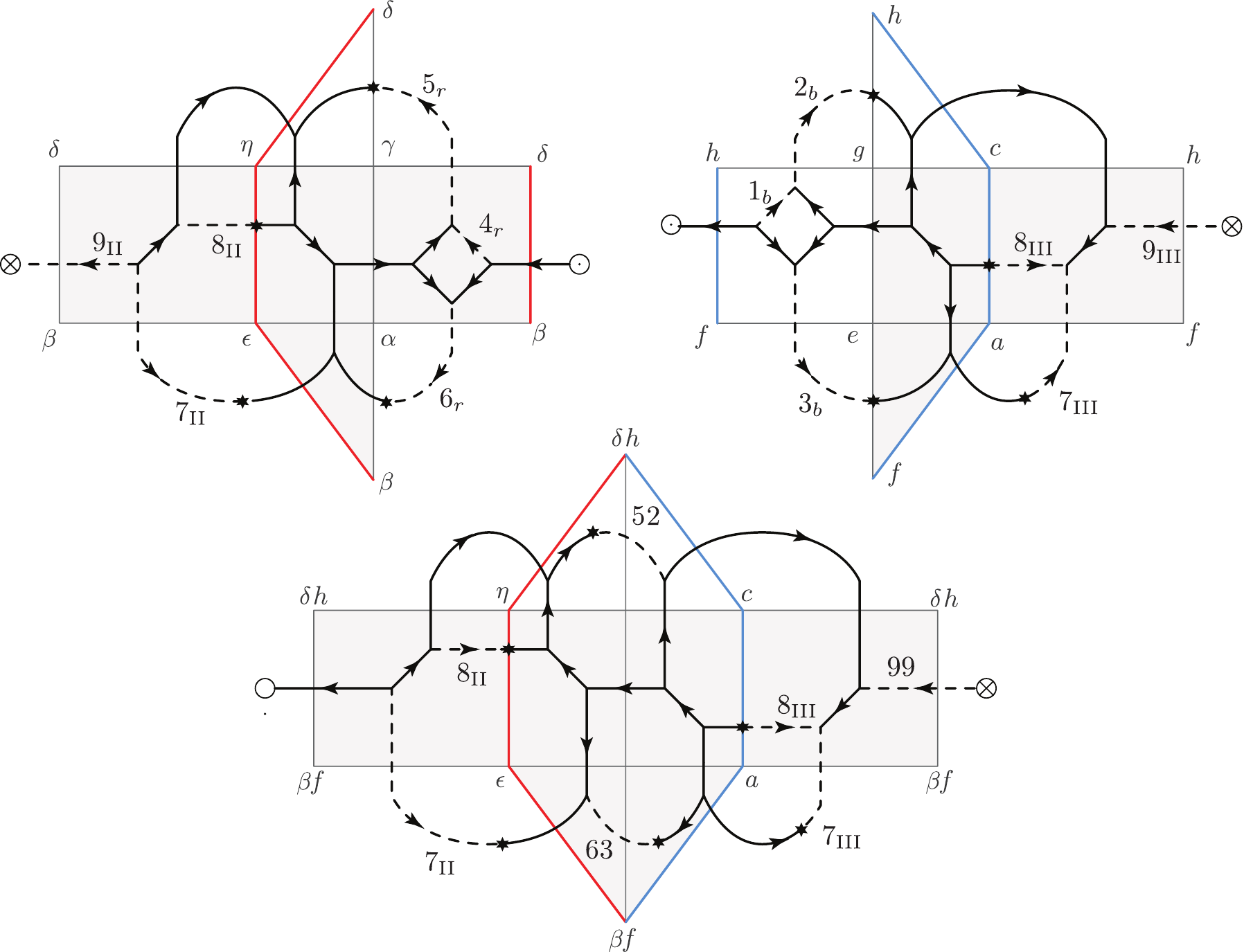}
	\caption{The prisms $\RN{2}$ and $\RN{3}$ are glued together along the face $(\gamma \, \delta \, \beta \, \alpha)$ to form the larger prism $\RN{2}{\sss \times}\RN{3}$. The pairs of leaves merged together are $(G_{5_r},G_{2_b})$, $(G_{6_r},G_{3_b})$ and $(G_{9_{\RN{2}}},G_{9_{\RN{3}}})$.}\vspace{-0.2em}
	\label{gluing1}
\end{figure}

\begin{figure}[!h]
	\centering
	\includegraphics[scale = 0.7]{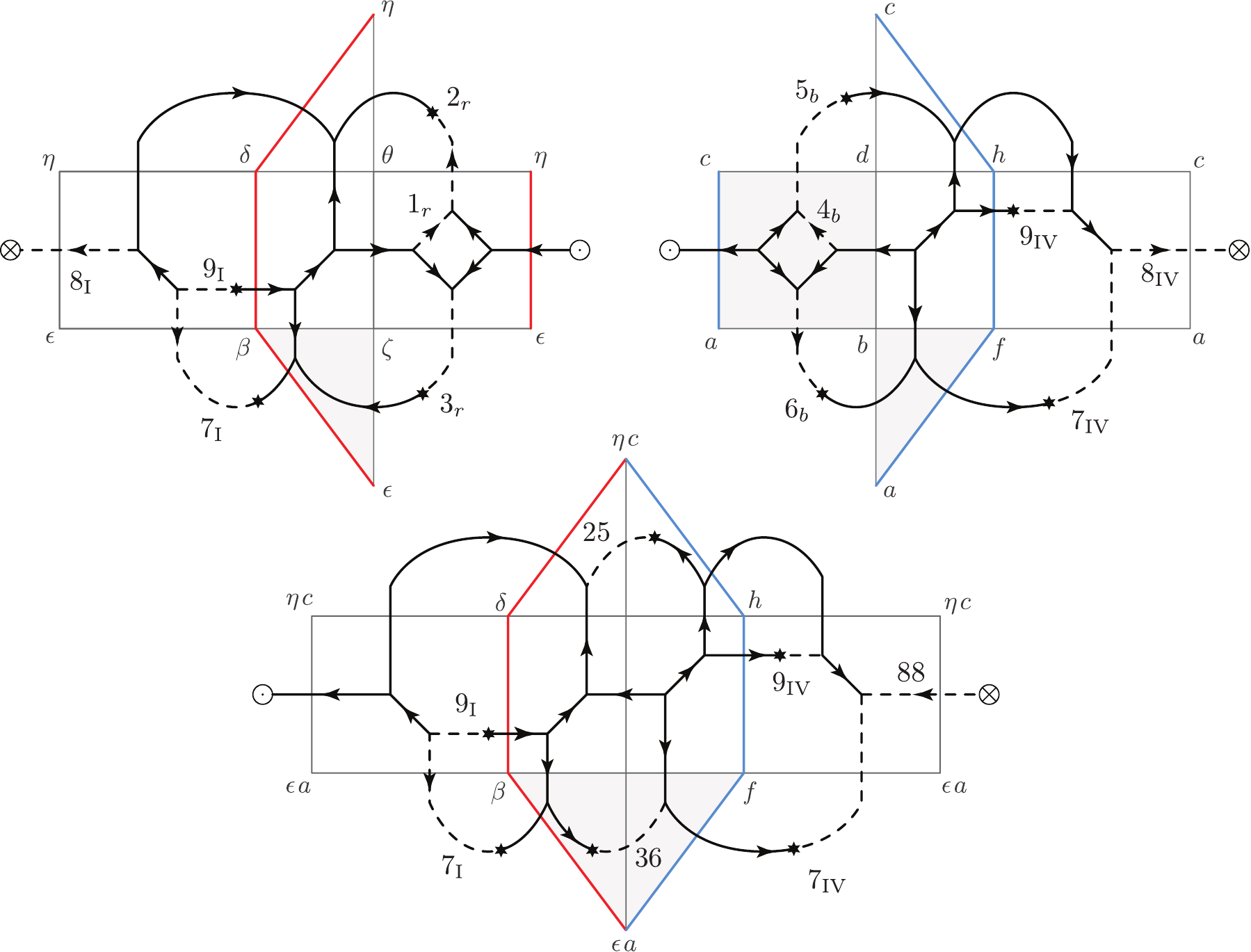}
	\caption{The prisms $\RN{1}$ and $\RN{4}$ are glued together along the face $(\theta \, \eta \, \epsilon \, \zeta)$ to form the larger prism $\RN{1}{\sss \times}\RN{4}$. The pairs of leaves merged together are $(G_{2_r},G_{5_b})$, $(G_{3_r},G_{6_b})$ and $(G_{8_{\RN{1}}},G_{8_{\RN{4}}})$.}
	\label{gluing2}
\end{figure}

Putting everything together we obtain the following gluing formula for the prism $\RN{2} {\sss \times}\RN{3}$

\begin{align}
	&\widetilde{\mathcal{A}}_{\RN{2}{\sss \times}\RN{3}}\big((\rho_{52},m_{2_b},m_{5_r}),(\rho_{63},m_{6_r},m_{3_b}),G_{7_\RN{2}},G_{7_\RN{3}}, G_{8_\RN{2}},G_{8_\RN{3}},(\rho_{99},k_{9_\RN{3}},k_{9_\RN{2}})\big) \\[0.2em] \nn 
	& \q =\sum_{\rho_{41},m_{41},n_{41} \atop p_{52},p_{63},p_{99}} \widetilde{\mathcal{A}}_{\RN{2}}\big( (\rho_{41},m_{41},n_{41}),(\rho_{52},m_{5_r},p_{52}),(\rho_{63},m_{6_r},p_{63}),G_{7_\RN{2}},G_{8_\RN{2}},(\rho_{99},p_{99},k_{9_\RN{2}})\big) \\[-0.7em] \nn
	& \q \q \q \q \q \;\;\, \times \widetilde{\mathcal{A}}_{\RN{3}}\big( (\rho_{41},m_{41},n_{41}),(\rho_{52},m_{2_b},p_{52}),(\rho_{63},m_{3_b},p_{63}),G_{7_\RN{3}},G_{8_\RN{3}},(\rho_{99},k_{9_\RN{3}},p_{99})\big) \\ \nn
	& \q \q \q \q \q \; \; \, \times (d_{\rho_{52}}d_{\rho_{63}}d_{\rho_{99}})^{-1/2} \; .
\end{align}
The corresponding conventions are depicted in figure \ref{gluing1}. We proceed similarly with the prisms ${\rm I}$ and ${\rm IV}$ and the operation is encoded in figure \ref{gluing2}.

\begin{figure}[h]
	\centering
	\includegraphics[scale = 0.7]{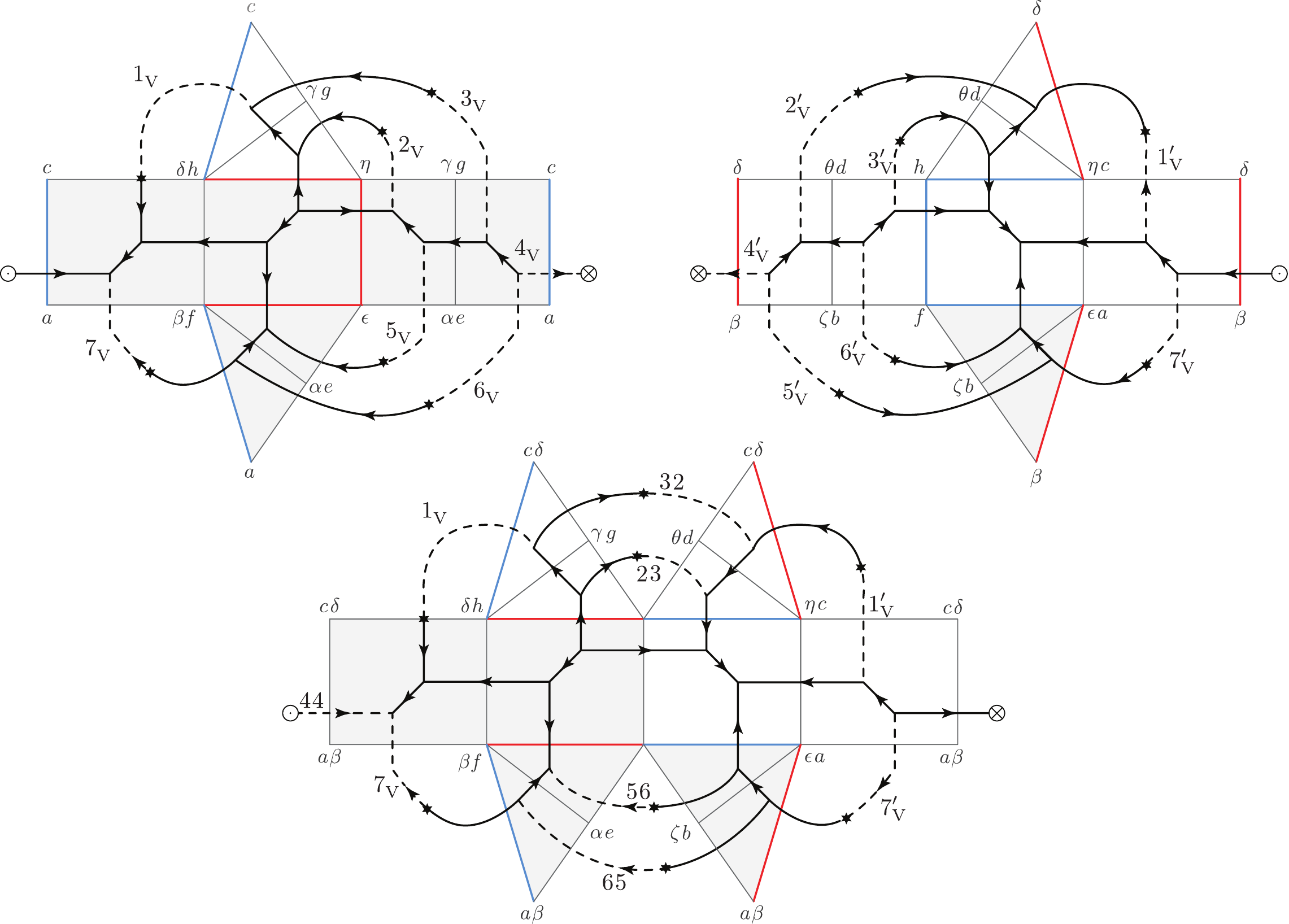}
	\caption{The prisms $\RN{2} {\sss \times}\RN{3}$ and $\RN{1} {\sss \times}\RN{4}$ obtained after tree transformation of the prisms resulting from the first gluing step are put together in order to get the new bigger cube.}
	\label{gluingfin}
\end{figure}

~\\
{\bf {\small  Gluing of the two larger prisms into a new cube:}}\\
The prisms $\RN{2} {\sss \times}\RN{3}$ and $\RN{1} {\sss \times}\RN{4}$ resulting from the first gluing operation are then glued together in order to obtain the coarse-grained cube. However it is first necessary to perform a tree transformation for both larger prisms in order to obtain the required number of shared leaves, namely five for this configuration. The loop--holonomies, for the prisms $\RN{2} {\sss \times}\RN{3}$ and $\RN{1} {\sss \times}\RN{4}$, obtained after the tree transformation are now denoted by $\{G_\RN{5}\}$ and $\{G_\RN{5}'\}$ respectively. Furthermore, the leaves which about to be glued are Fourier transformed. The gluing finally proceeds as before and is depicted in figure \ref{gluingfin}.

\begin{figure}[h]
	\centering
	\includegraphics[scale = 0.7]{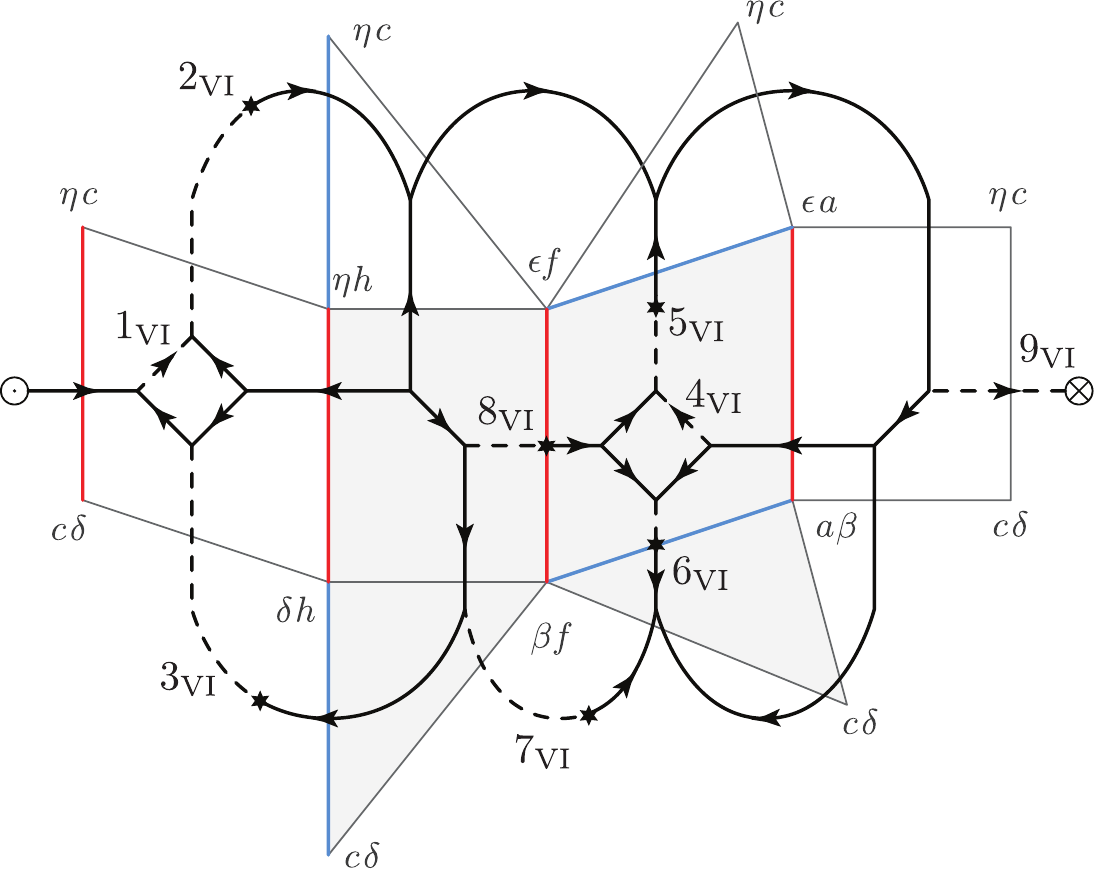}
	\caption{The cube obtained after the final gluing is rotated in preparation for the next iteration of the algorithm which is performed in a different plane. The spanning tree is modified in order to match exactly the one of the initial configuration.}
	\label{rotation}
\end{figure}

~\\
{\bf {\small  Final steps:}}\\
The last gluing step gives the amplitude for the coarse-grained cube, expressed with respect to a certain spanning tree. It is left to prepare this new cube for the next iteration which is performed in a different plane. We thus rotate the cube, Fourier transform all the leaves back to the `group' representation and then change the spanning tree to match the one of the initial cube. The corresponding loop--holonomies are denoted by $\{G_{\RN{6}}\}$. The result of the rotation as well as the tree transformation is presented figure \ref{rotation}. Up to irrelevant deformations of the 2d faces, this reproduces exactly the initial configuration depicted in figure \ref{blue_split}.

Finally we need to perform a gauge averaging in order to ensure that the global gauge invariance at the root is satisfied as well as a rescaling of the amplitude as described in the previous section. 

We can iterate this procedure a fixed number of times. The flow of the amplitude can be monitored by following the set of singular values, obtained in the splitting procedure. These singular valued are labeled by the representations attached to the three shared leaves. This indicates which representation channels are `excited', {\it i.e.} which representations appear in the (coarse-grained) partition function, and which not.  

Furthermore, monitoring the change of these sets of singular values (or spectra), indicates whether we reached a fixed point. The fixed point amplitudes typically represent amplitudes for topological field theories, {\it e.g.} BF theory or BF theory restricted to a normal subgroup.

\section{Application to finite group spin foam models}\label{sec_results}

\subsection{Parameterization of spin foam models}

Here we present the space of initial lattice gauge models, to which we will apply the Decorated Tensor Network algorithm. More details are presented in appendix \ref{app_param}.

As explained in section \ref{reformGauge} we consider 3D spin foam models. The cubical lattice allows us to also consider `proper spin foam' models, as we do have to consider four--valent intertwiners in this case. This agrees with the 4D models based on a triangulation.  To  test  the Decorated Tensor Network algorithm we choose as gauge group a finite group, here the group of permutations of three elements ${\cal S}_3$. 

We explained in section \ref{sec_latticegauge}, that a key dynamical ingredient of spin foam models is given by the invariant maps $\mathbb{P}'_{e^*}$, which are determined by the way the simplicity constraints are imposed. The space of all possible invariant maps is huge, we will therefore further restrict this space. Here we will follow the parametrization developed in \cite{holonomy1}, which covers the current spin foam models.  This will also allow us to compare the phase diagram obtained for the 3D spin foam models with corresponding phase diagrams for 2D spin net models \cite{merce}.

 First of all we impose a factorizing structure for the maps 
 \ba
 \mathbb{P}'_{e^*}: \text{Inv}(V_{\rho_1}\otimes \dots \otimes V_{\rho_4}) \rightarrow \text{Inv}(V_{\rho_1}\otimes \dots \otimes V_{\rho_4}) \; ,
\ea
namely we demand that these maps can be written as
\be\label{Efct2}
	 \mathbb{P}'_{e^*} = \mathbb{P}_{e^*} \cdot \big( \widetilde{E}(\rho_1)  \otimes \cdots \otimes \widetilde{E}(\rho_4)  \big) \cdot \mathbb{P}_{e^*}
\ee
where  we have introduced linear maps $\widetilde{E}(\rho): V_\rho \rightarrow V_\rho$.  The linear maps $\widetilde{E}(\rho)^n_m$ can be obtained from a group Fourier transform of a function $E$ on the gauge group ${\cal G}$.

A further requirement we will impose, is that the function $E$ is invariant under the adjoint action of a chosen subgroup. This again mimics the construction of the full models where the gauge group is ${\cal G}= \SU(2) \times \SU(2)$ and the subgroup is given by $\SU(2)$.

The $E$--functions are convenient to parametrize the choice of simplicity constraints and of how these are imposed. It is however hard to see how they change the Haar projector  $\mathbb{P}_{e^*}$ to a more general invariant map $\mathbb{P}'_{e^*}$.  To describe this change more directly we will parametrize this map via a $(\rho_1,\ldots,\rho_4)$--labelled family of matrices $C^{[\rho_1,\ldots,\rho_4]}(\iota,\iota')$, where $\{\iota\}$ labels a basis of orthonormal intertwiners on the space $V_{\rho_1}\otimes \dots \otimes V_{\rho_4}$ (see appendix \ref{app_param} for further details). To define the family of matrices we remind the reader that the Haar projector can be written as
\ba
\mathbb{P}_{e^*} \,=\, \sum_\iota |\iota \rangle \langle \iota|  \; .
\ea
Given the invariance properties of the map $\mathbb{P}'_{e^*}$, we can express it also in this basis of intertwiners
\ba
\mathbb{P}_{e^*}' \,=\, \sum_{\iota, \iota'}    C^{[\rho_1,\ldots,\rho_4]}(\iota,\iota') \,\,    |\iota \rangle \langle \iota'|  
\ea
where, for the form (\ref{Efct2}) of $\mathbb{P}'_{e^*}$ the $C$--matrices are given as
\ba
C^{[\rho_1,\ldots,\rho_4]}(\iota,\iota') \,=\,  \langle \iota |   \big( \widetilde{E}(\rho_1)  \otimes \cdots \otimes \widetilde{E}(\rho_4)  \big) |\iota' \rangle  \; .
\ea
For a multiplicity--free group, the tensor product of three representations has a unique intertwiner. In this case a basis of intertwiners on the tensor product of four representations can be constructed from a recoupling scheme. For instance we can couple  the pair $(\rho_1,\rho_2)$   to an intermediate representation $\rho_5$ and $(\rho_3,\rho_4)$ to the dual of this representation, that is $\rho_5^*$. To obtain an invariant tensor, labelled by $\rho_5$ one then couples $\rho_5$ and $\rho_5^*$ to the trivial representation. We can thus choose a basis $\{\iota\} \equiv \{\rho_5\}$ and the matrix $C$ can be completely labelled with representations $C^{[\rho_1,\ldots,\rho_4]}(\rho_5,\rho'_5)$. 

From the family of matrices $C$ we can more easily read off the behaviour of the map $\mathbb{P}'_{e^*}$. In particular we might encounter non--diagonal matrices $C$. These indicate a form of simplicity constraints that cannot be reabsorbed into a change of face weights. Note that (with our assumptions) we can always find an alternative basis of intertwiners for which the $C$ matrices are diagonal.

~\\
{\bf {\small The case of the symmetric group $\mathcal{S}_3$:}}\\
We now specify to the gauge group $\mathcal{S}_3$. The group $\mathcal{S}_3$ is the group of permutations of three elements. Denoting $e$ the unit element, $a = (12)$ the permutation of the first two elements and $b = (123)$ the cycle in the direction $1,2,3$, we obtain an expression for the set of six elements
\be
	\mathcal{S}_3 = \{e,a,bab^{-1},b^2ab^{-2},b,b^2\}.
\ee
 This group possesses two subgroups, namely $\mathbb{Z}_2 = \{e,a\}$ and $\mathbb{Z}_3 = \{ e,b,b^2\}$. In order to specify the $E$-function, we require invariance under adjoint action of one subgroup. Choosing $\mathbb{Z}_2$ as the subgroup (as it leads to a more interesting phase space), we obtain a space of $E$--functions 
\be
	E_{\mathbb{Z}_2 }(g) = \kappa \delta(g,e) + \alpha \, \delta(g,a) + \beta \big( \delta(g,bab^{-1})+\delta(g,b^2ab^{-2})\big)+\gamma \big( \delta(g,b)+\delta(g,b^2)\big)
	\label{Param1}
\ee
parametrized by four real numbers $\kappa, \alpha, \beta, \gamma$.

The group $\mathcal{S}_3$ has three irreducible representations, which we will denote by $\rho=0$ for the trivial representation, by $\rho=1$ the sign representation (assigning to a group element the sign of the permutation it describes), and by $\rho=2$ the two--dimensional representation.

With these definitions at hand, it is now possible to compute explicitly the family of $C$--matrices. The only case of an intertwiner space larger than one--dimensional arises for the quadruple of representations $[2,2,2,2]$. In this case the intertwiner space is three--dimensional, and can be described by an intermediate representation (for the coupling of two pairs) $\rho_5=0,1$ or $2$.   We will also give the $C$--`matrices for two one--dimensional intertwiner spaces, needed to fix the normalization condition for the projector $\mathbb{P}'_{e^*}$.
\begin{align}
	&C^{[0,0,0,0]}(0,0) = { \footnotesize \begin{pmatrix}
	(\alpha+2(\beta+\gamma)+\kappa)^4 )	\end{pmatrix}} \; , \q 
	C^{[1,1,1,1]}(0,0) = { \footnotesize \begin{pmatrix}
	(\alpha+2\beta-2\gamma-\kappa)^4)
	\end{pmatrix}} \\~ \nn \\
	& \q \q 
	C^{[2,2,2,2]}(\rho_5,\rho'_5) = { \footnotesize \begin{pmatrix}
	((\alpha-\beta)^2+(\gamma-\kappa)^2)^2 & 2\sqrt{2}(\alpha-\beta)^2(\gamma-\kappa)^2 & 0 \\ 2\sqrt{2}(\alpha-\beta)^2(\gamma-\kappa)^2 & (\alpha-\beta)^4+(\gamma-\kappa)^4 & 0 \\ 0 & 0 & (\alpha^2-2\alpha\beta+\beta^2+(\gamma-\kappa)^2)^2 
	\end{pmatrix}}.
\end{align}
The matrix $C^{[2,2,2,2]}$ is not diagonal, showing that the space of models is larger than standard lattice gauge theories (for which these matrices are diagonal). 

We look furthermore for a configuration that satisfies the projector conditions, {\it e.g.} for which $C \cdot C=C$. This is satisfied by
\be
	\kappa = \frac{2^{3/4}}{3^{5/4}}\; ,\q  \alpha = \frac{2^{3/4}}{3^{5/4}} \; , \q \beta = -\frac{1}{3\times 6^{1/4}} \; , \q \gamma = -\frac{1}{3 \times 6^{1/4}}.
	\label{nontriv}
\ee
Evaluating the matrix $C^{[2,2,2,2]}$ for these values gives
\be
	\label{Cmatrix1}
C^{[2,2,2,2]}(j_5,j_6) = { \footnotesize \begin{pmatrix}
		2/3 & \sqrt{2}/3 & 0 \\ \sqrt{2}/3 & 1/3 & 0 \\ 0 & 0 & 0 
\end{pmatrix}}.
\ee
This implements a projection on a one--dimensional intertwiner space spanned by  $(\sqrt{2},1)$.

\subsection{Results}

\begin{figure}[h]
	\centering
	\includegraphics[scale = 0.4]{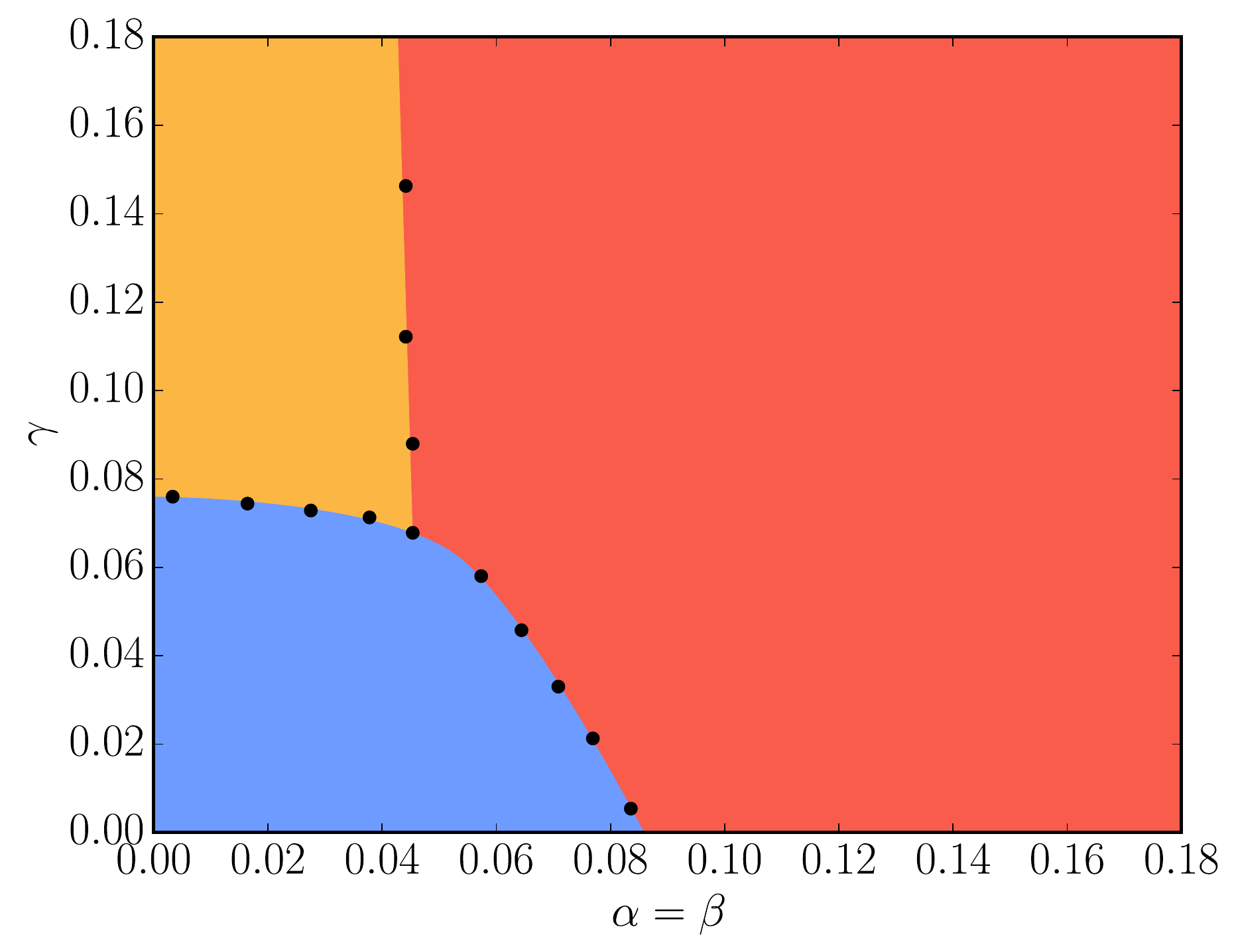}
	\includegraphics[scale = 1.25]{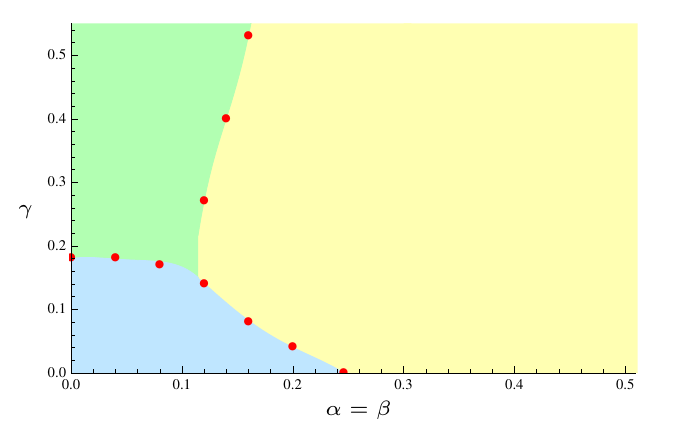}
	\caption{The left panel shows a phase diagrams using the $E$--function parametrization with $\alpha=\beta$. The blue phase (bottom--left) is associated to the BF fixed point, the red one phase (right) corresponds to the strong coupling phase while the orange one (top-left) is a BF phase  with respect to  ${\mathcal S}_3/\mathbb{Z}_3 \simeq \mathbb{Z}_2$. The black dots represent configurations that have been tested along the phase transition line. For the convenience of the reader we reprinted in the right panel the phase diagram for the same $E$--function parametrization for 2D spin nets from \protect\cite{merce}. There light blue indicates the BF phase, yellow the strong coupling phase and green the BF($\mathbb{Z}_2 $) phase.  }
	\label{fig:diagram1}
\end{figure}

\begin{figure}[h]
	\centering
	\includegraphics[scale = 0.4]{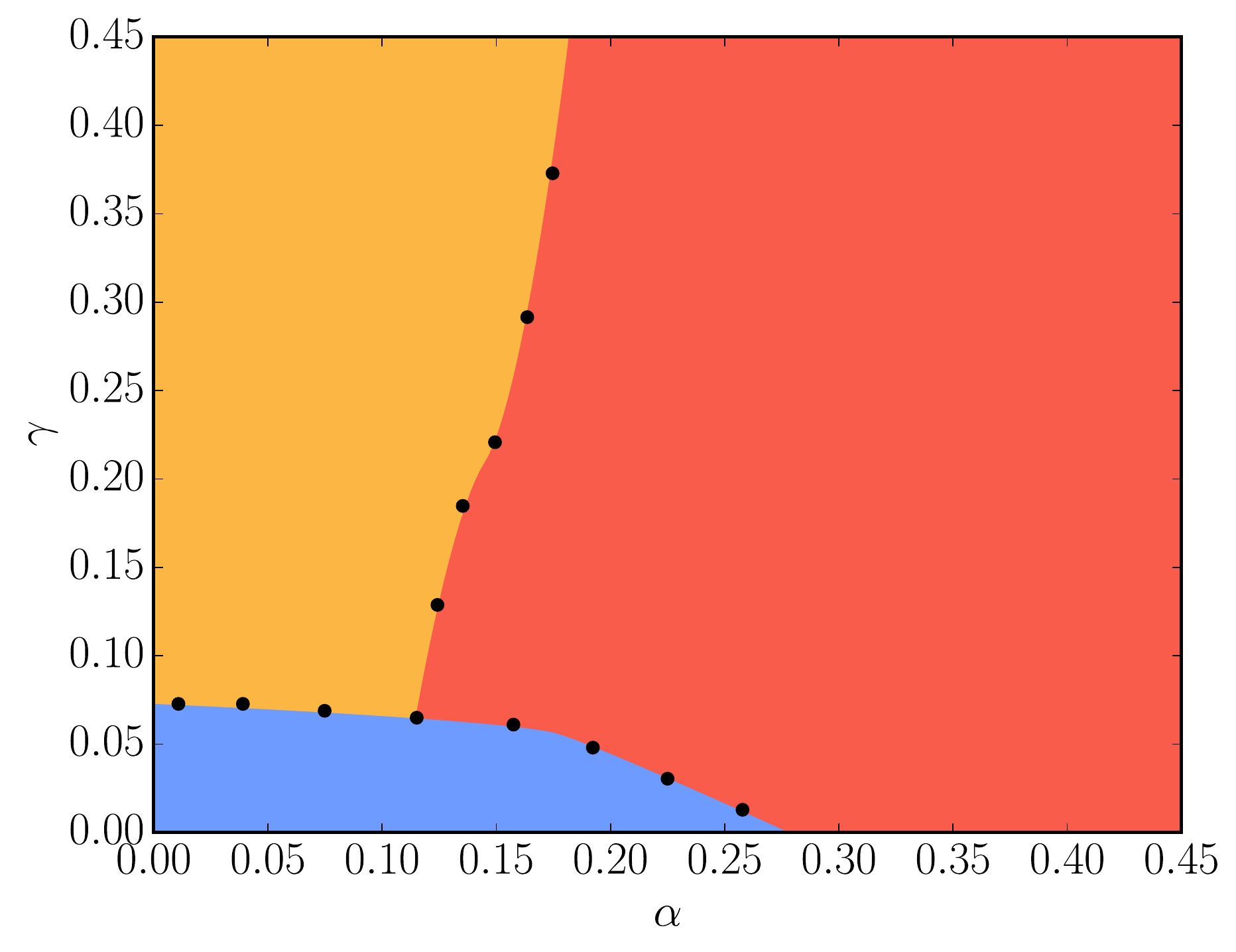}
	\includegraphics[scale = 1.25]{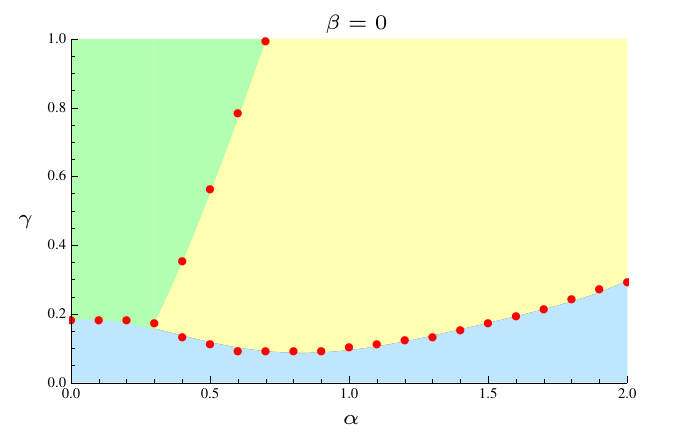}
	\caption{On the left a phase diagrams using the $E$--function parametrization with $\beta=0$. The blue phase (bottom-left) is associated to the BF fixed point, the red one phase (right) corresponds to the high temperature phase, while the orange one (top-left) is a BF phase  with respect to  ${\mathcal S}_3 / \mathbb{Z}_3 \simeq \mathbb{Z}_2$. On the right a corresponding phase diagram for 2D spin nets, reprinted from \protect\cite{merce}.}
	\label{fig:diagram2}
\end{figure}

In this section we present the first numerical results for the coarse--graining of 3D spin foam models for the symmetric group $\mathcal{S}_3$.  We explored the flow of a range of models encoded in the phase diagrams in figure \ref{fig:diagram1} and \ref{fig:diagram2}. In the parametrization \eqref{Param1} of the $E$--function via parameters $(\kappa,\alpha,\beta,\gamma)$ we choose to look at two two--dimensional planes given by $\kappa=1, \alpha=\beta$  and $\kappa=1,\beta=0$ respectively. That is we consider the $E$--functions
\begin{align}
	E_{\mathbb{Z}_2 }(g) &= \delta(g,e) + \alpha \big( \delta(g,a) +  \delta(g,bab^{-1})+\delta(g,b^2ab^{-2})\big)+\gamma \big( \delta(g,b)+\delta(g,b^2)\big)  \label{phaseE1}\\	
	E_{\mathbb{Z}_2}(g) &= \delta(g,e) + \alpha \, \delta(g,a) +\gamma \big( \delta(g,b)+\delta(g,b^2)\big).  \label{phaseE2}
\end{align} 
The face weights, as defined in \eqref{BF5}, are fixed to $\omega(\rho)=d_\rho$, the same as for the BF partition function. Thus we change in the BF partition function the Haar projector with $\mathbb{P}'_{e^*}$ which is determined by the $E$--functions through (\ref{Efct2}). Note that for $\alpha=\beta$ , {\it i.e.} \eqref{phaseE1} the $E$--function is invariant under conjugation of the full group (and not only $\mathbb{Z}_2$). In this case the exchange of the Haar projector with $\mathbb{P}'_{e^*}$ can also be absorbed into a change of face weights \cite{holonomy1}. This family of models does therefore define 'standard' lattice gauge theories. (Note that the  $C^{[2,2,2,2]}$ in (\ref{Cmatrix1}) is diagonal in this case.)  In contrast the family (\ref{phaseE2}) of models allows for non--trivial simplicity constraints, {\it e.g.} can in general not be rewritten into standard lattice gauge form.

Let us first focus on the phase diagram presented  in figure \ref{fig:diagram1}, displaying the case \eqref{phaseE1}. The phase diagram encodes to which fixed points the models are flowing to in the following way. Each point of the phase diagram corresponds to a pair of parameters $(\alpha=\beta, \gamma)$. These parameters define the initial amplitude of the coarse--graining procedure, namely the first blue cube. For this parametrization the system flows towards a given fixed point. Depending on this fixed point, we colour the point with coordinates $(\alpha = \beta, \gamma)$ of the phase diagram in the corresponding colour.  
There are three phases  associated to three distinct fixed points. More precisely, all the points of the phase diagram which share the same colour correspond to parameters for which the system flows towards the same fixed point. First there is the ordered $\mathcal{S}_3$ phase which corresponds to  BF theory with $\mathcal{S}_3$ group. In terms of representations, this is the fixed point for which all three irreducible representations are allowed. Then there is the strong coupling (or in analogy to the Ising model `unordered' or `high temperature')  fixed point for which only the trivial representation is allowed.  Finally the third phase  corresponds to the strong coupling phase with respect to a normal subgroup of $\mathcal{S}_3$, which is $\mathbb{Z}_3$. Alternatively it can be understood as BF theory  with respect to $\mathcal{S}_3/\mathbb{Z}_3 \simeq \mathbb{Z}_2$. Thus only the trivial and the sign representations, which are also representations of $\mathbb{Z}_2$, are allowed. 
Notice the presence of a triple point.


\begin{figure}[h]
	\centering
	\begin{minipage}[b]{1\textwidth}
		\centering
		\includegraphics[scale = 0.40]{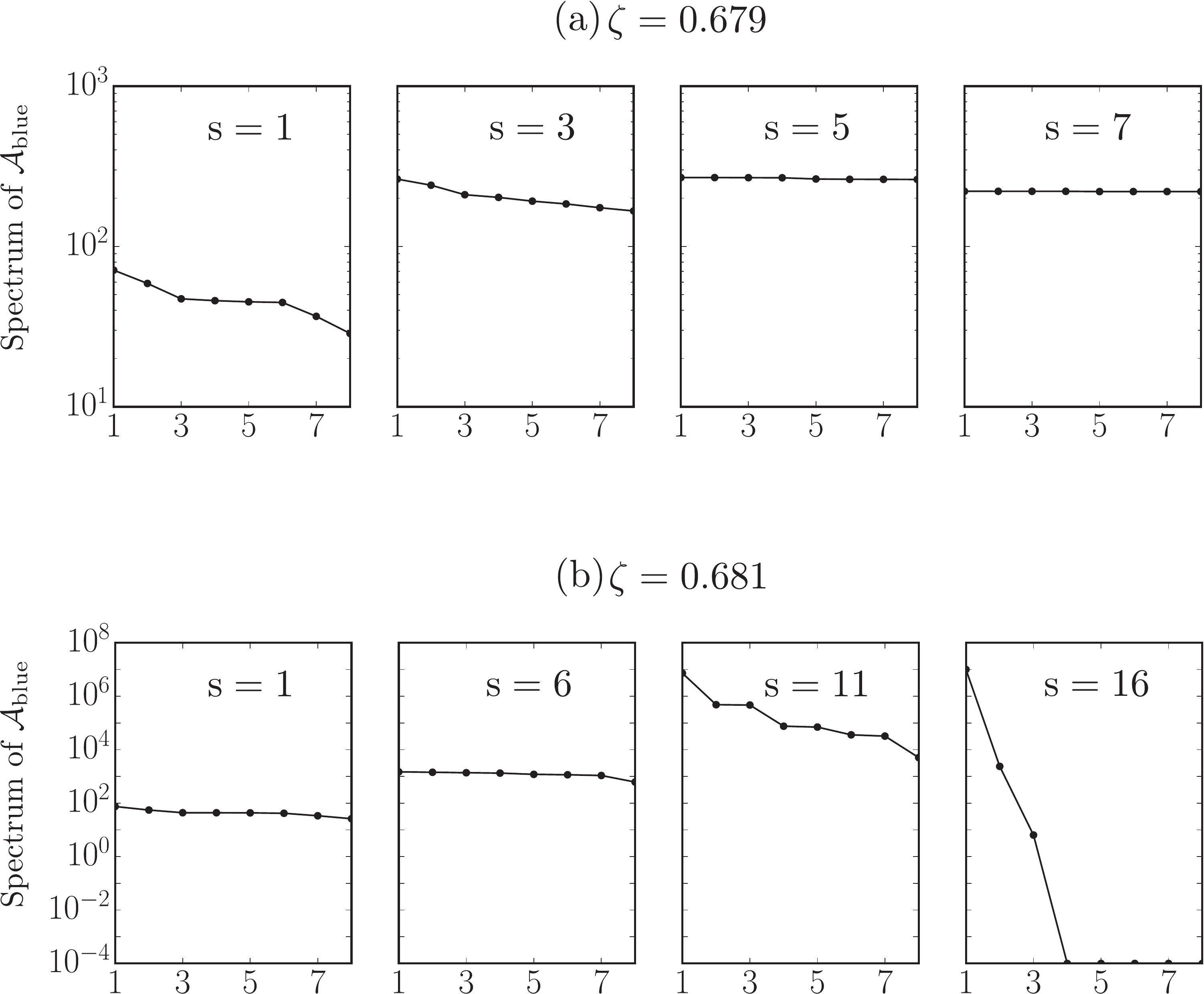}	
	\end{minipage}
	\caption{Partial spectrum of the blue cube amplitude $\mathcal{A}_{\text{blue}}$ for two different values of the parameter $\zeta$; only the singular values associated to the trivial and the sign representations (associated to the `shared' leaves and appearing as parameters for the SVD)  are displayed. The number of iterations is denoted by s. This shows an example of indication of phase transition between two points which respectively flow towards the BF fixed point (degenerate singular values) and the high temperature fixed point (only the configuration with the trivial representation  everywhere is non-vanishing). \label{spect}}
\end{figure}
This phase diagram is quite similar to the one found with spin nets in \cite{merce}, using the same parametrization via $E$--functions. The same three phases were found and also a triple point was present. 

Figure \ref{fig:diagram2} shows the phase diagram for the models described by (\ref{phaseE2}), which can in general not be rewritten into standard lattice gauge theory form. Again we have the same three phases appearing\footnote{In the case of 2D spin nets an additional fixed point was found along a phase transition line, that actually required fine tuning, and only appeared for certain (lower) bond dimensions. It is therefore most likely an artifact. Nevertheless it described a triangulation invariant theory for the spin net case. However, in the interpretation of spin nets as two spin foam (dual) vertices connected by many dual edges \cite{qgroup}, this fixed point described a complete decoupling of the two spin foam vertices. It is therefore not possible to define a spin foam version of this fixed point. We indeed found no sign of such fixed points in our investigations.}, and the form of the phase diagram is similar to the one found for 2D spin nets in \cite{merce}. A difference is, that the BF phase gives way to the high-temperature phase for $\gamma = 0$ as we increase $\alpha$, while it would always flow towards the BF fixed point in the case of 2D spin nets.

We also considered a family of models interpolating between the spin foam model \eqref{nontriv}, which satisfies the projector condition, and BF theory:
\ba
\kappa = 1+\zeta \big( \frac{2^{3/4}}{3^{5/4}}-1\big)\q ,\q 
\alpha = \zeta \frac{2^{3/4}}{3^{5/4}} \q , \q 
\beta = -\frac{\zeta}{3\times 6^{1/4}} \q , \q 
\gamma = -\frac{\zeta}{3 \times 6^{1/4}}
\ea
The BF fixed point is obtained for $\zeta=0$ whereas \eqref{nontriv} is reached for $\zeta=1$. The model \eqref{nontriv} flows to the strong coupling fixed point. (Changing the face weights to $\omega(\rho)=d_\rho^p$, with some parameter $p$, does not affect this fixed point. This is presumably due to the fact that there is only one representation with dimension larger than one.)  Varying the parameter $\zeta$ we change to the BF phase. The phase transition occurs around the value $\zeta=0.680$ and shows some behaviour typical of second order transitions, {\it i.e.} an increase in the number of iterations needed to converge to a fixed point.  Figure \ref{spect} shows some subset of the singular values that appear during the coarse--graining process, for configurations near the phase transition.  More reliable results about the phase transitions, which could confirm the existence of a conformal theory and determine its properties, could be obtained by implementing more involved algorithms, in particular by including an entanglement filtering process \cite{vidal-evenbly, evenbly2, looptnr}.  These still need to be developed however for higher dimensional (gauge) theories.

\section{Discussion}\label{discussion}

In this work we presented the first tensor network algorithm for coarse--graining 3D lattice gauge models with a non--Abelian structure group. This algorithm is based on decorated tensor networks, which are generalizations of tensor networks. These have been proposed in \cite{decorated} to obtain a more efficient algorithm to deal with (there Abelian) gauge models. This work has again shown that the decorated tensor networks are a flexible tool, allowing the encoding of different kind of boundary data for the building blocks.  This flexibility will also allow in future work to impose further truncations \cite{bahrsteinhausCubulations} or incorporate analytical techniques for evaluating the gluing of building blocks \cite{Jeff,simone}.

We have tested this Decorated Tensor Network algorithm on models with structure group ${\cal S}_3$ and have obtained phase diagrams. These encode the large scale behaviour of the models and their dependence on the chosen parametrization. We have confirmed the similarity of the phase diagrams of 3D spin foams with the 2D spin nets.  That also means that we have identified three phases, the strong coupling\footnote{This phase is in a gravitational context also known as Ashtekar--Lewandowski phase, due to its connection to the Ashtekar--Lewandowski vacuum. It describes a state peaked sharply on totally degenerate geometry.}  phase, in which only the trivial representation contributes to the partition function, the BF (over ${\cal S}_3$) phase, describing a topological theory, and the BF(${\mathbb Z}_2$) phase.  These phases can be all described within `standard' lattice gauge theories.  The BF(${\mathbb Z}_2$)  might be describe as carrying some notion of simplicity constraints as the two--dimensional representation does not contribute to the partition function. Thus we have not identified new phases, but this might be due to the small gauge group considered here.  (In the case of spin nets with  ${\cal S}_3$ structure group investigations did not show additional phases, but new phases have been identified in the case of  quantum group spin nets \cite{qgroup}.)

We have also tested a `proper' spin foam model, with a non--trivial (proper) projector replacing the Haar--projector, appearing in lattice gauge theories. This model itself  flows to the strong coupling fixed point.  Thus one can say that the simplicity constraints in this case are too strongly implemented: too many configurations are suppressed, so that the model flows to a fixed point, in which only one configuration contributes.  We defined a one--parameter family of models connecting this model to a BF model. We encountered a phase transition by varying this parameter, indicating the potential of a   continuum limit with propagating degrees of freedom. We can take this as evidence that the spin foam construction principle -- the implementation of simplicity constraints, starting from topological BF theory -- can lead to theories with propagating degrees of freedom. 

As we discussed, the non--Abelian structure group leads to a challenge for coarse--graining, as torsion (or charge) degrees of freedom might emerge which violate (the original form of) gauge invariance. Thus the method presented here, which projects onto gauge invariant amplitudes, is suited only for models in which one does not expect the emergence of such torsion degrees of freedom. This includes Yang Mills lattice gauge theories (due to the confinement conjecture, stating that on large scales free charges do not appear) and also gravitational theories for which one expects a vacuum with vanishing curvature. 

A case for which such an approach would fail is however general relativity with cosmological constant. For instance in \cite{improved1} it was shown on the classical level, that a  model with initially flat building blocks but with the action for gravity with cosmological constant, flows to a model with effective building blocks which are homogeneously curved. To show a corresponding flow for the quantum case is an outstanding task, but highly interesting as the quantization of 3D gravity with a cosmological constant is described by a quantum deformation of the underlying structure group \cite{turaev-viro, qgroupmodels1}.  However such a `condensation' of curvature would be also accompanied by a `condensation' of torsion degrees of freedom.  Similarly, for 4D gravity (without a cosmological constant), it is not clear what role curvature should play for the vacuum state. There are however many physical scenari in which curvature condensation is needed.

Thus we need an enlargement of the space of models considered so far, and also a corresponding enlargement of the (gauge invariant) Hilbert spaces attached to the boundary of the building blocks. Some suggestions have been made in \cite{eteraCG1}. Another approach replaces the spin network basis with the fusion basis \cite{fusbas,Delcamp:2016lux}, which has an inbuilt notion of coarse--graining and also allows for torsion degrees of freedom. It moreover also exist for quantum groups \cite{KKR, DGtqft}. A key feature of this fusion basis is a non--local organization of the degrees of freedom into finer and coarser ones. The challenge is to implement this feature into a (decorated) tensor network algorithm, which to some extend relies on a local gluing. Additionally one would like to also include a (short range) entanglement filtering procedure \cite{vidal-evenbly,evenbly2,looptnr}, in which the non--local features of the fusion basis can be actually helpful. 

Having developed a tensor network coarse--graining algorithm for 3D non--Abelian gauge models, the next step is to consider the 4D case.  This will require even more computational resources than in 3D. Thus designing an efficient algorithm will be essential.  The question is also whether one can implement more truncations -- although we have just discussed that one might have to consider torsion degrees of freedom and therefore a bigger model space. \cite{bahrsteinhausCubulations, BahrSteinhausPRL} implements  a very drastic but geometrically motivated truncation to models without curvature. (Thus avoiding to some degree the problem with effective violations of gauge invariance.) This reduces the summations in the gluing procedure of building blocks to zero (only matching conditions need to be satisfied), the only summations that do appear are when implementing the (pre--chosen) embedding maps. With this truncation it is possible to treat the full models. But a contribution from the action, and thus complex weights are avoided. Therefore only the face weights are tested, whose choice is encoded in one parameter. The coarse--graining flow is projected back on this one parameter and indicates a phase transition, at which a residual notion of diffeomorphism symmetry is restored.  These are encouraging results. One would wish however for a justification of these truncations from the dynamics of the system, in particular since the various truncations, including the truncation during the coarse--graining procedure, are not informed by the dynamics.  Decorated tensor network methods are a tool to do so and could also allow for a systematic test and relaxations of the truncations.

\acknowledgements
\noindent
BD thanks Asger Ipsen for participation at a early stage of the project. CD thanks Jonah Miller and Erik Schnetter for suggestions about the numerical implementation. The authors thank Benjamin Bahr, Marc Geiller, Sebastian Mizera, Sebastian Steinhaus and Guifre Vidal for discussions.
CD is supported by an NSERC grant awarded to BD.
This research was supported in part by Perimeter Institute for Theoretical Physics. Research at Perimeter Institute is supported by the Government of Canada through the Department of Innovation, Science and Economic Development Canada and by the Province of Ontario through the Ministry of Research, Innovation and Science.

\appendix
\section{Singular Value Decomposition \label{app_svd}}
For a real or complex $m \times n$ matrix $M$ with $m>n$, the Singular Value Decomposition (SVD) is a factorization of the form $U\Sigma V^{\ast}$ where $U$ and $V$ are unitary matrices of size $m\times n$ and $n \times n$ respectively, and $\Sigma$ a $n \times n$ diagonal matrix of non-negative real numbers. Making the indices explicit, the SVD of the matrix $M$ reads
\be
M_{AB} = \sum_{i=1}^{n} U_{A,i}\Sigma_{i}V_{B,i}^{\dagger}
\ee
where $\Sigma_i \equiv \Sigma_{ii}$. 

The entries of the matrix $\Sigma$ are called the `singular values' of the matrix $M$ and they are ordered from the largest one to the smallest one. We can define a `truncated singular value decomposition' by keeping only the $\chi$ largest singular values:
\be
M_{AB} \approx \sum_{i=1}^{\chi} U_{A,i}\Sigma_{i}V_{B,i}^{\dagger}
= \sum_{i=1}^{\chi}(U_{A,i}\sqrt{\Sigma_i})(\sqrt{\Sigma_i}V^{\dagger}_{B,i})
\equiv \sum_{i=1}^{\chi}(S_1)_{A,i}(S_2)_{B,i}
\label{truncated}
\ee
where we equally distributed the contribution from the singular values between the matrices $U$ and $V$. According to the Eckart-Young theorem \cite{Eckart1936} the result of \eqref{truncated} minimizes the Frobenius norm of the difference between $M$ and a matrix of rank $\chi$.

\section{Group Fourier transform \label{app_fourier}}

In this appendix, we provide additional details about the interplay between the different representations used throughout the algorithms, namely the `group' and the `Fourier transformed' representations. 

According to Peter-Weyl theorem, the Hilbert space $L^2(\mathcal{G})$ of square-integrable functions over $\mathcal{G}$ decomposes onto the orthogonal direct sum of irreducible representations. In particular for every function $\psi(g)\in L^2(\mathcal{G})$ we have the following relations
\be
\label{Ftrans}
\begin{dcases}
	\psi(g) = \sum_{\rho,m,n}\sqrt{d_{\rho}}D^{\rho}_{mn}(g)\widetilde{\psi}(\rho,m,n) \\
	\widetilde{\psi}(\rho,m,n) = \frac{1}{|\mathcal{{G}}|}\sum_{g}\sqrt{d_{\rho}}\, \overline{D^{\rho}_{mn}(g)}\psi(g)
\end{dcases}\; ,
\ee
which relies on the orthogonality of the irreducible representations matrices provided by
\be
\frac{1}{|\mathcal{G}|}\sum_g D^{\rho_1}_{m_1n_1}(g)\overline{D^{\rho_2}_{m_2n_2}(g)} = 
\frac{\delta_{\rho_1\rho_2}}{d_{\rho_1}}\delta_{m_1m_2}\delta_{n_1n_2}\; .
\label{ortho}
\ee
Formulae \eqref{Ftrans} define a notion of `group Fourier transform' for functions of $L^2(\mathcal{G})$ allowing to switch between the `group' representation and the `Fourier transformed' representation. For such a transformation rule, the Plancherel theorem applies and reads
\be
\frac{1}{|\mathcal{G}|}\sum_g \psi (g) \overline{\phi(g)} = \sum_{\rho,m,n}\widetilde{\psi}(\rho,m,n)\overline{\widetilde{\phi}(\rho,m,n)}
\label{Plancherel}
\ee
which follows from both the definition of the Fourier transform \eqref{Ftrans} and the orthogonality relation \eqref{ortho}. This relation defines the inner product for $L^2(\mathcal{{G}})$. It also corresponds to the operation performed when gluing the boundary spin networks of the cubes. The requirement we impose about the colour of the faces which are identified during the gluing is to ensure that both a representation matrix and its conjugate appears. By doing so, the contraction over the labels $(\rho,m,n)$ in the `Fourier transformed' representation corresponds to an inner product in the `group' representation via \eqref{Plancherel}.

Let us now consider a gauge invariant functional $\psi(\{g\})$ which depends on several group variables associated to a graph (without open links). We define the spin network basis as
\be
\text{SNW}\big(\{g\},\{\rho, \mathcal{I}\}\big)= \sum_{\{m,n\}}\Big(\prod_{e\,:\,\text{edges}}\sqrt{d_{\rho_e}}D^{\rho_e}_{m_en_e}(g_e)\Big)\mathcal{I}^{\{\rho\}}_{\{m,n\}} \q .
\label{SNW}
\ee
Here $\mathcal{I}^{\{\rho\}}_{\{m,n\}}$ are intertwiner associated to nodes of the graph and for the tensor product of links adjacent to these nodes. 
These intertwiners  can be decomposed into Wigner--$3\rho m$ symbols, which corresponds to a expansion of each node to a set of three--valent nodes. Here we assume a multiplicity free group for which the three--valent intertwiners are unique. In this case the intertwiners can be labelled also by representations (associated to the new links appearing in the expansion of the nodes to three--valent ones). We will therefore absorb the intertwiners $\{{\cal I}\}$ into the representation labels $\{\rho\}$. Equivalently we can also assume a three--valent graph in the following. 
The gauge invariant functional can then be expressed in the spin network basis as follows
\be
\psi(\{g\}) = \sum_{\{\rho\}}\widetilde{\psi}(\{\rho\})\,\text{SNW}\big(\{g\},\{\rho\}\big)
\label{switch}
\ee
which again defines a notion of Fourier transform such that the `Fourier transformed' representation is provided by
\be
\widetilde{\psi}(\{\rho\}) = \frac{1}{|\mathcal{G}|^{\#_{\rm edges}}} \sum_{\{g\}}\psi(\{g\})\,\overline{\text{SNW}\big(\{g\},\{\rho\}\big)} \; .
\ee
Let us present explicitly the case of the BF amplitude. For a spherical surface, the BF amplitude in the `group' representation can be represented by the functional
\be
\psi_{\text{BF}}(\{g\}) = \prod_{\ell \, : \, \text{loops}}\delta(g_{\ell})
\label{BFamp}
\ee
where $\{g_{\ell}\}$ denote the loop holonomies associated with a set of independent cycles of the underlying graph. The Dirac delta is defined by
\be
\delta (g_f) = \sum_{\rho}d_{\rho} \chi^{\rho}(g_f)
\ee
with $\chi_{\rho}$ the character of the irreducible representation $\rho$. The expansion of \eqref{BFamp} in the spin network basis gives the coefficients
\be
\widetilde{\psi}_{\text{BF}}(\{\rho\}) = \text{SNW}\big(\mathds{1},\{\rho\} \big)
\label{BFspin}
\ee
which corresponds to the evaluation of the spin network state where all the group elements are set to the identity. Equation \eqref{BFspin} defines the `Fourier transformed' representation of the BF amplitude.

\section{Parameterization of spin foam models \label{app_param}}
Starting from the Haar projector, simplicity constrains can be implemented by defining a new projector which projects onto a smaller subspace than the gauge invariant subspace. We will now see precisely how such a projector is defined and how we construct the corresponding spin foam amplitude.

~\\
{\bf {\small Definition of a new projector:}}\\
The Haar projector $\mathbb{P}_{e^{
		\ast}}$ projects onto the invariant subspace $\text{Inv}(V_{\rho_1}\otimes \dots \otimes V_{\rho_4})$ where $\{V_{\rho_i}\}$ are the representation spaces 
associated with the faces meeting at the edge $e^{\ast}$. Diagonalizing the projector onto the intertwiner basis, we obtain a formal expression of the form
\be
\big(\mathbb{P}_{e^{\ast}}\big)^{\{ n_{f^\ast} \}_{f^\ast \supset e^\ast}}_{\{ m_{f^\ast} \}_{f^\ast \supset e^\ast}} = \sum_{\iota} {^{\{ n_{f^\ast} \}_{f^\ast \supset e^\ast}}}| \iota \rangle \langle \iota |_{\{ m_{f^\ast} \}_{f^\ast \supset e^\ast}} \; , 
\ee
with $\{ | \iota \rangle \}$ a basis for the space $\text{Inv} \big( \bigotimes _{f^\ast \supset e^\ast}\big)$. Since the intertwiners are 4--valent, they can be decomposed into two 3--valent intertwiners given by the Wigner--$3\rho m$ symbols. In order to make the dependence on the representation labels explicit, we introduce the following intertwiner states
\be
| \rho_5 \rangle = 
\mathcal{I}
{ \footnotesize \begin{bmatrix}
		\rho_1, \cdots, \rho_4 \\ m_1, \cdots,  m_4	
\end{bmatrix}}(\rho_5)
\equiv  \sum_{m_5} 
{ \footnotesize \begin{pmatrix}
		\rho_1 & \rho_2 & \rho_5^* \\ m_1 & m_2 & m_5	
\end{pmatrix}}
{ \footnotesize \begin{pmatrix}
		\rho_3 & \rho_4 & \rho_5 \\ m_3 & m_4 & m_5	
\end{pmatrix}}
\ee
where $\rho^\ast$ denotes the representation contragradient to $\rho$. Using this notation, we rewrite the Haar projector
\be
\mathbb{P}_{e^\ast} = \sum_{\rho_5}
\mathcal{I}
{ \footnotesize \begin{bmatrix}
		\rho_1, \cdots, \rho_4 \\ m_1, \cdots,  m_4	
\end{bmatrix}}(\rho_5)
\, \overline{\mathcal{I}
	{ \footnotesize \begin{bmatrix}
			\rho_1, \cdots, \rho_4 \\ n_1, \cdots,  n_4	
	\end{bmatrix}}(\rho_5)}
\ee
which can be graphically represented as follows

\begin{align}
\mathbb{P}_{e^\ast} = 	\begin{array}{c} \includegraphics[scale = 0.85]{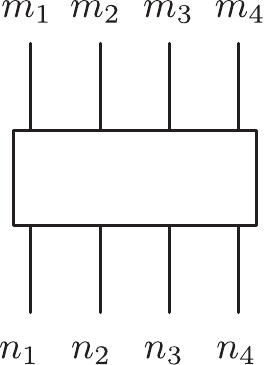}\end{array} = \sum_{\rho_5}
\begin{array}{c} \includegraphics[scale = 0.85]{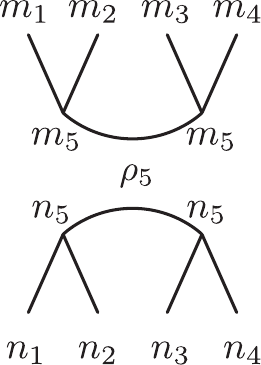}\end{array}
\end{align}
We are now looking for an invariant map $\mathbb{P}_{e^\ast}'$ which takes the following form
\be
\mathbb{P}_{e^\ast}' = \mathbb{P}_{e^\ast} \cdot \big( \widetilde{E}(\rho_1) \otimes \cdots \otimes \widetilde{E}(\rho_4) \big) \cdot \mathbb{P}_{e^\ast}
\ee
Making the magnetic indices explicit, we obtain
\begin{align}
\mathbb{P}_{e^\ast}'\big( \{\rho_a,m_a,n_a \}\big) &= \mathbb{P}_{e^\ast}\big( \{\rho_a,m_a,m_a' \}\big)\Big( \prod_{a}\widetilde{E}(\rho_a,m_a',n_a')\Big) \mathbb{P}_{e^\ast}\big( \{\rho_a,n_a',n_a \}\big) \\
&=\sum_{\rho_5,\rho_5'} \mathcal{I}
{ \footnotesize \begin{bmatrix}
	\rho_1, \cdots, \rho_4 \\ m_1, \cdots,  m_4	
	\end{bmatrix}}(\rho_5)
\, \overline{\mathcal{I}
	{ \footnotesize \begin{bmatrix}
		\rho_1, \cdots, \rho_4 \\ m_1', \cdots, m_4'	
		\end{bmatrix}}(\rho_5)} \\
&\q \q \q \times \Big( \prod_{a}\widetilde{E}(\rho_a,m_a',n_a')\Big)
\mathcal{I}{ \footnotesize \begin{bmatrix}
	\rho_1, \cdots, \rho_4 \\ n_1', \cdots,  n_4'	
	\end{bmatrix}}(\rho_5')
\, \overline{\mathcal{I}
	{ \footnotesize \begin{bmatrix}
		\rho_1, \cdots, \rho_4 \\ n_1, \cdots,  n_4	
		\end{bmatrix}}(\rho_5')} \\
&= \sum_{\rho_5,\rho_5'} \mathcal{I}
{ \footnotesize \begin{bmatrix}
	\rho_1, \cdots, \rho_4 \\ m_1, \cdots,  m_4	
	\end{bmatrix}}(\rho_5)
C^{[\rho_1, \dots,\rho_4]}(\rho_5,\rho_5')
\overline{\mathcal{I}
	{ \footnotesize \begin{bmatrix}
		\rho_1, \cdots, \rho_4 \\ n_1, \cdots,  n_4	
		\end{bmatrix}}(\rho_5')}
\end{align}
where
\be
C^{[ \rho_1,\dots,\rho_4]}(\rho_5,\rho_5') = \sum_{\{m,n\}} 
\overline{\mathcal{I}
	{ \footnotesize \begin{bmatrix}
			\rho_1, \cdots, \rho_4 \\ m_1, \cdots,  m_4	
	\end{bmatrix}}(\rho_5)}
\Big( \prod_{a}\widetilde{E}(\rho_a,m_a,n_a)\Big)
\mathcal{I}
{ \footnotesize \begin{bmatrix}
		\rho_1, \cdots, \rho_4 \\ n_1, \cdots,  n_4	
\end{bmatrix}}(\rho_5')\; .
\ee
As explained in the main text, non-diagonal matrices $C$ encode the implementation of simplicity constraints. On the contrary, if $C(\rho_5,\rho_5') = \delta_{\rho_5,\rho_5'}$, we recover the initial Haar projector. Note that in order for $\mathbb{P}_{e^\ast}'$ to be a projector, we require the matrix $C(\rho_5,\rho_5')$ to be idempotent. Furthermore, we can always find an alternative basis of intertwiners for which the matrices $C$ are diagonal. We will now construct such a basis. 

~\\
{\bf {\small Definition of a new basis of intertwiners:}}\\
Keeping the magnetic indices implicit, the new projector takes the simple form

\be
\mathbb{P}_{e^\ast}' = \sum_{\rho_5,\rho_5'}|\rho_5 \rangle C(\rho_5,\rho_5') \langle \rho_5' | \; .
\ee
The diagonalization of the matrix $C$ reads
\be
\widetilde{C}(i,i') = \sum_{\rho_5,\rho_5'} U(i,\rho_5)C(\rho_5,\rho_5')U^{\dagger}(\rho_5',i') \equiv \lambda_i \delta_{i,i'}
\ee
from which we deduce the expression
\be
C(\rho_5,\rho_5') = \sum_{i,i'}U^{\dagger}(\rho_5,i)\widetilde{C}(i,i')U(i',\rho_5').
\ee
Therefore we can rewrite the projector
\begin{align}
\mathbb{P}_{e^\ast}' &= \sum_{\rho_5,\rho_5' \atop i,i'} |\rho_5 \rangle U^{\dagger}(\rho_5,i)\widetilde{C}(i,i')U(i',\rho_5') \langle \rho_5' | \\
&= \sum_{\rho_5,\rho_5',i} | \rho_5 \rangle U^{\dagger}(\rho_5,i)\lambda_i U(i',\rho_5) \langle \rho_5 | 	\\
&= \sum_{\rho_5}\Big( \sum_{i,\rho_5' \atop i',\rho_5''} \sqrt{\lambda_i'}| \rho_5' \rangle U^{\dagger}(\rho_5',i')U(i',\rho_5)U^{\dagger}(\rho_5,i)U(i,\rho_5'')\langle \rho_5'' | \sqrt{\lambda_i} \Big) \\	
&\equiv \sum_{\rho_5}| \tilde{\rho}_5 \rangle \langle \tilde{\rho}_5 |
\end{align}
where 
\be
|\tilde{\rho}_5 \rangle = \sum_{i,\rho_5'} \sqrt{\lambda_i} |\rho_5' \rangle U^{\dagger}(\rho_5',i)U(i,\rho_5) = \widetilde{\mathcal{I}}
{ \footnotesize \begin{bmatrix}
		\rho_1, \cdots, \rho_4 \\ m_1, \cdots,  m_4	
\end{bmatrix}}(\rho_5) \; . 
\ee
This new basis of `deformed' intertwiners can be used to construct the spin foam amplitude in the spin representation.

~\\
{\bf {\small Boundary spin network state:}}\\
The definition of the spin foam amplitude in the `Fourier transformed' representation boils down to a contraction of `deformed' intertwiners, the contraction pattern depending on the underlying graph. The corresponding functional in the `group' representation can then be obtained according to the generalized Fourier transform \eqref{switch}. The first step therefore consists in defining the boundary spin network state using the `underformed' intertwiners. Since we eventually deal with gauge fixed amplitudes, we actually need to define the gauge fixed boundary spin network state.

\begin{figure}[h]
	\centering
	\includegraphics[scale = 0.7]{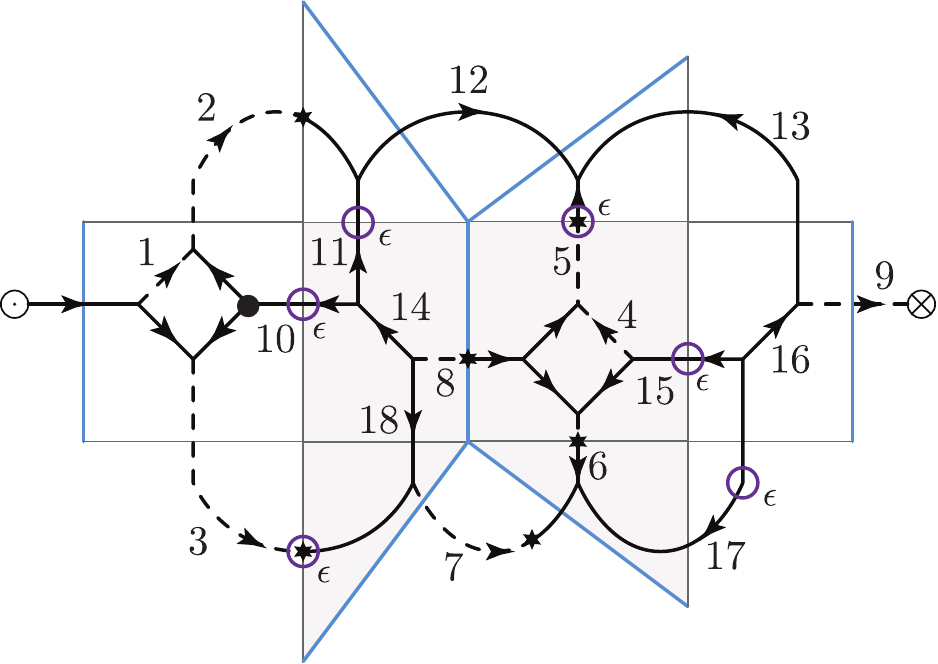}
	\caption{The parametrization of the system requires the explicit evaluation of  the boundary spin network of the initial blue cube. All the edges are labeled and we denote by a $\epsilon$ the edges whose end-points are not of the same colour. The spin network amplitude associated with such edges involves an $\epsilon$-matrix.  }
	\label{blue2}
\end{figure}

The amplitude we will construct is the one associated to the initial `blue' cube of the algorithm. Therefore the underlying graph of the spin network state is presented figure \ref{blue2}. 
The general formula for the spin network state is given by \eqref{SNW}, however there are two subtleties. Firstly, one has two conventions for the Fourier transform depending on the colour of the face. Similarly, there are two conventions for the ordering of the links meeting at a node. We choose anti-clockwise for white faces and clockwise for grey faces. In order to make the influence of the orientation of the links explicit, we need to introduce a few additional definitions. To do so, we will focus on the case of $\SU(2)$. In the next appendix, we give more details about the symmetric group $\mathcal{S}_3$ which we formulate so as to be as similar to $\SU(2)$ as possible. We consider the invariant group element
\be
\epsilon = \begin{pmatrix} 0 & 1 \\ -1 & 0\end{pmatrix} \q , \q \epsilon^{-1} = \epsilon^t = \epsilon^\dagger = - \epsilon 
\ee
which is such that  $ [\;\epsilon\;]^{\rho}_{mn} \equiv D^\rho_{mn}(\epsilon) = (-1)^{\rho-m}\delta_{m,-n}$ and $ [\;\epsilon^{-1}\;]^{\rho}_{mn} \equiv D^\rho_{mn}(\epsilon^{-1}) = (-1)^{\rho+m}\delta_{m,-n}$. This tensor defines an intertwining map and realises a unitary transformation between the matrix representation $D^\rho$ and its contragradient $(D^\rho)^*$ 
\begin{align} \nn
D^\rho_{mm'}(g^{-1}) &= \overline{D^\rho_{m'm}(g)} \\ &= \big[ [\; \epsilon \;]^\rho D^\rho(g) [\; \epsilon^{-1} \; ]^\rho \big]_{m'm}
\end{align}
Therefore the intertwiner associated to a node with two incoming links $l_1,l_2$ and one outgoing link $l_3$ is given by
\be
\label{def-dual}
{ \footnotesize \begin{pmatrix}
		\rho_1^* & \rho_2^* & \rho_3 \\ m_1 & m_2 & n_3	
\end{pmatrix}}
\equiv \sum_{m_1',m_2'} D^{\rho_1}_{m_1m_1'}D^{\rho_2}_{m_2m_2'}(\epsilon)
{ \footnotesize \begin{pmatrix}
		\rho_1 & \rho_2 & \rho_3 \\ m_1' & m_2' & n_3	
\end{pmatrix}}
\ee
Using the ordering convention stated above we have for instance the following correspondence for a white face
\be
{ \footnotesize \begin{pmatrix}
		\rho_2 & \rho_1^* & \rho_3^* \\ n_1 & m_2 & m_3	
\end{pmatrix} }\quad \longleftrightarrow \quad
\begin{array}{c}
	\includegraphics[scale =0.7]{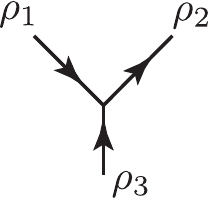}
\end{array}
\ee

Finally we have to take into account the fact that some leaves connect two faces of different colour. In that case the two half--links of the leaf will not transform according to the same definition. In such a case the expression for the corresponding spin network amplitude involves an $\epsilon$ matrix. On figure \ref{blue2} we have labeled with a $\includegraphics[scale =0.9]{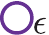}$ the edges for which this difficulty appears. We distinguish two situations depending on the orientation of the edge. From a white face to a grey face, we obtain
\be \nn
\begin{array}{c}
	\includegraphics[scale =0.85]{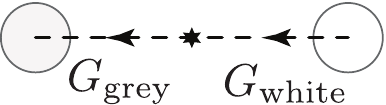}
\end{array}
\ee
\be
\Psi \big( G_{\text{white}},G_{\text{grey}}\big) = \sqrt{d_\rho}\,D^\rho_{mn'}(G_{\text{white}}) [\;\epsilon\;]^{\rho}_{n' m'} \overline{D^\rho_{m' n}(G_{\text{grey}})}
\ee
and similarly from a grey face to a white face
\be \nn
\begin{array}{c}
	\includegraphics[scale =0.85]{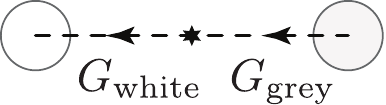}
\end{array}
\ee
\be
\Psi \big( G_{\text{grey}},G_{\text{white}}\big) = \sqrt{d_\rho}\, \overline{D^\rho_{mn'}(G_{\text{white}})} [\;\epsilon^{-1}\;]^{\rho}_{n' m'}D^\rho_{m' n}(G_{\text{grey}}).
\ee
Putting everything together the gauge fixed spin network amplitude is given by
\begin{align} 
\text{SNW}_{\Gamma}\big( \{\rho\}, \{ G_{\ell}\}\big)_{|\text{g.f.}} = \sum_{\{m\},\{n\} \atop \{\widetilde{m}\}} 
& D^{\rho_2}_{m_2n_2}(G_2)\,D^{\rho_3}_{\widetilde{m}_3n_3}(G_3)\,D^{\rho_3}_{\widetilde{m}_5n_5}(G_5)\,D^{\rho_6}_{m_6n_6}(G_6)\\[-1em] \nn
\times & \overline{D^{\rho_7}_{m_7n_7}(G_7)}\, \overline{D^{\rho_8}_{m_8n_8}(G_8)}\,D^{\rho_9}_{m_9n_9}(G_9) \\[0.4em] \nn 
\times & [\;\epsilon\;]^{\rho_1}_{m_1 \widetilde{m}_1}\, [\;\epsilon\;]^{\rho_2}_{m_2 \widetilde{m}_2}\, [\;\epsilon\;]^{\rho_4}_{m_4 \widetilde{m}_4}\, [\;\epsilon\;]^{\rho_6}_{m_6 \widetilde{m}_6}\, [\;\epsilon\;]^{\rho_7}_{m_7 \widetilde{m}_7}\, [\;\epsilon\;]^{\rho_8}_{m_8 \widetilde{m}_8}\, [\;\epsilon\;]^{\rho_9}_{m_9 \widetilde{m}_9}\\[-0.2em]
\nn \times &[\;\epsilon\;]^{\rho_{12}}_{m_{12} \widetilde{m}_{12}}\, [\;\epsilon\;]^{\rho_{13}}_{m_{13} \widetilde{m}_{13}}\, 
[\;\epsilon\;]^{\rho_{14}}_{m_{14} \widetilde{m}_{14}}\, [\;\epsilon\;]^{\rho_{16}}_{m_{16} \widetilde{m}_{16}}\, [\;\epsilon\;]^{\rho_{18}}_{m_{18} \widetilde{m}_{18}} \prod_{\rho=1}^{18}\sqrt{d_\rho}\\ \nn
\times &	{ \footnotesize\begin{pmatrix}
	\rho_1 & \rho_2 & \rho_9 \\ \widetilde{m}_1 & n_2 & \widetilde{m}_9
	\end{pmatrix} }
{ \footnotesize\begin{pmatrix}
	\rho_{11} & \rho_{12} & \rho_{2} \\ \widetilde{m}_{11} & m_{12} & \widetilde{m}_2
	\end{pmatrix} }
{ \footnotesize\begin{pmatrix}
	\rho_{3} & \rho_{18} & \rho_{7} \\ \widetilde{m}_3 & \widetilde{m}_{18} & n_7
	\end{pmatrix} }
{ \footnotesize\begin{pmatrix}
	\rho_{4} & \rho_{5} & \rho_{15} \\ \widetilde{m}_4 & n_5 & \widetilde{m}_{15}
	\end{pmatrix} } \\ \nn 
\times &{ \footnotesize\begin{pmatrix}
	\rho_{5} & \rho_{13} & \rho_{12} \\ \widetilde{m}_5 & \widetilde{m}_{13} & \widetilde{m}_{12}
	\end{pmatrix} } 
{ \footnotesize\begin{pmatrix}
	\rho_{7} & \rho_{6} & \rho_{17} \\ \widetilde{m}_7 & \widetilde{m}_6 & \widetilde{m}_{12}
	\end{pmatrix} } 
{ \footnotesize\begin{pmatrix}
	\rho_{6} & \rho_{8} & \rho_{4} \\ n_6 & \widetilde{m}_8 & m_4
	\end{pmatrix} }
{ \footnotesize\begin{pmatrix}
	\rho_{10} & \rho_{1} & \rho_{3} \\ \widetilde{m}_{10} & m_1 & n_3
	\end{pmatrix} } \\ \nn
\times &{ \footnotesize\begin{pmatrix}
	\rho_{14} & \rho_{10} & \rho_{11} \\ \widetilde{m}_{14} & \widetilde{m}_{10} & \widetilde{m}_{11}
	\end{pmatrix} } 
{ \footnotesize\begin{pmatrix}
	\rho_{17} & \rho_{16} & \rho_{15} \\ \widetilde{m}_{17} & m_{16} & \widetilde{m}_{15}
	\end{pmatrix} }
{ \footnotesize\begin{pmatrix}
	\rho_{9} & \rho_{13} & \rho_{16} \\ n_9 & m_{13} & \widetilde{m}_{16}
	\end{pmatrix} }
{ \footnotesize\begin{pmatrix}
	\rho_{8} & \rho_{18} & \rho_{14} \\ n_8 & m_{18} & m_{14}
	\end{pmatrix} }\; .
\end{align}

~\\
{\bf {\small Spin foam amplitude:}}\\
The spin foam amplitude is provided by the contraction of the `deformed' intertwiners following the same pattern and conventions as for the spin network state:
\begin{align}
\widetilde{\psi}_{\text{SF}}\big( \{\rho\} \big) = \sum_{\{m\},\{n\} \atop \{\widetilde{m}\}} 
& [\;\epsilon\;]^{\rho_2}_{m_2 \widetilde{m}_2}\,  [\;\epsilon\;]^{\rho_6}_{m_6 	\widetilde{m}_6}\,  [\;\epsilon\;]^{\rho_8}_{m_8 \widetilde{m}_8}\, [\;\epsilon\;]^{\rho_9}_{m_9 \widetilde{m}_9}\, [\;\epsilon\;]^{\rho_{13}}_{m_{13} \widetilde{m}_{13}} \,
[\;\epsilon\;]^{\rho_{18}}_{m_{18} \widetilde{m}_{18}}\prod_{\rho=1}^{18}\sqrt{d_\rho} \\ \nn
\times 
&\widetilde{\mathcal{I}}
{ \footnotesize \begin{bmatrix}
	\rho_2\,,\, \rho_9\,,\,\rho_3\,,\,\rho_{10} \\ n_2,\widetilde{m}_9,m_3,  m_{10}	
	\end{bmatrix}}(\rho_1)
\widetilde{\mathcal{I}}
{ \footnotesize \begin{bmatrix}
	\rho_{10}\,,\, \rho_{11}\,,\,\rho_8\,,\,\rho_{18} \\ m_{10},m_{11},n_8,  m_{18}	
	\end{bmatrix}}(\rho_{14})
\widetilde{\mathcal{I}}
{ \footnotesize \begin{bmatrix}
	\rho_5\,,\, \rho_{15}\,,\,\rho_6\,,\,\rho_{8} \\ m_5,m_{15},n_6,  \widetilde{m}_{8}	
	\end{bmatrix}}(\rho_4) \\ \nn ~\\ \nn
\times &\widetilde{\mathcal{I}}
{ \footnotesize \begin{bmatrix}
	\rho_9\,,\, \rho_{13}\,,\,\rho_{15}\,,\,\rho_{17} \\ n_9,m_{13},m_{15},  m_{17}	
	\end{bmatrix}}(\rho_{16})
\widetilde{\mathcal{I}}
{ \footnotesize \begin{bmatrix}
	\rho_5\,,\, \rho_{13}\,,\,\rho_2\,,\,\rho_{11} \\ m_5,\widetilde{m}_{13},\widetilde{m}_2,  m_{11}	
	\end{bmatrix}}(\rho_{12})
\widetilde{\mathcal{I}}
{ \footnotesize \begin{bmatrix}
	\rho_3\,,\, \rho_{18}\,,\,\rho_6\,,\,\rho_{17} \\ m_3,\widetilde{m}_{18},\widetilde{m}_6,  m_{17}	
	\end{bmatrix}}(\rho_7) \; .
\end{align}
The group representation of the spin foam amplitude in the gauge fixed basis is finally deduced by identifying and summing over the irreducible representations according to equation (\ref{switch}). The result of this last operation is the initial amplitude for the blue cube denoted by $\mathcal{A}_{\text{blue}}(\{G\})$.

\section{Recoupling theory of $\mathcal{S}_3$}
In this appendix, we briefly summarise the recoupling theory of the symmetric group $\mathcal{S}_3$. The group $\mathcal{S}_3$ is generated by two elements denoted $a$ and $b$, respectively a reflection of order 2 and a rotation by ${2\pi/3}$ of order 3, which satisfy the defining relation
\be
a^2 = b^3 = (ab)^2 = 1.
\ee
The group contains six elements and their expression in terms of $a$ amd $b$ is $\{ e,a,bab^{-1},b^2ab^{-2},b,b^2\}$. The group possesses three irreducible representations. We denote $\rho=0$ the trivial representation, $\rho=1$ the sign representation and $\rho=2$ the two-dimensional standard representation. 

We want the recoupling theory of $\mathcal{S}_3$ to be as close as possible as to the one of $\SU(2)$. In order to do  we define Clebsch-Gordan coefficients which mimic the ones of the Lie group $\SU(2)$. Therefore we make a choice of basis in which the standard representation matrix of $b$ is diagonal. In this basis, the standard representation of the group elements $a$ and $b$ is
\be
D^2(a) = { \footnotesize\begin{pmatrix} 
		0 & 1 \\ 1 & 0
\end{pmatrix}} \q , \q 
D^2(b) = { \footnotesize\begin{pmatrix} 
		\text{exp}(\frac{2\pi i}{3}) & 0 \\ 0 & \text{exp}(-\frac{2\pi i}{3})
\end{pmatrix}}.
\ee 
The representations for each group element are summarized in the following table
\be
\begin{tabular}{l*{6}{c}r}
	$D^\rho(g)$              & $e$ & $a$ & $bab^{-1}$ & $b^2ab^{-2}$ & $b$  &$ b^2$ \\
	\hline
	$\rho=0$		 & 1 & 1 & 1 & 1 & 1 & 1   \\
	$\rho=1$            & 1 & -1 & -1 & -1 &  1 & 1   \\
	$\rho=2$            & ${ \footnotesize\begin{pmatrix} 1 & 0 \\ 0 & 1\end{pmatrix}}$ & 
	${ \footnotesize\begin{pmatrix} 0 & 1 \\ 1 & 0\end{pmatrix}}$ &
	${ \footnotesize\begin{pmatrix} 0 & \overline{\omega} \\ \omega & 0\end{pmatrix}}$ &
	${ \footnotesize\begin{pmatrix} 0 & \omega \\ \overline{\omega} & 0\end{pmatrix}}$ &
	${ \footnotesize\begin{pmatrix} \omega & 0 \\ 0 & \overline{\omega}\end{pmatrix}}$ &
	${ \footnotesize\begin{pmatrix} \overline{\omega} & 0 \\ 0 & \omega\end{pmatrix}}$ \\
\end{tabular}
\ee
where $\omega = \text{exp}(\frac{2\pi i}{3})$. Furthermore, the invariant map $\epsilon$ which performs the transformation between a representation and its contragradient is identified with the group element $a$. 
We can then define the Clebsch-Gordan coefficients as follows
\be
| \rho_3,m_3 \rangle = \sum_{m_1,m_2}C^{\rho_1\rho_2\rho_3}_{m_1m_2m_3}|\rho_1,m_1\rangle|\rho_2,m_2\rangle
\ee
Let us for instance look at the tensor product $2 \otimes 1 \simeq 2$. Denoting $(1 \; 0)^t$ as $| m= +1 \rangle$ and $(0 \; 1)^t$ as $| m= -1 \rangle$, we have the following relations
\begin{align}
&D^{2 \otimes 1}(b)|m=+1\rangle = \text{exp}(\tfrac{2\pi i }{3}) |m=+1\rangle \\
&D^{2 \otimes 1}(b)|m=-1\rangle = \text{exp}(\tfrac{-2\pi i }{3}) |m=-1\rangle \\
&D^{2 \otimes 1}(a)|m=+1\rangle = - |m=-1\rangle \\
&D^{2 \otimes 1}(a)|m=-1\rangle = - |m=+1\rangle .
\end{align}
We can summarise the equations above into
\be
D^{2 \otimes 1}(\, . \,) = 
{ \footnotesize\begin{pmatrix} 
		1 & 0 \\ 0 & -1
\end{pmatrix}}
D^2(\, . \,)
{ \footnotesize\begin{pmatrix} 
		1 & 0 \\ 0 & -1
\end{pmatrix}}
\ee
The choice of transformation matrix made here is motivated by the wish to obtain 3$jm$-symbols invariant under cyclic permutation. With this choice, the Clebsch-Gordan coefficients are given by
\begin{align}
&C^{122}_{011} = -C^{212}_{101} = +1 \\
&C^{122}_{0-1-1} = -C^{121}_{-10-1} = -1
\end{align}
Similarly we deduce the Clebsch-Gordan coefficients associated with the tensor products $1 \otimes 1 = 0$, $1 \otimes 0 = 1$ and $2 \otimes 2 = 0 \oplus 1 \oplus 2$.
We finally define the invariant tensors according to the rules
\be
{ \footnotesize\begin{pmatrix}
		\rho_1&\rho_2 & \rho_3 \\ m_1 & m_2 & m_3
\end{pmatrix}} = 
\begin{cases}
	C^{\rho_1\rho_2\rho_3}_{m_1m_2m_3} \q \text{if} \q \rho_3 = 0 \\[0.5em]
	-C^{\rho_1\rho_2\rho_3}_{m_1m_2m_3} \q  \text{if} \q  \rho_3= 1 \\[0.5em]
	\frac{1}{\sqrt{d_{\rho_3}}}C^{\rho_1\rho_2\rho_3}_{m_1m_2-m_3} \q \text{if} \q \rho_3 = 2
\end{cases}
\ee
such that the non-vanishing ones are given by
\begin{align*}
&{ \footnotesize\begin{pmatrix}
	0 & 0 & 0 \\0 & 0 & 0
	\end{pmatrix} } = 1 \q, &  \q
&{ \footnotesize\begin{pmatrix}
	1 & 2 & 2 \\ 0 & 1 & -1
	\end{pmatrix} } = \frac{1}{\sqrt{2}} \q,  &  \q
&{ \footnotesize\begin{pmatrix}
	2 & 1 & 2 \\ 1 & 0 & -1
	\end{pmatrix} } = -\frac{1}{\sqrt{2}} \q,  &  \q
&{ \footnotesize\begin{pmatrix}
	2 & 1 & 2 \\ -1 & 0 & 1
	\end{pmatrix} } = \frac{1}{\sqrt{2}} \\
&{ \footnotesize\begin{pmatrix}
	2 & 2 & 1 \\ -1 & 1 & 0
	\end{pmatrix} } = -\frac{1}{\sqrt{2}} \q,  &  \q
&{ \footnotesize\begin{pmatrix}
	2 & 2 & 1 \\ 1 & -1 & 0
	\end{pmatrix} } = \frac{1}{\sqrt{2}} \q,  &  \q
&{ \footnotesize\begin{pmatrix}
	1 & 2 & 2 \\ 0 & -1 & 1
	\end{pmatrix} } = -\frac{1}{\sqrt{2}} \q,  &  \q
&{ \footnotesize\begin{pmatrix}
	1 & 1 & 0 \\ 0 & 0 & 0
	\end{pmatrix} } = -1  \\
&{ \footnotesize\begin{pmatrix}
	1 & 0 & 1 \\ 0 & 0 & 0
	\end{pmatrix} } = -1 \q,  &  \q
&{ \footnotesize\begin{pmatrix}
	0 & 1 & 1 \\ 0 & 0 & 0
	\end{pmatrix} } = -1 \q,  &  \q
&{ \footnotesize\begin{pmatrix}
	2 & 2 & 0 \\ 1 & -1 & 0
	\end{pmatrix} } = \frac{1}{\sqrt{2}} \q,  &  \q
&{ \footnotesize\begin{pmatrix}
	2 & 0 & 2 \\ -1 & 0 & 1 
	\end{pmatrix} } = \frac{1}{\sqrt{2}} \\
&{ \footnotesize\begin{pmatrix}
	0 & 2 & 2 \\ 0 & 1 & -1
	\end{pmatrix} } = \frac{1}{\sqrt{2}} \q,  &  \q
&{ \footnotesize\begin{pmatrix}
	2 & 0 & 2 \\ 1 & 0 & -1
	\end{pmatrix} } = \frac{1}{\sqrt{2}} \q,  &  \q
&{ \footnotesize\begin{pmatrix}
	0 & 2 & 2 \\ 0 & -1 & 1
	\end{pmatrix} } = \frac{1}{\sqrt{2}} \q,  &  \q
&{ \footnotesize\begin{pmatrix}
	2 & 2 & 2 \\ 1 & 1 & 1
	\end{pmatrix} } = \frac{1}{\sqrt{2}} \\
&{ \footnotesize\begin{pmatrix}
	2 & 2 & 2 \\ -1 & -1 & -1
	\end{pmatrix} } = \frac{1}{\sqrt{2}} \; .
\end{align*}

\bibliography{biblio}

\end{document}